%% file: ttp98-41.tex
\documentclass[11pt,twoside]{article}
\usepackage{epsf,feynmp}
\usepackage[nosort]{cite}
\input{header}

\begin{document}
\input{title}
\tableofcontents
\input{introduction}
\input{tools}
\input{algebraic}
\input{implement_intro}

\input{survey}

\input{selected_intro}

\input{matmin}

\input{explmp}

\input{geficom}

\input{feynarts}

\input{comphep}

\input{applications_intro}

\input{polfunc}
\input{zbb}

\input{beta}

\input{hgg}
\input{strongW}

\input{delta_r}
\input{acknowledge}

\end{document}

%% file: header.tex
\newcommand{\captionwidth}{14cm}

\newcommand{\agt}{\,\rlap{\lower 3.5 pt \hbox{$\mathchar \sim$}} \raise 1pt
 \hbox {$>$}\,}
\newcommand{\alt}{\,\rlap{\lower 3.5 pt \hbox{$\mathchar \sim$}} \raise 1pt
 \hbox {$<$}\,}

\newcommand{\gsim}{\;\rlap{\lower 3.5 pt \hbox{$\mathchar \sim$}} \raise 1pt
 \hbox {$>$}\;}
\newcommand{\lsim}{\;\rlap{\lower 3.5 pt \hbox{$\mathchar \sim$}} \raise 1pt
 \hbox {$<$}\;}
\newcommand{\dd}{{\rm d}}                  
\newcommand{\bld}[1]{\boldmath{$#1$}}      
\newcommand{\sprod}[2]{#1\hspace{-.1em}\cdot\hspace{-.1em}#2}   
\newcommand{\llangle}{\left\langle}        
\newcommand{\rrangle}{\right\rangle}

\renewcommand{\Im}{{\rm Im}}
\newcommand{\msbar}{$\overline{\rm MS}$}

\newcommand{\logqmms}{l_{qm}}

\newcommand{\logqmums}{l_{q\mu}}
\newcommand{\logmum}{l_{\mu m}}

%% file: title.tex
\title{
  \vspace{1em}
  \begin{flushright}
    {\bf\normalsize
    TTP98-41, BUTP-98/28, hep-ph/9812357}\\[-.5em]
    {\bf\normalsize December 1998}\\[1em]
  \end{flushright}
  \Large \sc
  Automatic Computation of Feynman Diagrams
  }

\author{{\sc R. Harlander}$^a$ and {\sc M. Steinhauser}$^b$
  \\[3em]
  {\it a) Institut f\"ur Theoretische Teilchenphysik,
      Universit\"at Karlsruhe, D-76128 Karlsruhe, Germany}
  \\[.5em]
  {\it b) Institut f\"ur Theoretische Physik, Universit\"at
      Bern, CH-3012 Bern, Switzerland}
}

\date{}
\maketitle

\begin{abstract} 
  \noindent
  Quantum corrections significantly influence the quantities observed in
  modern particle physics. The corresponding theoretical computations
  are usually quite lengthy which makes their automation mandatory.
  This review reports on the current status of automatic calculation
  of Feynman diagrams in particle physics.  The most important
  theoretical techniques are introduced and their usefulness is
  demonstrated with the help of simple examples. A survey over
  frequently used programs and packages is provided, discussing their
  abilities and fields of applications.  Subsequently, some powerful
  packages which have already been applied to important physical
  problems are described in more detail. The review closes with the
  discussion of a few typical applications for the automated computation
  of Feynman diagrams, addressing current physical questions like
  properties of the $Z$ and Higgs boson, four-loop corrections to
  renormalization group functions and two-loop electroweak corrections.
\end{abstract}

\begin{center}
  {\bf\small Keywords}\\
  {\small Feynman diagrams, computer algebra, calculational techniques,
  radiative corrections, quantum chromodynamics, electroweak physics}
\end{center}

%% file: introduction.tex
\section{Introduction}
\subsection{Quantum field theory in particle physics}
%
The huge amount of data collected in the recent years mainly at the
Large Electron Positron Collider (LEP) at CERN (Geneva) and the
Stanford Linear Collider (SLC)
as well as at the proton--anti-proton machine TEVATRON 
at FERMILAB (Chicago) has
resulted in an impressive experimental precision for many parameters in
particle physics. For example, the mass and width of the $Z$ boson are
measured with an accuracy which is of the same order of magnitude as
for Fermi's constant~\cite{Kar98,Hol98}.
Therefore it is essential to also improve the
accuracy of predictions following from theoretical models. It is the
purpose of this article to give an overview of some of the technical
tools to perform the corresponding calculations.

At present, the most favoured theoretical model in particle physics is
the so-called Standard Model. It incorporates all known quarks and
leptons and describes their interactions by the gauge group
SU(3)$\times$SU(2)$\times$U(1). Here, the SU(3) generates the strong
interactions among the quarks by the exchange of eight massless gluons,
while the SU(2)$\times$U(1), usually called the electroweak sector,
comprises electromagnetic and weak interactions, carried by the
massless photon and the massive charged $W^\pm$ and neutral $Z$ boson,
respectively.

Up to now all theoretical predictions deduced from the Standard Model
have found perfect confirmation in experiment. For example, the
existence of the top quark was predicted after the discovery
of the $\Upsilon$-resonances in 1977 \cite{Herb77} merely out of the
need for an SU(2) doublet partner for the bottom quark. At least equally
impressive, however, was the accurate prediction of its mass, $M_t$, by
comparing precision measurements of certain observables at LEP and SLC
to their theoretical values. The top quark mass dependence in these
observables only appears as a higher order effect through the so-called
radiative corrections, typical for quantum field theories. So when the
first evidence of the discovery of the top quark was announced by the
CDF and D0 collaborations~\cite{TOP95}, the agreement of directly and
indirectly measured values for the top mass served as a very convincing
argument.  Today statistics have improved considerably and the original
signal has been impressively confirmed~\cite{CDFD098}.

A similar situation exists with the Higgs boson, at the present time
being the only particle of the Standard Model not yet discovered. Its
existence, however, is by no means guaranteed as was the top quark after
$b$ discovery.  Furthermore, predictions for the Higgs mass $M_H$ suffer
from much larger uncertainties because the dependence of the relevant
radiative corrections on $M_H$ is only logarithmic~\cite{Vel77} while it
is quadratic on $M_t$.  Nevertheless, it is possible to set the limits
to $M_H=76^{+85}_{-47}$~GeV with an upper bound of $M_H< 262$~GeV at
$95\%$ confidence level~\cite{Teu98}.  The current lower limit for $M_H$
from the direct search at LEP reads $M_H\agt 89.8$~GeV~\cite{Mh_LEP}.

From the theoretical point of view the Standard Model does not look like
a ``final'' theory. First of all, it contains a lot of free parameters
and any attempt to predict their values needs a generalization of the
model.  Furthermore, there is no way to find unification of the strong,
weak and electromagnetic coupling constant within the Standard Model.  A
very promising and theoretically attractive extension is the Minimal
Supersymmetric Standard Model (MSSM) which not only allows for gauge
unification but even solves the problem of quadratic divergences and
many others. Phenomenological calculations, however, are even more
difficult here since the particle content is approximately doubled.
Moreover, some important technical tools are no longer applicable.

Let us now turn to the question how to extract physical information like
cross sections and decay rates from the Standard Model, leading to
predictions like the values for $M_t$ and $M_H$ mentioned above. At the
moment the vast majority of the calculations are based on perturbation
theory, represented by the extremely depictive Feynman diagrams.
Together with the Feynman rules, which constitute the translation rules
from the graphical to a mathematical notation, they are in complete
one-to-one correspondence to the individual terms of the perturbative
series. As long as one is interested only in perturbative results, one
may formulate the theory in terms of the Feynman rules alone, i.e., by
providing the list of possible vertices, describing particle
interactions, and propagators. Any physical process may then be
evaluated by connecting the vertices and propagators in all possible
ways to give the desired initial and final states.  The specific order
in perturbation theory is reflected in the number of vertices of the
corresponding Feynman diagram. Given a fixed initial state and
increasing the number of vertices leads to the appearance either of
closed loops or of additional particles in the final state. In the
former case, this implies an integration over the corresponding loop
momentum, while the latter case results in a more complicated phase
space integral when calculating the rate. Usually, diagrams with more
than one closed loop are called multi-loop, those with more than two
particles (``legs'') in the final state are called multi-leg diagrams.
The technology, in particular the automation of the computations in the
multi-loop sector is much further developed than for multi-leg diagrams.
Furthermore, to a certain extent it is possible to relate multi-leg
diagrams to multi-loop ones via the optical theorem. Thus our main focus
in this article lies in the calculation of loop diagrams.

\subsection{Motivation for automatic computation of Feynman diagrams}
%
The first realization of the idea to pass purely algebraic or, more
precisely, symbolic operations to a computer indeed was driven by
particle physics, when in 1967 M.~Veltman developed the program {\tt
  SCHOONSCHIP} \cite{VelSS}, mainly to control the evaluation of fermion
traces, i.e.\ traces of gamma matrices. Today, {\tt SCHOONSCHIP} to a
large extent is superceded by other systems which have, however, taken
over many of its basic ideas.  Some important examples from the early
days that also have their roots in particle physics are {\tt
  MACSYMA}~\cite{macsyma}, {\tt REDUCE}~\cite{reduce} and others. The by
now highly developed algebraic system {\tt Mathematica}~\cite{Wolfram}
actually is a derivative of the program {\tt SMP}, created in 1980 by
the particle physicist S.~Wolfram.  While some of these programs claim
to be fairly general, the most direct descendant of {\tt SCHOONSCHIP} is
{\tt FORM}~\cite{form}, a system mainly tailored to high energy physics.
In Section~\ref{secalgprg} a few representatives of the most important
algebraic systems will be introduced.  The main concern of this review
are not these algebraic programs, but rather their applications, i.e.,
packages based on these systems.  It should be clear, however, that they
are very important prerequisites for the automation of Feynman diagram
calculations.

The higher the order of perturbation theory under consideration, the
larger is the number of contributing Feynman diagrams and the more
complicated is their individual evaluation. This makes it desirable and
even unavoidable to develop efficient algorithms which can be
implemented on a computer in a simple way.  Therefore, large efforts in
the field of Feynman diagram evaluation are devoted to find algorithms
for the specific types of operations arising in a typical calculation.
One may characterize these operations by analyzing the steps necessary
for a perturbative field theoretic calculation: given a set of Feynman
rules, first the contributing Feynman diagrams for the process under
consideration have to be generated.  This, being mainly a question of
combinatorics, is clearly a well suited problem for computerization, and
meanwhile very effective algorithms have been implemented with emphasis
on slightly diverse purposes. The next step is the evaluation of the
corresponding mathematical expressions for the diagrams which can be
divided into an algebraic and an analytic part. The algebraic part
consists of operations like contracting Lorentz indices and calculating
fermion traces.  Nowadays many algebraic systems provide functions
optimized for this kind of manipulations.

The analytic part of a Feynman diagram calculation is concerned with the
Feynman integrals, and usually this is the point where computers fail to
be applicable without human intervention.  However, Feynman integrals
establish a very special class of integrals, and for huge subclasses
algorithms exist that {\em algebraically} reduce each of them to basic
sets of meanwhile tabulated expressions.  Two of the most important
techniques of this kind are the so-called tensor
reduction-~\cite{PasVel79} and the integration-by-parts
algorithm~\cite{CheTka81}.  The first one reduces any Lorentz structure
in the integrand to invariants and has been worked out explicitly at the
one-loop level for the general case and up to two loops for propagator
diagrams. Using the integration-by-parts algorithm, on the other hand,
recurrence relations can be derived which express diagrams with
non-trivial topologies through simpler ones at the cost of increasing
their number and the degree of the denominators.  The practical
realization of these algorithms would be unfeasible in realistic
calculations without the use of powerful computer systems.

From the considerations of the previous paragraphs one can see that as
long as one is interested in the {\it exact} evaluation of Feynman
diagrams, the number of problems that can completely be passed to
computers is rather limited which is mainly due to the relatively small
number of exactly solvable Feynman integrals. There are, however, many
processes where some kind of approximate result may be equally helpful.
This becomes clear by recalling that, on the one hand, working at fixed
order in perturbation theory is an approximation anyway, and, on the
other hand, experimental results are not free of uncertainties
either.
Therefore, it is enormously important to have approximation procedures
for Feynman diagrams. For instance, the numerical evaluation of
complicated Feynman integrals, which is a typical task for a computer,
is a possibility to obtain results with finite accuracy.
This approach is currently favoured for the automation of tree- and one-loop
calculations involving lots of different mass parameters like in the
Standard Model or even the MSSM. For recent developments in the
numerical evaluation of Feynman diagrams beyond the one-loop level
we refer to~\cite{numcalc}.

A different kind of approximation may be obtained by expanding the
integrals with respect to ratios of different scale parameters. Here the
so-called asymptotic expansions of Feynman diagrams are becoming more
and more popular, since they allow any multi-scale diagram to be reduced
to single-scale ones, at least in principle. While the evaluation of the
former may be a hard job even at one-loop level, the latter ones can be
computed up to three loops using the integration-by-parts algorithm
mentioned before.  The required efforts for a manual application of
asymptotic expansions increase steeply with the number of loops.
Meanwhile, however, the most important variants have been implemented
and have also been applied to the calculation of three-loop
contributions to several important physical observables.

Performing huge calculations with computers inevitably leads to seemingly
unrelated difficulties of purely organisational nature: Computers
nowadays are not yet as stable as one would like them to be. In this way
they force the user to cover on possible breakdowns, keeping the loss
minimal. In the case where the long run-time is mainly due to a huge
number of diagrams, each of them taking only a few minutes to be
evaluated, the solution to this problem is rather straightforward.
Saving each diagram on disk after computation naturally guarantees that
the loss will be of only a few minutes (apart from more severe problems
like disk crashes or the like).  For more complicated diagrams, however,
it is helpful if the algebra program itself is structured in a way that
intermediate results will not be lost after system crashes.

All the algorithms mentioned above allow the calculation of Feynman
diagrams to be automated to more or less high degree, where one will
certainly choose different combinations of them for different purposes.
It is quite difficult to give an exhaustive definition of automation.
For instance, often the computer system either applies only for one
special problem (or a very restricted class of problems), nevertheless
saving the user a great deal of effort and preventing otherwise unavoidable
errors. In other cases the system is rather flexible but requires a
certain amount of human intervention for each specific problem.  The
final goal of automation where a typical screen snapshot would look like
\begin{verbatim}
  model: SM 
  loops: 3 
  initial-state: ep,em 
  final-state: t,tb 
  parameters: default 
  observable: sigma_tot 
  energy: 500 GeV

    sigma = 0.8 pb

\end{verbatim}
is at the moment still out of range.
Nevertheless, many of the recently performed calculations in high energy
physics would never have been possible without a high degree of
automation. Let us, at the end of this introduction, list a few of these
applications that will be addressed in more detail in
Section~\ref{secapplications}.

\subsection{Examples for physical applications}
%
Most of the multi-loop and multi-leg computations have been performed
for the QCD sector of the Standard Model. On the one hand they are
numerically most important simply because the coupling constant is
relatively large. On the other hand they are simpler to evaluate due to
the fact that very often only one dimensionful scale appears like the
external momentum or a large internal mass.

The ``classical'' example which is currently known to order $\alpha_s^3$
are QCD corrections to the hadronic R ratio,
$R(s)=\sigma(e^+e^-\to\mbox{hadrons})/\sigma(e^+e^-\to\mu^+\mu^-)$.
Already in 1979 ${\cal O}(\alpha_s^2)$ corrections have been
computed~\cite{CheKatTka79DinSap79CelGon80} in the massless limit.  A
large part of the calculations was carried out without using computers
and lots of efforts had to be spent on nowadays trivial things like
fermion traces.  It took more than ten years until the next order was
completed~\cite{GorKatLar91SurSam91,Che97R}.  Even then the manipulation
of the diagrams was largely done ``by hand''. As a consequence,
important checks like gauge parameter independence could not be
performed.  Only recently the ${\cal O}(\alpha_s^3)$ result was checked
by a completely independent calculation which used the powerful methods
of automatic generation and computation of diagrams~\cite{Che97R}.

The full mass dependence of the ${\cal O}(\alpha_s^2)$ corrections to
$R(s)$ has become available~\cite{CheKueSte96} only very recently, when
high moments of the polarization function were evaluated.  In this
calculation the problem was not in the number of diagrams which is less
than 20, but in the huge expressions occurring in intermediate steps.  (For
some diagrams up to four giga-bytes of disk space were required.)  For
such computations it is indispensible that enormously powerful algebra
systems are available and that their use is completely automated to
avoid any possible human error in intermediate steps.

The number of diagrams to be considered for the ${\cal O}(\alpha_s^3)$
corrections to $R(s)$ at $m=0$ is of the order of one hundred and thus
it is still possible to organize the calculation by hand.  The situation
is different for the decay of the Higgs boson into gluons, for example.
Among other reasons this is a very important process because the inverse
reaction of gluon fusion is the dominant production mechanism of Higgs
bosons at the future Large Hadron Collider (LHC). Since the leading
order QCD corrections to the loop-induced process $H\to gg$ amount to
roughly $68\%$~\cite{InaKubOka83,DjoSpiZer91}, it was necessary to
evaluate the next-to-leading order contribution. However, the number of
three-loop diagrams to be computed in this case is of the order one
thousand.  This in combination with the large expressions in the
intermediate calculational steps is the reason why one has to rely on a
high degree of automation as will be described in
Section~\ref{sec:applic:hgg}.

An even larger number of diagrams had to be dealt with when the
renormalization group functions $\beta$ and $\gamma_m$, governing the
running of the strong coupling constant and the quark masses,
respectively, were computed at four-loop level.  Roughly $50\,000$
diagrams for $\beta$~\cite{RitVerLar97}
and $2\,000$ for 
$\gamma_m$~\cite{Che97VerLarRit97}
contributed, and it is
quite clear that such calculations would never have been possible
without the intensive use of extremely efficient computer algebra
systems in combination with administrative software concerned with
book-keeping.

So far, none of the examples above really had to rely on one of the
approximation procedures like numerical integration or asymptotic
expansion mentioned above. This becomes relevant for the decay of the
$Z$ boson into quarks at ${\cal O}(\alpha\alpha_s)$, for example. Here,
the bottom quark channel is of special interest since the top quark
appears as a virtual particle.  With the present technology, a
calculation of the full $M_t$-dependence is out of reach because up to
four different mass scales are involved.  However, asymptotic expansions
provide a very promising method to get a result which is almost
equivalent to the full answer. In the approach which was used to tackle
the ${\cal O}(\alpha\alpha_s)$ terms to this process~\cite{HarSeiSte97},
the 69 contributing diagrams are split into 234 subdiagrams.  The manual
application of the method of asymptotic expansions was therefore
impossible and the existence of program packages that apply this
procedure automatically was important.

Meanwhile, experimental accuracy allows the electroweak sector of the
Standard Model to be tested even at the two-loop level, and several
groups have tackled the corresponding calculations.  As was already
outlined, the strategy here is to reduce all integrals to scalar ones.
The latter depend on a lot of different mass scales, such that their
evaluation must be done numerically.  In Section~\ref{sec:applic:deltar}
we will describe the computation~\cite{BauWei97} of two-loop corrections
induced by the Higgs boson to $\Delta r$ which enters the relation
between $G_F, M_Z, M_W$ and $\alpha$.

It is not only since the increase of the center-of-mass energy at LEP
above the $W$ boson production threshold that processes involving four
(or more) particles in the final state became very topical.  In such
reactions even the contribution from lowest order perturbation theory may
pose a serious problem as quite a lot of diagrams are involved and a
highly non-trivial phase space integration has to be performed. In
Section~\ref{sec:applic:strongW} the scattering of vector bosons 
is discussed in the background of a potentially large coupling among the
bosons at high energies~\cite{BHKPYZ98}.

Other applications that will not be discussed in more detail but shall
further substantiate the success of computer algebra in Feynman diagram
calculations are, for example, QCD corrections to
$\Delta\rho$~\cite{Avd95,CheKueSte951} and $\Delta
r$~\cite{CheKueSte952}, moments of QCD structure
functions~\cite{LarTkaVer91,LarRitVer94,LarNogRitVer96}, top mass
effects in the decay of intermediate~\cite{LarRitVer95CheKwi96} and
heavy~\cite{HarSte97} Higgs bosons as well as in $e^+e^-$ collisions
\cite{HarSte98}.  Involved one-loop calculations concerning radiative
corrections to the gauge boson scattering were considered
in~\cite{DenDitHah97}.  The muon anomalous magnetic
moment~\cite{CzaKraMar96} and the neutron anomalous electric
moment~\cite{CzaKra97} are important examples of two-loop calculations
in the electroweak sector.  In~\cite{FriKniKreRie96} the main difficulty
was the evaluation of two-loop vertex diagrams in order to obtain
corrections to the decay of a heavy Higgs boson.  Investigations of $b$
decay~\cite{CzaMel97,CzaMel98b} and threshold production of heavy
quarks~\cite{CzaMel98,BenSmi98,HoaTeu98,BenSigSmi98} are examples for
the developments of new algorithms and their implementation.  A large
list of results obtained with the help of programs that automatically
compute tree-level processes involving many particles can be found
in~\cite{comphep}.

Many more important calculations for physical observables in the
multi-loop and multi-leg sector have been performed --- mainly within
the last few years, but let us close the list at this point.

The outline of this review is as follows: Section~\ref{sec:tools}
introduces the most important technical tools that allow the automation
of Feynman diagram calculations. This includes on the one hand some of
the required algorithms, and on the other hand the most frequently used
algebraic programming languages to implement these algorithms.  Concrete
examples for such implementations will be described in
Section~\ref{secimpl}. This splits into a surveying part whose main
purpose is to collect a list of the most important programs, and a more
specific part concerned with the actual realization in the light of a
representative set of selected packages.  Finally,
Section~\ref{secapplications} gives a list of physical applications that
have been considered in the literature and that would not have been
feasible without the help of a certain degree of automation. Some of
these examples actually were the driving force for some of the
afore-mentioned packages to be developed. Others are real applications
in the sense that existing programs could successfully be used to tackle
as yet unsolved problems even by people that were not involved in the
development of the programs.

%
%

%% file: tools.tex
%
\section{Theoretical tools\label{sec:tools}}

\subsection{Multi-loop diagrams}\label{secmultiloop}
%
The main field for multi-loop calculations certainly is QCD in the
perturbative regime. The reason is, on the one hand, that in general the
complexity of a given Feynman diagram increases rapidly with the number
of dimensional parameters involved.  In the Standard Model, for example, with
its relatively large amount of different particles and scales, with the
currently available technology it is hardly possible to compute
processes for electroweak phenomena beyond two-loop level. In QCD, by
contrast, the gauge bosons are massless, and the quark masses are such
that for many physical reactions one may either neglect them completely
or consider only one quark flavour as massive and all others as massless
particles.  On the other hand, in QED, the fine structure constant is
very small, rendering the higher order corrections negligible in
general.  The coupling constant in QCD is about ten times larger, but
still small enough to play the role of an expansion parameter.


\subsubsection{Dimensional Regularization and Minimal Subtraction}
%
The momentum integrations of loop diagrams are in general not convergent
in four-dimensional space-time.  This requires the introduction of the
concept of renormalization, rendering physical quantities finite and
attributing a natural interpretation to the parameters and fields of the
underlying theory.  In order to isolate the divergent pieces one
introduces a so-called regularization scheme, the most popular one being
dimensional regularization~\cite{tHoVel72} at that time. The strategy is
to replace the four-dimensional loop integrals appearing in Feynman
diagrams by ``$D$-dimensional'' ones, obeying the basic relations of
standard convergent integrals like linearity {\it etc}.  $D$ is
considered to be a complex parameter. The limit of integer $D$, if it
exists, is required to reproduce the value for the standard integral.
All manipulations are those of convergent integrals, and only {\it
  after} integration one takes the limit $D\to 4$.  The divergences are
then reflected as poles in $1/(D-4)$.

The concept of renormalization consists in compensating these
divergences by adding suitable divergent terms, so-called counter-terms,
to the Lagrange density. Since the only requirement on
the counter-terms is to cancel the divergences, there obviously is a
freedom in choosing their finite parts. A particular choice is called
a renormalization scheme.

A convenient renormalization scheme in combination with dimensional
regularization is the so-called minimal subtraction or
MS scheme~\cite{tHo73}.  It prescribes to precisely subtract the poles in
\begin{eqnarray}
\varepsilon &\equiv& (4-D)/2\,.
\end{eqnarray}
Even more popular is the
\msbar~scheme (modified MS scheme)~\cite{BarBurDukMut78} which is
based on the observation that the poles in $1/\varepsilon$
appear only in the combination
\begin{equation}
\Delta = {1\over \varepsilon} - \gamma_{\rm E} + \ln 4\pi
\end{equation}
and that therefore it is convenient to subtract $\Delta$ instead of
$1/\varepsilon$. In the following we will use the \msbar~scheme throughout
unless stated otherwise. If divergent parts will be quoted explicitly,
the $\gamma_{\rm E}$ and $\ln 4\pi$ terms will be omitted.

Dimensional regularization, like any other regularization
scheme, introduces an arbitrary mass parameter, usually denoted
by $\mu$. This becomes clear by considering, for example, the
one-loop integral
\begin{eqnarray}
  \frac{1}{i}\int {\dd^Dp\over (2\pi)^D} 
  {1\over  (-\left(q-p\right)^2)^a 
           \left(-p^2\right)^b} &=& {(-1/q^2)^{a+b-D/2}\over (4\pi)^{D/2}}
  \frac{\Gamma(a+b-D/2)\Gamma(D/2-a)\Gamma(D/2-b)}
       {\Gamma(D-a-b)\Gamma(a)\Gamma(b)}
\,,
\label{eqml1loop}
\end{eqnarray}
where $\Gamma(x)$ is Euler's gamma function. Note that $a$
and $b$ are not necessarily integer.
The Laurent-series of the r.h.s.\ with respect to $\varepsilon$
leads to logarithms with dimensional arguments. Therefore, one multiplies
any quantity by an appropriate $D$-dependent power of $\mu$ such that
the mass dimension of the whole object is independent of $D$. It is clear
that only logarithmic $\mu$-dependence may arise by this procedure.
In terms of the Lagrangian, the artificial mass scale manifests itself
in a coupling constant with $D$-dependent mass dimension.

The combination of dimensional regularization and minimal subtraction
has many computational consequences. One can show, for example, that within
the framework of dimensional regularization massless tadpoles,
i.e.~integrals that do not carry any dimensional parameter except the
integration momenta, may be set to zero consistently. On the other hand,
minimal subtraction guarantees that any renormalization constant is a
series in the coupling constant alone, without explicit dependence on
any dimensional quantity like masses or momenta.  Many calculations can
be considerably simplified by exploiting one of these properties.
Indeed, one of the most powerful tools for the computation of multi-loop
diagrams, the so-called algorithm of integration-by-parts,
strongly resides on the properties of dimensional regularization. It
will be described in more detail below.

There are also certain drawbacks of dimensional regularization as well
as the \msbar~scheme.  One of them, of course, is the lack of any
reference to physical intuition, as one has it for regularization
schemes like introducing cut-offs or a discrete space-time lattice, and
for renormalization schemes like the on-shell scheme.  Besides this,
dimensional regularization generally causes problems whenever explicit
reference to four dimensions is made. For example, the anti-commuting
definition of $\gamma_5$ leads to inconsistencies when working in $D$
dimensions.  A consistent definition of $\gamma_5$ was given
in~\cite{tHoVel72} and formalized in~\cite{BreMai77}. It defines
$\gamma_5$ to anti-commute with $\gamma_0,\ldots,\gamma_3$ and to
commute with all the remaining $\gamma$-matrices. This definition
obviously breaks Lorentz covariance and requires the introduction of so-called
evanescent operators which makes practical calculations quite tedious.
Also in the Supersymmetric Standard
Model, defined in four-dimensional space time, one runs into problems,
but we will not dwell on them here since this will not concern what
follows.


\subsubsection{Integration-by-parts\label{sec::IP}}
%
Let us now describe one of the benefits of dimensional regularization
in more detail, namely the integration-by-parts algorithm.  It uses the
fact that the $D$-dimensional integral over a total derivative is equal
to zero:
\begin{eqnarray}
  \int{\rm d}^D p {\partial\over \partial p^\mu} f(p,\ldots) &=& 0\,.
  \label{eqipgen}
\end{eqnarray}
By explicitly performing the differentiations one obtains recurrence
relations connecting a complicated Feynman integral to several simpler
ones.  The proper combination of different recurrence relations allows
any Feynman integral (at least single-scale ones) to be reduced to a
small set of so-called master integrals. The latter ones have to be
evaluated only once and for all, either analytically or numerically.

The integration-by-parts algorithm was initially introduced for 
massless two-point
functions up to three loops \cite{CheTka81}, where two non-trivial
master integrals were to be evaluated. Further
on, the technique was applied to those massive tadpole integrals
contributing to the three-loop QCD corrections of the photon
polarization function in the limit $q^2\ll m^2$~\cite{Bro92}, where $q$
is the external momentum of the correlator and $m$ is the mass of the
heavy quark. One non-trivial master integral results in this case.  The
procedure was extended to apply to the three-loop QCD corrections for
the $\rho$ parameter~\cite{Avd95,CheKueSte951} where two more master
integrals had to be evaluated.

The recurrence relations for all possible three-loop tadpole integrals
with a single mass were derived in~\cite{Avd97} and the (most
complicated) master integrals were calculated in~\cite{Bro98}.  At
four-loop level the integration-by-parts technique was applied to
completely massive
tadpole diagrams, only aiming at their divergent parts, however.  This
restricted problem leads to two four-loop master
integrals~\cite{RitVerLar97}.

The integration-by-parts technique equally well applies to on-shell 
integrals, the
complexity at $n$-loop level being comparable to the tadpole case at
$n+1$ loops, however.  The recurrence relations at two loops were
derived some time ago~\cite{GraBroGraSch90} and were applied to the
fermion propagator in order to determine the relation between the
on-shell and $\overline{\rm MS}$ mass in QCD~\cite{GraBroGraSch90}, and
the wave function renormalization constant~\cite{BroGraSch91}.  The
three-loop on-shell integrals contributing to the anomalous magnetic
moment of the electron could be reduced to 18 master
integrals with the help of the integration-by-parts algorithm, and
thus an 
analytic
evaluation of this quantity could be performed~\cite{LapRem96}.  Very
recently ${\cal O}(\alpha^2)$ corrections of the $\mu$ decay were
calculated~\cite{RitStu98}, and the integration-by-parts method was
used to 
determine the
pole part of the corresponding four-loop on-shell integrals.
\begin{figure}[h]
  \begin{center}
  \leavevmode
      \epsfxsize=2.5cm
      \epsffile[189 293 423 500]{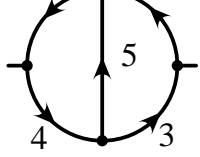}\\
    \parbox{\captionwidth}{
      \caption[]{\label{figtriangle}\sloppy
        Two-loop master diagram. The arrows denote the direction of
        momentum flow.
        }      }
  \end{center}
\end{figure}

To demonstrate the power of the integration-by-parts method let us
consider the scalar two-loop diagram of Fig.~\ref{figtriangle}.
The corresponding Feynman integral shall be denoted by
\begin{equation}
I(n_1,\ldots,n_5) = \int{\dd^D p\over (2\pi)^D}{\dd^D k\over (2\pi)^D}
{1\over (p_1^2 + m_1^2)^{n_1}\cdots(p_5^2+m_5^2)^{n_5}}\,,
\end{equation}
where $p_1,\ldots,p_5$ are combinations of the loop momenta $p,k$ and
the external momentum $q$ (we work in Euclidean space here).
$n_1,\ldots,n_5$ are called the indices of the integral.  Consider the
subloop defined by lines 2, 3 and 5, and take its loop momentum to be
$p=p_5$. If we then apply the operator $\sprod{(\partial/\partial
  p_5)}{p_5}$ to the {\it integrand} of $I$, we obtain a relation of the
form (\ref{eqipgen}), where
\begin{equation}
f(p_5,\ldots) = \frac{p_5^{\mu}}
  {(p_5^2+m_5^2)^{n_5} (p_2^2+m_2^2)^{n_2} (p_3^2+m_3^2)^{n_3}}
\,.
\end{equation}
Performing the differentiation and using momentum conservation at each
vertex one derives the following equation:
\begin{eqnarray}
\Big[ 
    - n_3 {\bf 3^+}\left({\bf 5^-}-{\bf 4^-}+m_4^2-m_5^2-m_3^2\right)
    - n_2 {\bf 2^+}\left({\bf 5^-}-{\bf 1^-}+m_1^2-m_5^2-m_2^2\right)
\nonumber\\
    + D-2n_5-n_3-n_2+2n_5 m_5^2 {\bf 5^+} 
\Big]\, I(n_1,\ldots,n_5) &=& 0
\,,
\label{eqtrianglerec}
\end{eqnarray}
where the operators ${\bf 1^{\pm}}, {\bf 2^{\pm}}, \ldots$ are used in
order to raise and lower the indices: ${\bf I^{\pm}}I(\ldots,n_i,\ldots)
= I(\ldots,n_i\pm 1,\ldots)$.  In Eq.~(\ref{eqtrianglerec}), generally
referred to as the triangle rule, it is understood that the operators to
the left of $I(n_1,\ldots,n_5)$ are applied {\em before} integration.  If the
condition $m_5=0, m_3=m_4$ and $m_1=m_2$ holds, increasing one index
always means to reduce another one.  Therefore this recurrence relation
may be used to shift the indices $n_1$, $n_4$ or $n_5$ to zero which leads
to much simpler integrals.

\begin{figure}[th]
\leavevmode
 \begin{center}
 \begin{tabular}{cccccc}
   \epsfxsize=2.5cm
   \parbox{1cm}{\epsffile[189 293 423 499]{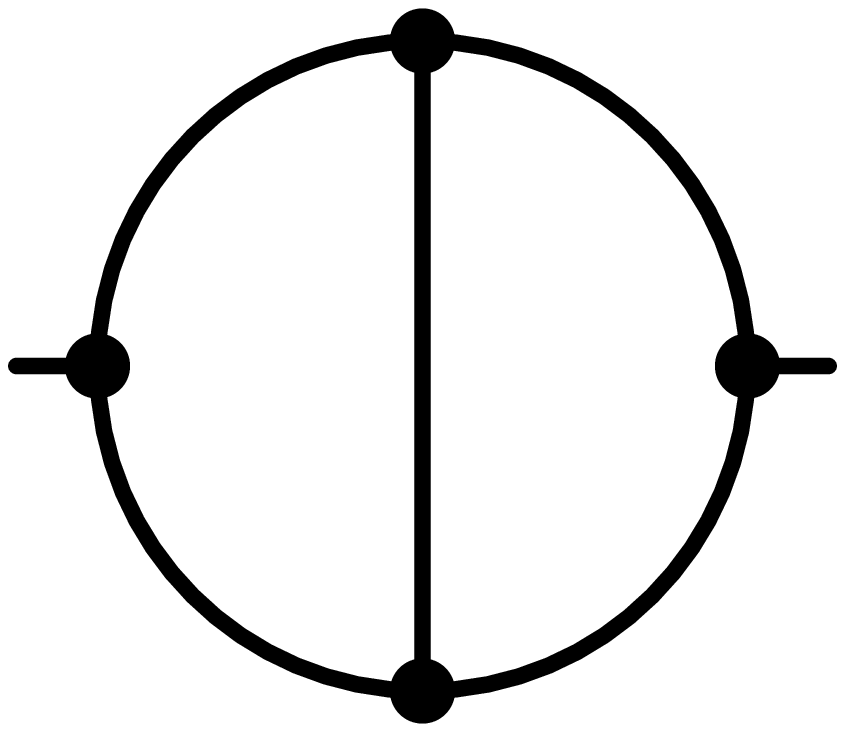}}
&
$\displaystyle =\frac{1}{\varepsilon}\Bigg[$
&
   \epsfxsize=2.5cm
   \parbox{1cm}{\epsffile[189 293 423 499]{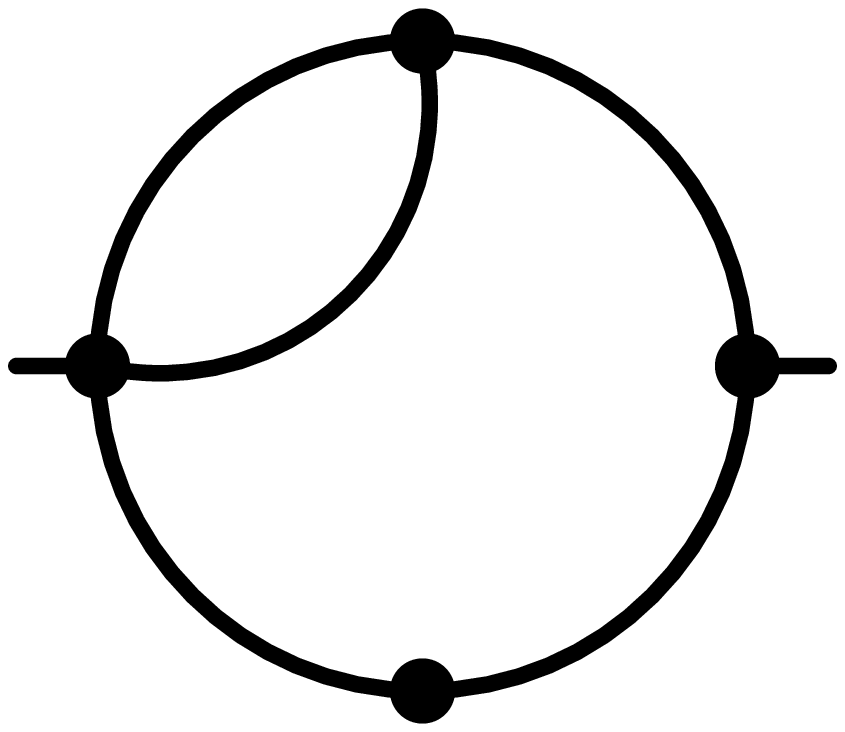}}
&
---
&
   \epsfxsize=2.5cm
   \parbox{1cm}{\epsffile[189 293 423 499]{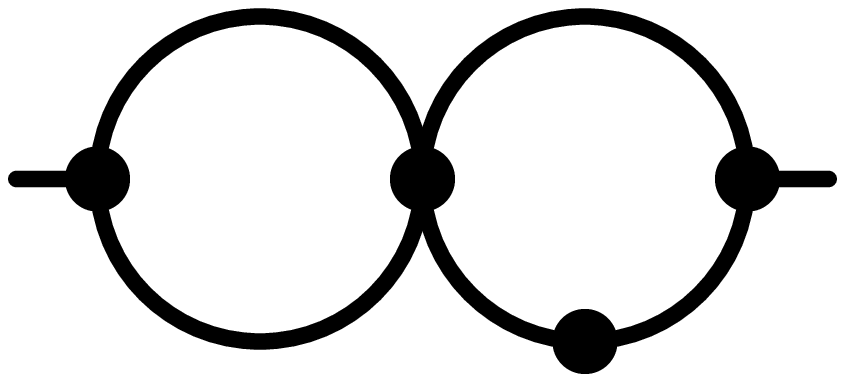}}
&
$\displaystyle \Bigg]$
 \end{tabular}
\parbox{\captionwidth}{
 \caption[]{\label{fig2loopIP}\sloppy
          Symbolic equation resulting from Eq.~(\ref{eqrecI})
          applied to the diagram $I(1,1,1,1,1)$. The dot indicates that
          the respective denominator appears twice.
         }
}
 \end{center}
\end{figure}

The triangle rule constitutes an important building block for the
general recurrence relations. The strategy is to combine several
independent equations of the kind~(\ref{eqtrianglerec}) in order to
arrive at relations connecting one complicated integral to a set of
simpler ones. For example, while the direct evaluation of even the
completely massless case for the diagram in Fig.~\ref{figtriangle} is
non-trivial, application of the triangle rule (\ref{eqtrianglerec})
leads to
\begin{equation}
I(n_1,\ldots,n_5) \,=\, \frac{1}{D-2n_5-n_2-n_3}
\Big[
n_2{\bf 2^+}\left({\bf 5^-} - {\bf 1^-}\right)
+
n_3{\bf 3^+}\left({\bf 5^-} - {\bf 4^-}\right)
\Big]\,I(n_1,\ldots,n_5)
\,.
\label{eqrecI}
\end{equation}
Repeated application of this equation reduces one of the indices $n_1$,
$n_4$ or $n_5$ to zero. For example, for the simplest case
($n_1=n_2=\ldots=n_5=1$) one obtains the equation pictured in
Fig.~\ref{fig2loopIP}: The non-trivial diagram on the l.h.s.\ is
expressed as a sum of two quite simple integrals which can be solved by
applying the one-loop formula of Eq.~(\ref{eqml1loop}) twice. This
example also shows a possible trap of the integration-by-parts
technique. In general its application introduces artificial
$1/\varepsilon$ poles which cancel only after combining all terms. They
require the expansion of the individual terms up to sufficiently high
powers in $\varepsilon$ in order to obtain, for example, the finite part
of the original diagram.  This point must carefully be respected in
computer realizations of the integration-by-parts algorithm: One must
not cut the series at too low powers because then the result goes wrong;
keeping too many terms, on the other hand, may intolerably slow down the
performance.

In our example, the l.h.s.\ in Fig.~\ref{fig2loopIP} is finite, each term
on the r.h.s., however, develops $1/\varepsilon^2$ poles. The first
three orders in the expansion for $\varepsilon\to 0$ cancel, and the
${\cal O}(\varepsilon)$ term of the square bracket, together with the
$1/\varepsilon$ in front of it, leads to the well-known result:
$I(1,1,1,1,1)=6\zeta(3)/q^2$, where $q$ is the external momentum.

In general, the successive application of recurrence relations generates
a huge number of terms out of a single diagram.  Therefore, a calculation
carried out by hand becomes very tedious and the use of computer algebra
is essential.

\begin{figure}[h]
  \begin{center}
    \leavevmode
    \epsfxsize=2.5cm
    \epsffile[212 280 400 500]{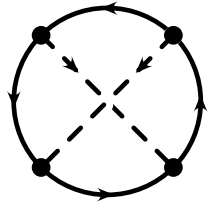}\\
    \parbox{\captionwidth}{
      \caption[]{\label{fig3ltad}\sloppy
        Three-loop topology for which the system of recurrence relations
        was explicitely solved. Solid lines carry a common mass $M$, dashed
        lines are massless.
        }
      }
  \end{center}
\end{figure}

At the end of this section let us describe an alternative approach which
tries to avoid the explicit use of recurrence relations and thus the
large number of terms in intermediate steps of the calculation.  The
crucial observation is that an arbitrary integral is expressible as
linear combination of the master integrals where the coefficients simply
depend on the dimension $D$ and the indices of the original integral.
Therefore, an attempt to explicitly solve the system of recurrence
relations in terms of integral representations was made
in~\cite{Bai96,BaiSte98}.  It was even possible to derive additional
recurrence relations over the space-time dimension $D$ in this approach.
It was successfully applied to the class of three-loop tadpoles pictured
in Fig.~\ref{fig3ltad}, where a significant reduction of CPU time could
be achieved.  Further developments in this direction look quite
promising.


\subsubsection{Tensor decomposition and tensor reduction\label{sectensdec}}
%
The calculation of Feynman diagrams for realistic field theories
inevitably leads to tensor integrals, i.e.\ Feynman integrals carrying
loop momenta with free Lorentz indices in the numerator. Since allowing
for tensor structure largely increases the number of possible integrals,
it is important to have algorithms that reduce them to a set of basis
integrals. The algorithm described in the present section is referred to
as the Passarino-Veltman method~\cite{PasVel79}.  At one loop-level, it
reduces any tensor integral to integrals with unity in the numerator.
At two-loop level this is no longer true, except for two-loop
propagator-type integrals~\cite{WeiSchBoe94}.  We will briefly introduce
the corresponding technique at the end of this section.

Consider an arbitrary 1-loop integral carrying tensor
structure in the integrand,
\begin{equation}\label{eq::tmunuN}
  I_{\mu\nu\rho\cdots}^N = \mu^{4-D} \int{\dd^D k\over (2\pi)^D}{k_\mu
    k_\nu k_\rho \cdots \over D_0 D_1\cdots D_{N-1}} =
\llangle{k_\mu k_\nu k_\rho \cdots \over D_0 D_1\cdots D_{N-1}}\rrangle_D\,,
\end{equation}
with
\begin{equation}\label{eq::Dnotation}
D_0 = k^2 - m_0^2, \qquad D_i = (k+p_i)^2 - m_i^2
\end{equation}
(the $i\epsilon$ in the denominator is suppressed here).  The statement
is \cite{PasVel79} that it can be expressed in terms of the set of
scalar integrals defined as
\begin{equation}\label{eq::scalint}
I_0^N = \llangle{1\over D_0 D_1\cdots D_{N-1}}\rrangle_D\,.
\end{equation}
The strategy is more or less straightforward and relies on the
decomposition of the integral under consideration into covariants, built
out of $g_{\mu\nu}$ and the external momenta. The general solution to
this problem in terms of a recursive algorithm can be found in
\cite{Denner:habil}. It reduces the rank of the tensor in the numerator
by one in each step. Here we only want to give an example to get an idea
of how the algorithm works.
Consider the three-point second-rank tensor integral
\begin{equation}\label{eq::tmunu3}
I^3_{\mu\nu} = \llangle {k_\mu k_\nu\over D_0 D_1 D_2} \rrangle_D\,,
\end{equation}
with its tensor decomposition
\begin{equation}\label{eq::tensdec}
I^3_{\mu\nu} = C_{00} g_{\mu\nu} +
C_{11} p_{1\mu} p_{1\nu} + C_{22} p_{2\mu} p_{2\nu} + C_{12} p_{1\mu} p_{2\nu}
+ C_{21} p_{2\mu} p_{1\nu}\,.
\end{equation}
The $C_{ij}$ are often called {\it Passarino-Veltman} coefficients and
are named by $A,B,C,\ldots$, corresponding to the value of $N=1,2,3,\ldots$ in
Eq.~(\ref{eq::tmunuN}).
Defining
\begin{equation}\label{eq::rimu}
R_i^\mu = \llangle{k^\mu(\sprod{p_i}{k})\over D_0 D_1 D_2}\rrangle_D\,,\quad
  i=1,2\,,\quad \mbox{and}\qquad R_{00} = \llangle {k^2\over D_0 D_1
  D_2}\rrangle_D\,,
\end{equation}
we may write the tensor decomposition of those quantities as
\begin{equation}\label{eq::Rtensdec}
R_i^\mu = r_{i1}\, p_1^\mu + r_{i2}\, p_2^\mu\,,\qquad R_{00} = r_{00}\,.
\end{equation}
Contracting (\ref{eq::tensdec}) by $p_{i,\nu}$ $(i=1,2)$ one obtains
a set of four equations by separately comparing coefficients of
$p_1^\mu$ and $p_2^\mu$:
\begin{equation}\label{eq::rij}
\begin{array}{lllllll}
r_{11}&=&C_{00} & + & C_{11}\, p_1^2 & + & C_{12}\, \sprod{p_1}{p_2}\,,\\[.2ex]
r_{12}&=&       &   & C_{22}\, \sprod{p_1}{p_2} & + & C_{21}\, p_1^2\,,\\[.2ex]
r_{22}&=&C_{00} & + & C_{22}\, p_2^2 & + & C_{21}\, \sprod{p_1}{p_2}\,,\\[.2ex]
r_{21}&=&      &   & C_{11}\, \sprod{p_1}{p_2} & + & C_{12}\, p_2^2\,.
\end{array}
\end{equation}
Contraction of (\ref{eq::tensdec}) with $g_{\mu\nu}$, on the other hand,
yields
\begin{equation}\label{eq::r00}
r_{00} = D\,C_{00} + C_{11}\,p_1^2 + C_{22}\,p_2^2 +
C_{12}\,\sprod{p_1}{p_2} + C_{21}\, \sprod{p_1}{p_2}\,. 
\end{equation}
Together, these are five equations for the five unknowns $C_{ij}$
(actually there are only {\it four} unknowns, since $C_{12} = C_{21}$
because of the symmetry $\mu \leftrightarrow \nu$ in $I^3_{\mu\nu}$; this
redundancy may serve as a useful check in the end).
Combining (\ref{eq::r00}) with the sum of the first and third equation
in (\ref{eq::rij}), one immediately gets the solution for $C_{00}$:
\begin{equation}\label{eq::C00sol}
C_{00} = {1\over D-2}\left(r_{00} - r_{11} - r_{22}\right)\,.
\end{equation}
The remaining coefficients may be determined by rewriting
(\ref{eq::rij}) as two sets of systems of linear equations:
\begin{equation}\label{eq::Cij}
\left(
  \begin{array}{c}
    r_{11} - C_{00} \\
    r_{21}
  \end{array}
\right) =
{\bf X}
\left(
  \begin{array}{c}
    C_{11} \\
    C_{12}
  \end{array}
\right)\,,\quad
\left(
  \begin{array}{c}
    r_{12} \\
    r_{22} - C_{00}
  \end{array}
\right) =
{\bf X}
\left(
  \begin{array}{c}
    C_{21} \\
    C_{22}
  \end{array}
\right)\,,
\end{equation}
where
\begin{equation}
{\bf X} = \left(
\begin{array}{cc}
  p_1^2        & \sprod{p_1}{p_2} \\
  \sprod{p_1}{p_2} & p_2^2
\end{array}
\right)\,.
\end{equation}
So, if ${\bf X}$ is invertible, Eq.~(\ref{eq::C00sol}) and the inverse
of Eqs.~(\ref{eq::Cij}) determine the $C_{ij}$ in terms of the $r_{ij}$.

In turn, by rewriting
\begin{equation}\label{eq::scalprod}
\sprod{k}{p_i} = {1\over 2}\left[D_i - D_0 - f_i\right]\,,\quad
\mbox{with } f_i = p_i^2 - m_i^2 + m_0^2\,,
\end{equation}
and inserting this into the first equation of (\ref{eq::rimu}), one
arrives at
\begin{eqnarray}
r_{11}\,p_1^\mu + r_{12}\,p_2^\mu &=& {1\over 2}\left[
\llangle{k_\mu\over D_0 D_2}\rrangle_D -
\llangle{k_\mu\over D_1 D_2}\rrangle_D -
f_1 \llangle{k_\mu\over D_0 D_1 D_2}\rrangle_D\right]\,,
\nonumber\\
r_{21}\,p_1^\mu + r_{22}\,p_2^\mu &=& {1\over 2}\left[
\llangle{k_\mu\over D_0 D_1}\rrangle_D -
\llangle{k_\mu\over D_1 D_2}\rrangle_D -
f_2 \llangle{k_\mu\over D_0 D_1 D_2}\rrangle_D\right]\,,
\end{eqnarray}
which allows to compute the $r_{ij}$ through first-rank tensor
integrals.  Thus, the first run-through of recurrence is done. The
second one, in turn, is only concerned with at most first-rank tensors.
In that way, any tensor integral can be reduced to the basic set of
scalar integrals defined in (\ref{eq::scalint}), provided that ${\bf X}$
is invertible.  Several improvements to the original algorithm concerned
with the problem of vanishing $\det{\bf X}$ and also with the numerical
evaluation of the scalar integrals have been worked
out~\cite{Stu88,OldVer89}. An overview and a complete list of references
can be found in~\cite{Stu98}.

At two-loop level this strategy no longer reduces numerators of the
integrals to unity. The reason is that in general one encounters
``irreducible numerators'', i.e.\ scalar products that are not
expressible in terms of the denominator via relations like
(\ref{eq::scalprod}).  But in the case of propagator-type integrals,
i.e.\ those with only one external momentum $p$, the full reduction may
still be achieved by applying the mechanism above first to a
subloop~\cite{WeiSchBoe94}.  For example, consider the diagram in
Fig.~\ref{tensred2l.ps}, and a corresponding first-rank tensor-integral:
\begin{figure}
  \begin{center}
    \leavevmode
    \epsfxsize=6.cm
    \epsffile[110 245 465 560]{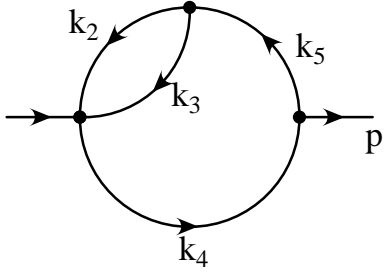}
    \hfill
    \parbox{\captionwidth}{
    \caption[]{\label{tensred2l.ps}\sloppy Two-loop propagator-type
      diagram. The momenta $k_i$ are linear combinations of the loop
      momenta $l,k$ and the external momentum $p$.  }}
  \end{center}
\end{figure}
\begin{equation}
S_\mu = \int{\dd^D k\over (2\pi)^D}{\dd^D l\over (2\pi)^D}{k_\mu\over
  D_k D_l D_{k+l} D_{l+p}} \,.
\end{equation}
The notation here is a bit different from the one in
(\ref{eq::Dnotation}):
\begin{equation}
D_{k_i} = k_i^2 - m_i^2\,,
\end{equation}
with
\begin{equation}
k_2 = k\,,\qquad k_3 = k+l\,,\qquad k_4 = l+p\,,\qquad k_5 = l\,.
\end{equation}
Direct
application of the Passarino-Veltman algorithm through the ansatz $S_\mu
= p_\mu \,S(p^2)$ and contracting this equation by $p^\mu$ leads to the
irreducible numerator $\sprod{p}{k}$, so that no simplification results
in this way.

The idea is instead to perform the tensor reduction of the subintegral
over $k$ first, considering $l$ as its external momentum:
\begin{eqnarray}
S_\mu &=&
\int{\dd^D l}\, {1\over D_l D_{p+l}}
\int{\dd^D k}\, {k_\mu\over D_k D_{k+l}} =
\int{\dd^D l}\, {1\over D_l D_{p+l}}\cdot{l_\mu\over l^2}
\int{\dd^D k}\, {\sprod{k}{l}\over D_k D_{k+l}}\nonumber\\&=&
\int{\dd^D l}\, {\sprod{l}{p}\over l^2 D_l D_{p+l}} 
\int{\dd^D k}\, {\sprod{k}{l}\over D_k D_{k+l}} 
\cdot{p_\mu\over p^2} \nonumber\\&=&
 {p_\mu\over 4 p^2} \,\int{\dd^D k}\int{\dd^D l}\, 
 {\left[D_{k+l} - D_k - l^2 - m_3^2 + m_2^2\right]
 \left[D_{p+l} - D_l - p^2 - m_5^2 + m_4^2\right]\over
   l^2 D_k D_{k+l} D_l D_{p+l}}\,.
\end{eqnarray}
After canceling common factors, the numerator has no dependence on loop
momenta any more which is what we were aiming for. The application to
arbitrary propagator-type tensor integrals can be found
in~\cite{WeiSchBoe94}.



\subsubsection{Tensor reduction by shifting the space-time dimension
  \label{sec::tredtar}}
%
As already noted in Section~\ref{sectensdec}, the Passarino-Veltman
method reduces the tensor structure in the numerator to unity only at
one-loop level, the reason being ``irreducible numerators'' appearing at
higher loop order. An alternative approach that circumvents this problem
has been worked out at one-loop level in \cite{Dav91} and generalized to
an arbitrary number of loops in \cite{Tar96,Tar97}. However, at two-loop
level it has only been applied to propagator type integrals up to now
(see, e.g., \cite{FleJegTarVer99}).
The basic idea is to express tensor integrals in $D$ dimensions through
scalar ones with a shifted value of $D$.  Again we do not want to
present the algorithm in its full generality here, but try to shed some
light on the main ideas by giving a concrete example.

Consider again the integral (\ref{eq::tmunu3}). By introducing an
auxiliary vector $a_\mu$, it may be written as
\begin{eqnarray}\label{eq::tart}
\llangle {k_\mu k_\nu\over D_0 D_1 D_2}\rrangle_D =
\left({1\over i}{\partial\over \partial a^\mu}\right)
\left({1\over i}{\partial\over \partial a^\nu}\right)
\llangle {1\over D_0 D_1 D_2} e^{i\sprod{a}{k}}\rrangle_D \bigg|_{a=0}\,.
\end{eqnarray}
Using the Schwinger-parameterization for propagators,
\begin{equation}
{1\over k^2-m^2+i\epsilon} = {1\over i}\int_0^\infty\dd\alpha
e^{i\alpha(k^2-m^2+i\epsilon)}\,,
\end{equation}
one finds
\begin{eqnarray}
  && \llangle {1\over D_0 D_1 D_2} e^{i\sprod{a}{k}}\rrangle_D =
  \int_0^\infty\dd\vec{\alpha} \int{\dd^D k\over (2\pi)^D}
  \,e^{-i\left[\alpha_0m_0^2 + \alpha_1m_1^2 + \alpha_2m_2^2\right]}\,
  \times\nonumber\\&&\mbox{\hspace{1ex}} \exp\left\{i\left[(\alpha_0 +
  \alpha_1 + \alpha_2)k^2 + 2\,\left(\alpha_1 p_1 + \alpha_2 p_2 +
  {a\over 2}\right)\!\cdot\!k + \alpha_1 p_1^2 + \alpha_2
  p_2^2\right]\right\}\,,
\end{eqnarray}
where $\dd\vec\alpha = \dd\alpha_0\dd\alpha_1\dd\alpha_2$, and with the
help of
\begin{equation}
\int {\dd^D k\over (2\pi)^D}\,\exp\left[i(A\,k^2 + 2\sprod{p}{k})\right] =
i{e^{-i\,{p^2\over A}}\over (4\pi i A)^{D/2}}\,,
\end{equation}
one obtains
\begin{eqnarray}\label{eq::Ifinal}
&& \llangle {1\over D_0 D_1 D_2} e^{i\sprod{a}{k}}\rrangle_D =
{i\over (4\pi i)^{D/2}}\int_0^\infty \dd\vec\alpha \,
{e^{-i\left(\alpha_0m_0^2 + \alpha_1m_1^2 + \alpha_2m_2^2\right)} \over 
  (\alpha_0 + \alpha_1 + \alpha_2)^{D/2}}  
\exp\bigg\{-i\Big[-\alpha_1(\alpha_0 + \alpha_2)p_1^2 
\nonumber\\&& \mbox{\hspace{3ex}} -
  \alpha_2(\alpha_0 + \alpha_1)p_2^2 +
  2\,\alpha_1\alpha_2
  \sprod{p_1}{p_2} + \alpha_1\,\sprod{p_1}{a} +
  \alpha_2\,\sprod{p_2}{a} + a^2/4\Big]/(\alpha_0 
  + \alpha_1 + \alpha_2)\bigg\}\,.
\end{eqnarray}
Inserting (\ref{eq::Ifinal}) into (\ref{eq::tart}) and explicitly
performing the differentiations yields
\begin{eqnarray}
&&\llangle{k_\mu k_\nu\over D_0 D_1 D_2}\rrangle_D =
{i\over (4\pi i)^{D/2}}\int\dd \vec\alpha\,
{e^{-i\left(\alpha_0m_0^2 + \alpha_1m_1^2 + \alpha_2m_2^2\right)}
  \over
  (\alpha_0 + \alpha_1 + \alpha_2)^{D/2+2}}
(\alpha_1 p_1 + \alpha_2 p_2)_\mu (\alpha_1 p_1 + \alpha_2 p_2)_\nu
\, \times\nonumber\\&&\mbox{\hspace{6ex}}
\exp\bigg\{-i\Big[-\alpha_1(\alpha_0 + \alpha_2)p_1^2 -
\alpha_2(\alpha_0 + \alpha_1)p_2^2 + 2\,\alpha_1\alpha_2\,
  \sprod{p_1}{p_2} \Big]/(\alpha_0 + \alpha_1 + \alpha_2)\bigg\}\,.
\label{eq::alpha-param}
\end{eqnarray}
The technique to derive Eq.~(\ref{eq::alpha-param}) and its
generalization to arbitrary multi-loop diagrams is known since long (see,
e.g., \cite{BreMai77}).  However, instead of differentiating with
respect to $a$ we may equally well apply the following operator on the
scalar integral $\llangle{1\over D_0 D_1 D_2}\rrangle_D$, in that way
getting rid of the auxiliary vector $a$:
\begin{eqnarray}
\lefteqn{T_{\mu\nu}\left(\{p_1,p_2\},\left\{{\partial\over \partial
  m_1^2},{\partial\over \partial m_2^2}\right\}, {\bf d^+}\right) \equiv } 
\nonumber\\&\equiv&
\left({1\over i}{\partial\over \partial a^\mu}\right)
\left({1\over i}{\partial\over \partial a^\nu}\right)
  \exp\left[-i(\alpha_1 \sprod{p_1}{a} + \alpha_2\sprod{p_2}{a})
  \rho\right]\bigg|_{a=0,\,\alpha_j = 
    i{\partial\over \partial m_j^2},\,
    \rho    = 4\pi i {\bf d^+}}
  \nonumber\\&=& -\left(p_{1\mu}{\partial\over \partial m_1^2} + 
  p_{2\mu}{\partial\over \partial m_2^2}\right)
  \left(p_{1\nu}{\partial\over \partial m_1^2} + 
  p_{2\nu}{\partial\over \partial m_2^2}\right)\,(4\pi i\,{\bf d^+})^2\,,
\end{eqnarray}
where the operator ${\bf d^+}$ increases the space-time dimension by
$2$, i.e.\ ${\bf d^+} \llangle \cdots\rrangle_D = \llangle
\cdots\rrangle_{D+2}$. Finally, we have
\begin{eqnarray}
  &&\llangle {k_\mu k_\nu\over D_0 D_1 D_2}\rrangle_D = 
  T_{\mu\nu}\llangle{1\over D_0 D_1 D_2}\rrangle_D = (4\pi)^2 \Bigg[
  2\,p_{1\mu} p_{1\nu}\llangle{1\over D_0 D_1^3 D_2}\rrangle_{D+4} +
  \nonumber\\&&\mbox{\hspace{6ex}} +
  2\,p_{2\mu} p_{2\nu}\llangle{1\over D_0 D_1 D_2^3}\rrangle_{D+4} +
  (p_{1\mu} p_{2\nu} + p_{1\nu} p_{2\mu}) 
  \llangle{1\over D_0 D_1^2 D_2^2}\rrangle_{D+4}\Bigg]\,.
\end{eqnarray}

The algorithm formally applies to an arbitrary number of loops and
external legs.  But this means only that any tensor integral can be
reduced to scalar integrals with a shifted number of space-time
dimension, the latter ones remaining still to be evaluated. A strategy
to cope with these diagrams is to use generalized recurrence
relations~\cite{Tar97}.  At the one-loop level they have been worked out
for arbitrary $n$-point functions in \cite{Tar96}. At the two-loop
level, however, so far they are only published for propagator-type
diagrams~\cite{Tar97}.  Since we feel that these generalized recurrence
relations are beyond the scope of this review, let us refer the
interested reader to the literature \cite{Tar97,TarRhein98}.
The algorithm described above was used in \cite{FleJegTarVer99} for the
computation of the two-loop QCD corrections to the fermion propagator.

\subsection{\label{subasymp}Asymptotic expansion of Feynman diagrams}

%
\subsubsection{Generalities}
As was outlined in Section~\ref{secmultiloop}, the
integration-by-parts algorithm was 
successfully applied to single-scale integrals, i.e.\ massive
tadpole, massless propagator-type, or on-shell integrals.
For an arbitrary multi-scale diagram it is in general rather difficult to
solve recurrence relations.

However, if the scales involved follow a certain hierarchy, a
factorization is possible.  For example, consider the operator product
expansion for the correlator of currents $j(x) = \bar \psi(x) \Gamma
\psi(x)$ in the limit $Q^2 \to \infty$, where $\Gamma$ is some Dirac
matrix and $\psi$ a quark field with mass $m$:
\begin{equation}
i\int \dd x e^{iqx} {\rm T} j(x) j(0) \stackrel{Q^2\to \infty}{\simeq}
\sum_n C_n {\cal O}_n\,,
\end{equation}
where T denotes the time ordered product.  It was realized
in~\cite{CheGorTka82,Tka83,BroGen84,CheSpi87} that if one adopts the
minimal subtraction regularization scheme and abandons normal ordering
of the operators, then $Q$ appears only in the coefficient functions
$C_n$ while the mass $m$ is completely absorbed into the operators
${\cal O}_n$ and appears only in the matrix elements.

The attempts to find an algorithm that produces this factorization for
arbitrary Feynman integrals finally resulted in the prescriptions for
the asymptotic expansion of Feynman
diagrams~\cite{PivTka84,GorLar87,CheSmi87,Smi91,Smi95} (see
also~\cite{Davetal}). These prescriptions provide well defined recipes
that are completely decoupled from any field theoretic derivation and
even lack a rigorous proof, as in the cases (iii) and (iv) below.
However, their success in practical applications justifies them a
posteriori.

At the moment the following procedures are used in the calculations:
\renewcommand{\labelenumi}{(\roman{enumi})}
\begin{enumerate}
\item Large-Momentum Procedure: $Q \gg q,m$
\item Hard-Mass Procedure: $M \gg q,m$
\item Threshold Expansion
\item Expansion with the external momenta on the mass shell.
\end{enumerate}
\renewcommand{\labelenumi}{(\Roman{enumi})}
The first two of them will be considered more closely in the next
subsection.  The presentation will be rather informal, explaining the
procedures in a ready-to-use form.  Currently, the technical apparatus
for the latter two cases is much less developed. For this reason they
will be touched upon only briefly here.

One may treat large-momentum and hard-mass procedure on the same
footing. Thus, in what follows we only present the general formulae in
the case of large external momenta --- the transition to the hard-mass
procedure is straightforward.  The prescription for the large-momentum
procedure is summarized by the following formula:
\begin{eqnarray}
\Gamma(Q,m,q) & \stackrel{Q\to \infty}{\simeq} &
\sum_\gamma \Gamma/\gamma(q,m)
\,\,\star\,\, 
T_{\{q_\gamma,m_\gamma\}}\gamma(Q,m_\gamma,q_\gamma)
\,.
\label{eqasexp}
\end{eqnarray}
Here, $\Gamma$ is the Feynman diagram under consideration, $\{Q\}$
($\{m,q\}$) is the collection of the large (small) parameters, and the
sum goes over all subgraphs $\gamma$ of $\Gamma$ with masses $m_\gamma$
and external momenta $q_\gamma$, subject to certain conditions to be
described below.  $T_{\{q,m\}}$ is an operator performing a Taylor
expansion in $\{q,m\}$ {\em before} any integration is carried out.  The
notation $\Gamma/\gamma\star T_{\{q,m\}}\gamma$ indicates that the
subgraph $\gamma$ of $\Gamma$ is replaced by its Taylor expansion which
should be performed in all masses and external momenta of $\gamma$ that
do not belong to the set $\{Q\}$.  In particular, also those external
momenta of $\gamma$ that appear to be integration momenta in $\Gamma$
have to be considered as small. Only after the Taylor expansions have
been carried out, the loop integrations are performed.  In the following
we will refer to the set $\{\gamma\}$ as {\em hard subgraphs} or simply {\em
  subgraphs}, to $\{\Gamma/\gamma\}$ as {\em co-subgraphs}.

The conditions for the subgraphs $\gamma$ are different for
the hard-mass and the large-momentum procedure\footnote{
  Actually they are very similar and it is certainly possible to merge
  them into one condition using a more abstract language. For our
  purpose, however, it is more convenient to distinguish the two procedures.}.
For the large-momentum procedure, $\gamma$ must
\begin{itemize}
\item contain all vertices where a large momentum enters or leaves the
  graph
\item be one-particle irreducible after identifying these
  vertices.
\end{itemize}
From these requirements it is clear that the hard subgraphs become
massless integrals where the scales are given by the large momenta. In
the simplest case of one large momentum one ends up with propagator-type
integrals.  The co-subgraph, on the other hand, may still contain small
external momenta and masses. However, the resulting integrals are
typically much simpler than the original one.

In the case of hard-mass procedure, $\gamma$ has to
\begin{itemize}
\item contain all propagators carrying a large mass
\item be one-particle irreducible in its connected parts after
  contracting the heavy lines.
\end{itemize}
Here, the hard subgraphs reduce to tadpole integrals with the large masses
setting the scales. The co-subgraphs are again simpler to
evaluate than the initial diagram.

The large-momentum and hard-mass procedure provide expansions for {\it
  off-shell} external momenta which are either large or small as
compared to internal masses. Recently a procedure allowing the
asymptotic expansion of Feynman integrals near threshold was suggested.
Graphical prescriptions similar to those for the large-momentum and
hard-mass procedure have been worked out and applied to
the cross section of quark production at $e^+e^-$ colliders near
threshold~\cite{BenSigSmi98,CzaMel98}. The expansion parameter in this case
is given by the velocity of the produced quarks.

A method to expand on-shell Feynman diagrams was developed in
\cite{Smi97,CzaSmi97}.  Two typical limits for a large external momentum
on a mass shell were considered: one where the mass shell is itself
large, and the other one where the mass shell is zero.
The latter case, called the Sudakov
limit, is a purely Minkowskian phenomenon. This distinguishes it from
the cases described so far which can be formulated completely in
Euclidean space, simplifying rigorous proofs.  The ``philosophy'' of
these expansions is very similar to the hard-mass and large-momentum
procedure.  However, apart from the criteria on the subgraphs, the way of
performing the expansion of propagators is also different. For example,
the expansion of lines carrying large masses and not belonging to a
one-particle-irreducible component of the subgraph (so-called cut lines)
looks as follows:
\begin{eqnarray}
T_\kappa\frac{1}{\kappa k^2 + 2Qk}\Bigg|_{\kappa=1}
\,.
\end{eqnarray}
$k$ is an integration momentum and $Q$ a large external momentum.  The
graphical representation of the procedure becomes much less transparent
because of that.

In~\cite{CzaSmi97} the two-loop master integral with two different
masses and on-shell external momentum was considered: $m\ll M$,
$Q^2=M^2$. The first 19 terms of the expansion in $m/M$ were evaluated.
Similar computations have also been performed for the fermion
propagator~\cite{AvdKal97}.  Two-loop vertex diagrams with external
momenta $p_1$ and $p_2$ obeying the Sudakov limit $p_i^2=0$ and
$(p_1+p_2)^2\to-\infty$ were examined in \cite{Smi97_2}, and an
expansion in $m^2/(p_1+p_2)^2$ was obtained.

Let us finally emphasize that due to analyticity the obtained expansions
often provide valuable information also in other regions of the
parameter space. This knowledge was used to reconstruct the photon
polarization function by combining the results of asymptotic expansions
in different limits (see also Section~\ref{sec::polfunc}).


\subsubsection{\label{subsubexa}Examples}
\begin{figure}[t]
  \begin{center}
    \parbox{\captionwidth}{
  \leavevmode
  \epsfxsize=2.5cm
  \epsffile[150 260 420 450]{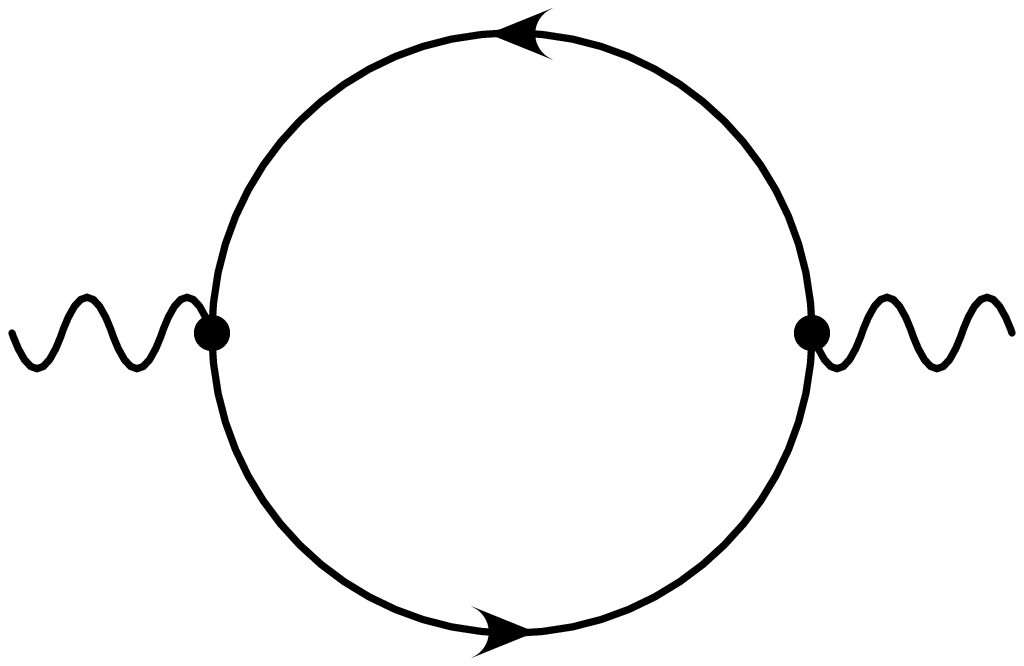}\hspace{1em}
  \raisebox{2.1em}
  {\Large $= \ \ 1\ \star\,\, $}
  \epsfxsize=2.5cm
  \epsffile[150 260 420 450]{d1q.ps}\hspace{1em}
  \raisebox{2.1em}
  {\Large $+ \ \ 2\times \!\!\!\!$}
  \epsfxsize=2cm
  \raisebox{.5em}{\epsffile[150 260 420 450]{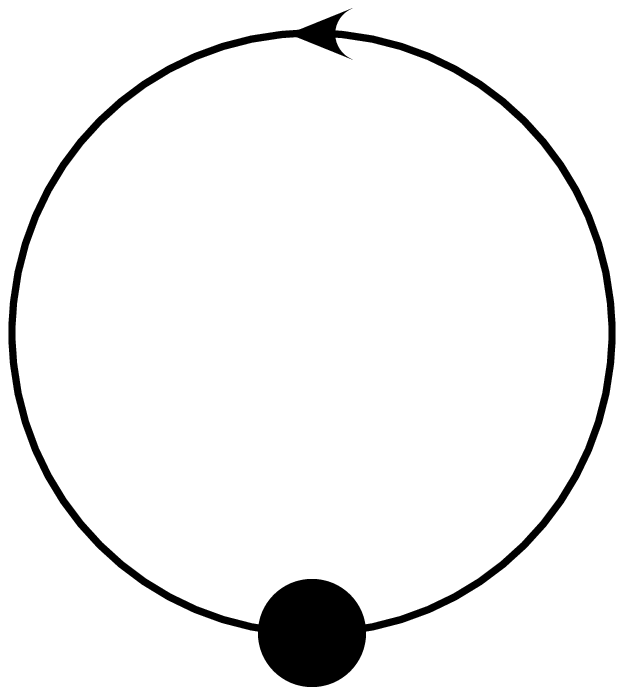}}\hspace{-.5em}
  \raisebox{2.1em}{\Large $\star\,\,$}
  \epsfxsize=2.5cm
  \epsffile[150 260 420 450]{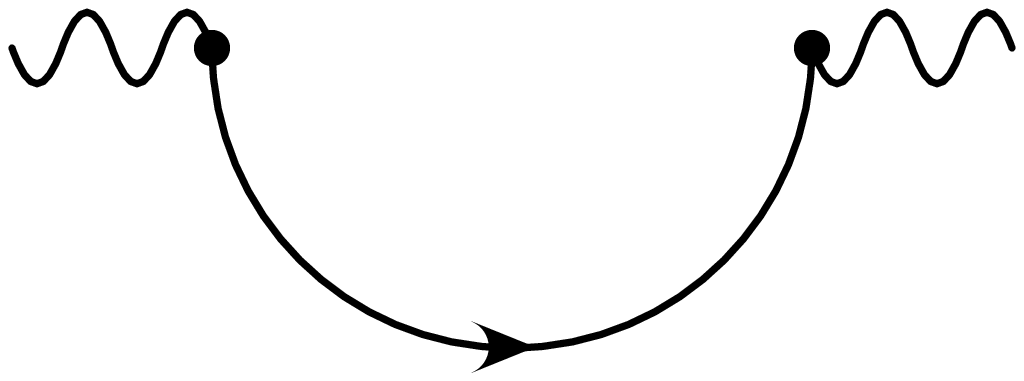}\hspace{1em}}
    \parbox{\captionwidth}{
      \caption[]{\label{fig::lmp1l}\sloppy
        Large-momentum procedure for the one-loop photon polarization
        function.
        }      }
  \end{center}
\end{figure}

Let us first consider the one-loop contribution to the photon propagator
shown on the l.h.s.\ of the diagrammatic equation in
Fig.~\ref{fig::lmp1l}.  Both fermion lines are supposed to carry the
same mass, $m$, and $q$ is the external momentum.  The application of
the large-momentum procedure leads to the subdiagrams shown on the
r.h.s.\ of this equation.  The first one represents a simple Taylor
expansion w.r.t.\ $m$, thus leading to massless one-loop integrals which
can be solved with the help of Eq.~(\ref{eqml1loop}). Starting from a
certain order in $m^2$, this subdiagram develops infra-red poles which
are absent in the original diagram. They are due to the massless
denominators arising in the Taylor expansion.  However, they cancel
against twice the ultra-violet poles of the second subgraph.  The factor
of two arises because a symmetric subgraph should be considered as well.
To make this cancellation more transparent we present the first four
terms of an expansion in $m^2$ for the transverse part of the one-loop
polarization function (cf.\ Section~\ref{sec::polfunc}):
\begin{eqnarray}
  \Pi^{(0)}_{\rm bare}(q^2)  
&\stackrel{q^2\gg m^2}{=}&
{3\over 16\pi^2}\,\Bigg\{
  {4\over 3 \varepsilon}
  + {20\over 9} 
  - {4\over 3}\,\logqmums
  + 8\,{m^2\over q^2}
  + \left({m^2\over q^2}\right)^2\,\bigg( 
    - {8\over\varepsilon} 
    - 8
    + 8\,\logqmums
  \bigg)
  \nonumber\\&&\mbox{\hspace{3em}}
  + \left({m^2\over q^2}\right)^3\,\bigg( 
    - {32\over 3 \varepsilon} 
    - {80\over 3} + {32\over 3}\,\logqmums
  \bigg)
  \nonumber\\&&\mbox{\hspace{1em}}
  + 2\,\bigg[\left({m^2\over q^2}\right)^2\,\bigg(
    {4\over \varepsilon}
    + 6
    + 4\,\logmum
  \bigg)
  + \left({m^2\over q^2}\right)^3\,\bigg( 
    {16\over 3 \varepsilon}
    + {88\over 9} 
    + {16\over 3}\,\logmum
  \bigg)
  \bigg]
  + \ldots
  \Bigg\}= 
  \nonumber\\
  &=&
  {3\over 16\pi^2}\,\Bigg\{  
  {4\over 3\varepsilon}
  + {20\over 9} - {4\over 3}\,\logqmums
  + 8 {m^2\over q^2}
  + \left({m^2\over q^2}\right)^2\,\bigg( 
    4 
    + 8\,\logqmms
  \bigg)
  \nonumber\\&&\mbox{\hspace{3em}}
  + \left({m^2\over q^2}\right)^3\,\bigg( 
    - {64\over 9} 
    + {32\over 3}\,\logqmms 
  \bigg)
+\ldots
  \Bigg\}\,,
  \label{eq::lmp1l}
\end{eqnarray}
with $\logqmums=\ln(-q^2/\mu^2)$, $\logqmms=\ln(-q^2/m^2)$ and
$\logmum=\ln(\mu^2/m^2)$.  The first two lines correspond to the first
subgraph of Fig.~\ref{fig::lmp1l}, and the terms in square brackets are
due to the second subgraph.  The fourth and fifth lines show the sum of
all subgraphs which corresponds to the consistent expansion of the full
one-loop diagrams.  The remaining pole in $1/\varepsilon$ is the
ultra-violet divergency of the full diagram and is usually removed by
renormalization (see below). On the other hand, all spurious poles
cancel in the sum.

Let us now analyze the diagram on the l.h.s.\ of Fig.~\ref{fig::lmp1l} in
the limit $q^2\ll m^2$. In this case the hard-mass procedure applies. It
leads to a trivial Taylor expansion, and one ends up with bubble
integrals.  The first few terms read:
\begin{eqnarray}
\Pi^{(0)}_{\rm bare}(q^2) 
&\stackrel{q^2\ll m^2}{=}&
{3\over 16\pi^2}\,\Bigg\{
  {4\over 3 \varepsilon}
  + {4\over 3}\,\logmum
  + \frac{4}{15}\frac{q^2}{m^2}
  + \frac{1}{35}\left(\frac{q^2}{m^2}\right)^2
  + \frac{4}{945}\left(\frac{q^2}{m^2}\right)^3
  + \ldots
\Bigg\}\,.
\end{eqnarray}
Note that the $1/\varepsilon$ pole is the same as in
Eq.~(\ref{eq::lmp1l}). In the case of the photon propagator
the pole is usually removed by requiring that the polarization
function vanishes for $q^2=0$.
Finally the one-loop polarization function in the two limiting cases
reads:
\begin{eqnarray}
\Pi^{(0)}(q^2) 
&\stackrel{q^2\gg m^2}{=}&
  {3\over 16\pi^2}\,\Bigg\{  
    {20\over 9} - {4\over 3}\,\logqmms
  + 8 {m^2\over q^2}
  + \left({m^2\over q^2}\right)^2\,\bigg( 
    4 
    + 8\,\logqmms
  \bigg)
  \nonumber\\&&\mbox{\hspace{3em}}
  + \left({m^2\over q^2}\right)^3\,\bigg( 
    - {64\over 9} 
    + {32\over 3}\,\logqmms 
  \bigg)
+\ldots
  \Bigg\}\,,
\nonumber\\
\Pi^{(0)}(q^2)
&\stackrel{q^2\ll m^2}{=}&
{3\over 16\pi^2}\,\Bigg\{
    \frac{4}{15}\frac{q^2}{m^2}
  + \frac{1}{35}\left(\frac{q^2}{m^2}\right)^2
  + \frac{4}{945}\left(\frac{q^2}{m^2}\right)^3
  + \ldots
\Bigg\}\,.
\end{eqnarray}

\begin{figure}[t]
\begin{center}
\begin{tabular}{ccc}
\leavevmode
\epsfxsize=5.0cm
\epsffile[142 267 470 525]{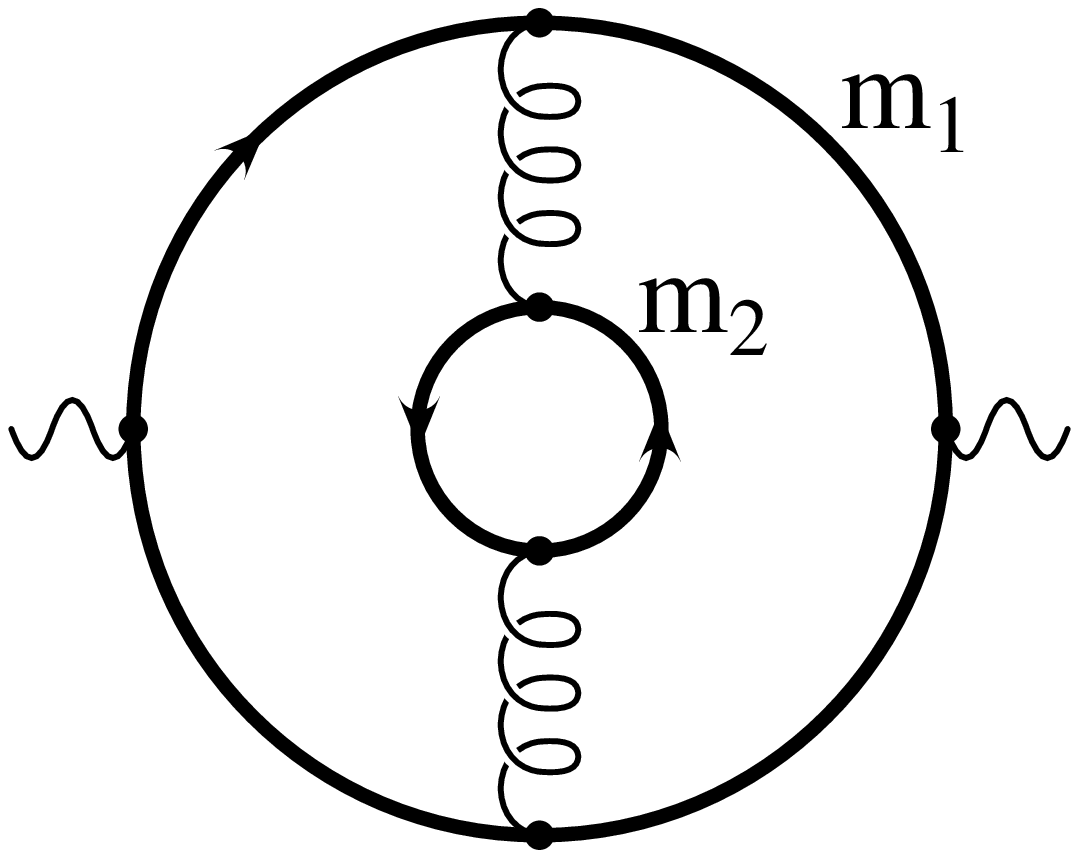}
&\hspace{2em}&
\leavevmode
\epsfxsize=5.0cm
\epsffile[142 267 470 525]{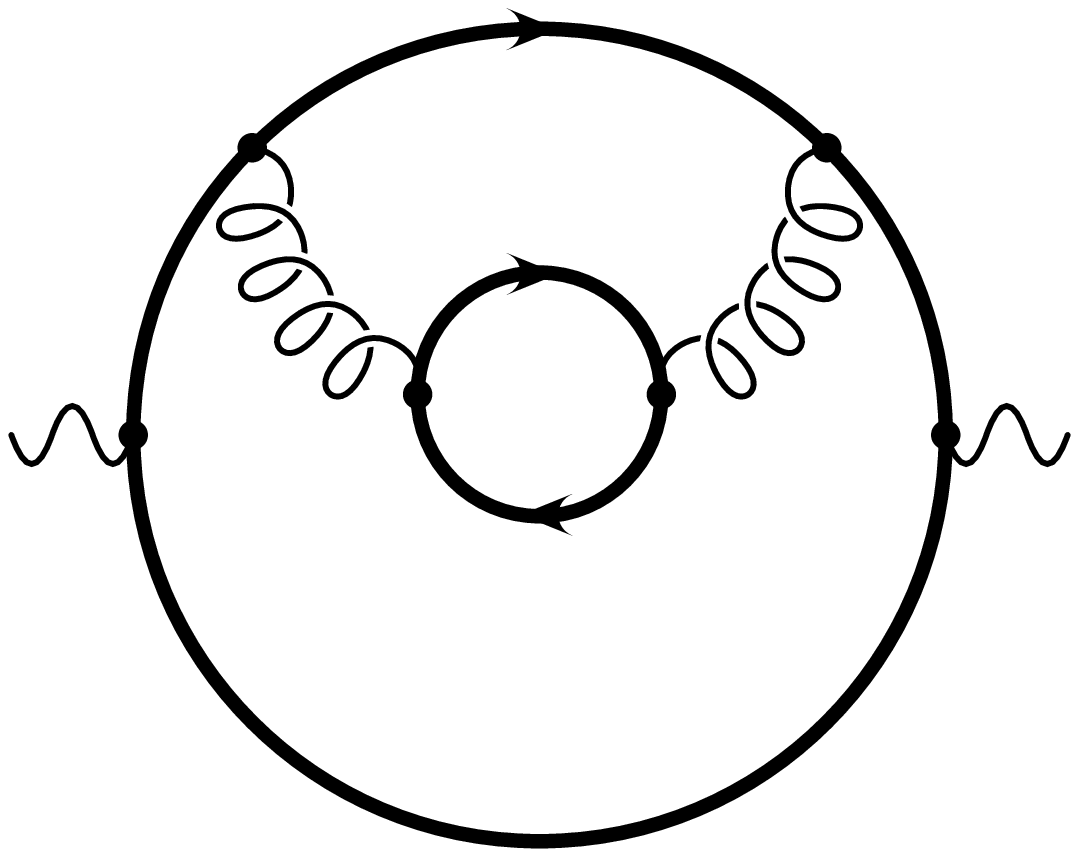} \\
(a) &\hspace{2cm}& (b)
\end{tabular}
\parbox{\captionwidth}{\sloppy
\caption[]{\label{figdb} 
Fermionic double-bubble diagrams with generic masses $m_1$ and $m_2$.
}}
\end{center}
\end{figure}

Consider now the the case of the double-bubble diagrams pictured in
Fig.~\ref{figdb}. They provide a gauge invariant subclass of all
three-loop graphs contributing to the photon polarization function.
Note, however, that the particle type is irrelevant for the procedures
described in the previous section; only the mass and momentum
distribution is important.  The outer and the inner mass are denoted by
$m_1$ and $m_2$, respectively, and $q$ is the external momentum.

Of special interest is the high energy expansion of the current
correlators, i.e., $q^2\gg m_1^2,m_2^2$, for example the case where $m_1=0$ and
$m_2=m$.  The imaginary part leads to ${\cal O}(\alpha_s^2)$ corrections
to $R(s)$ in the energy range sufficiently large as compared to the mass of
the produced quarks. One can think of the production of light quarks
($u,d,s$) accompanied by a pair of charm quarks ($m=M_c$) at energies
$\sqrt{s}\gsim5$~GeV, taking into account charm mass effects.

\begin{figure}[t]
  \begin{center}
  \leavevmode
   \epsfxsize=3cm
   \epsffile[150 260 420 450]{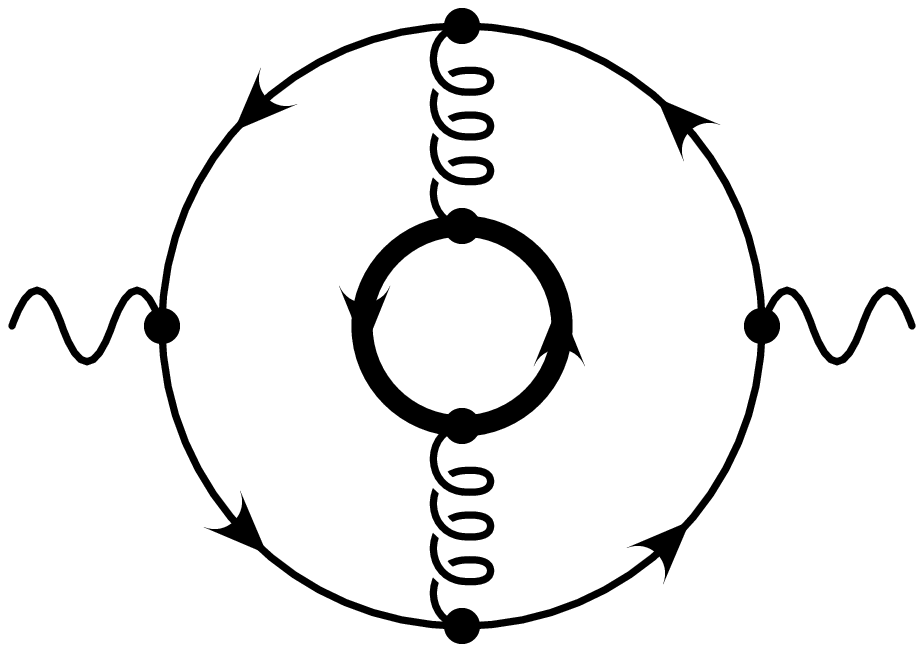}\hspace{1em}
   \raisebox{2.8em}
   {\Large $= \ \ 1\ \star $}
   \epsfxsize=3.cm
   \epsffile[150 260 420 450]{db.ps}\hspace{0em}
   \raisebox{2.8em}
   {\Large $+ \ \ 2\times \!\!$}
   \epsfxsize=1.5cm
   \raisebox{1.7em}{\epsffile[150 260 420 450]{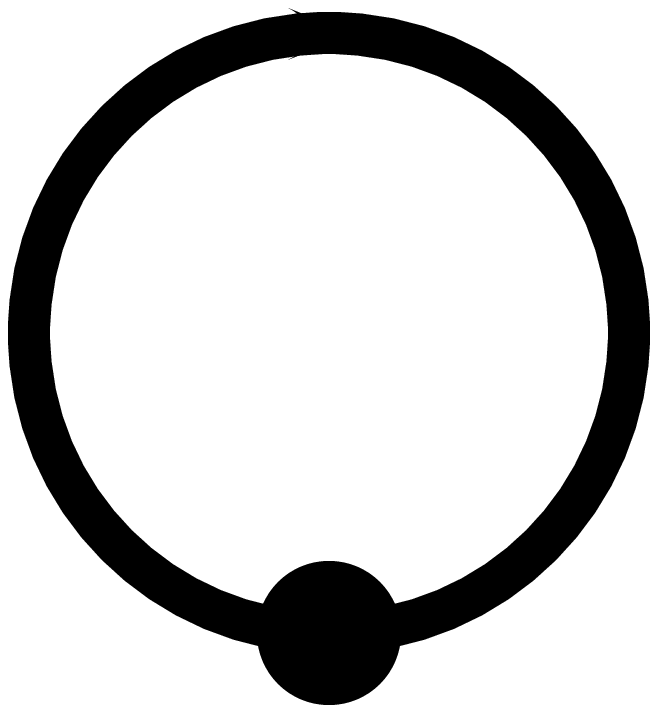}}\hspace{0em}
   \raisebox{2.8em}{\Large $\star$}
   \epsfxsize=3.cm
   \epsffile[150 260 420 450]{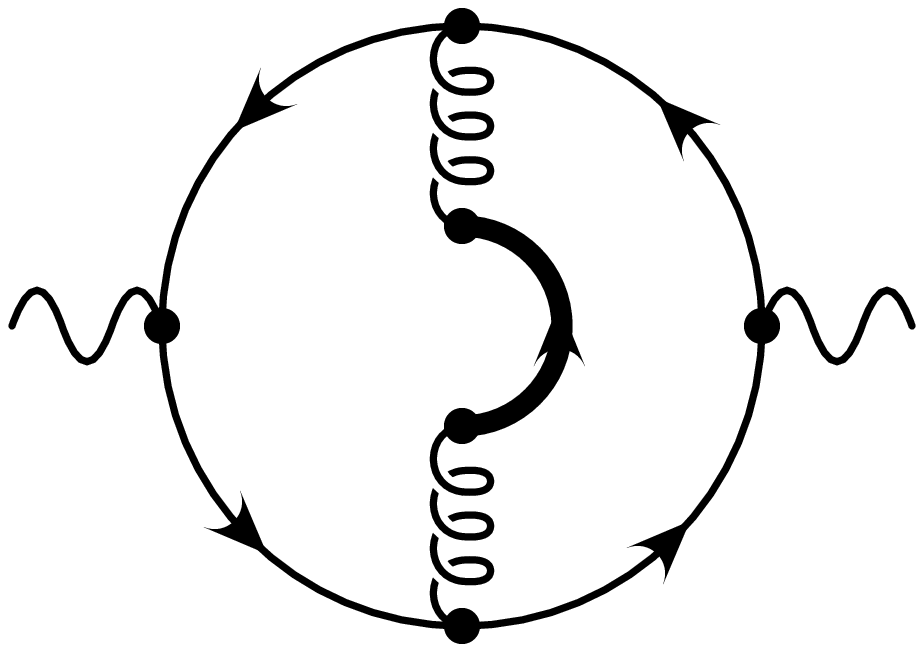}\hspace{1em}\\[1em]
   \mbox{\hspace{1em}}\raisebox{2.8em}
   {\Large $+$}
   \epsfxsize=1.5cm
   \raisebox{1.7em}{\epsffile[150 260 420 450]{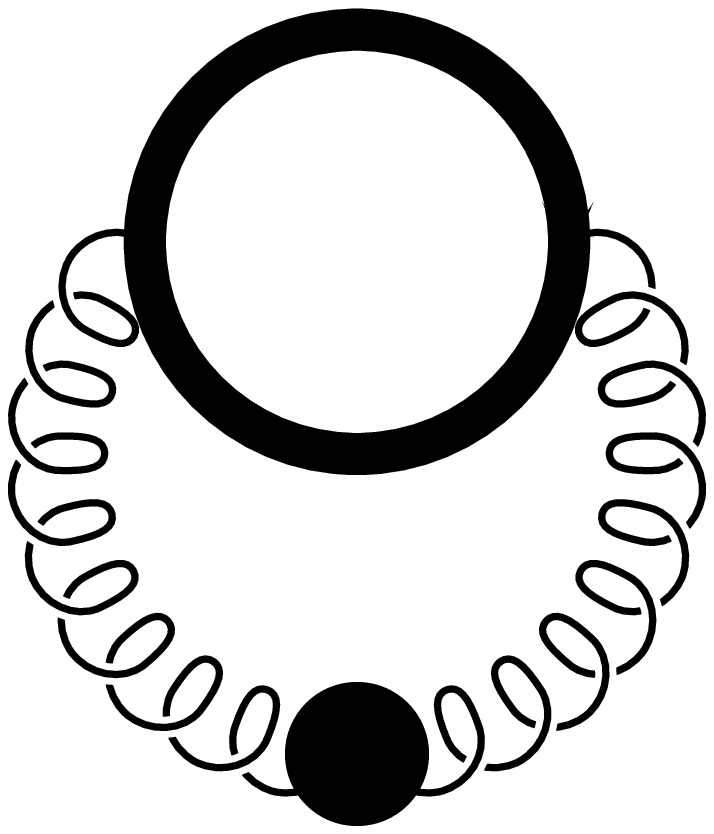}}
   \epsfxsize=3.cm
   \raisebox{2.8em}{\Large $\star$}
   \epsffile[150 260 420 450]{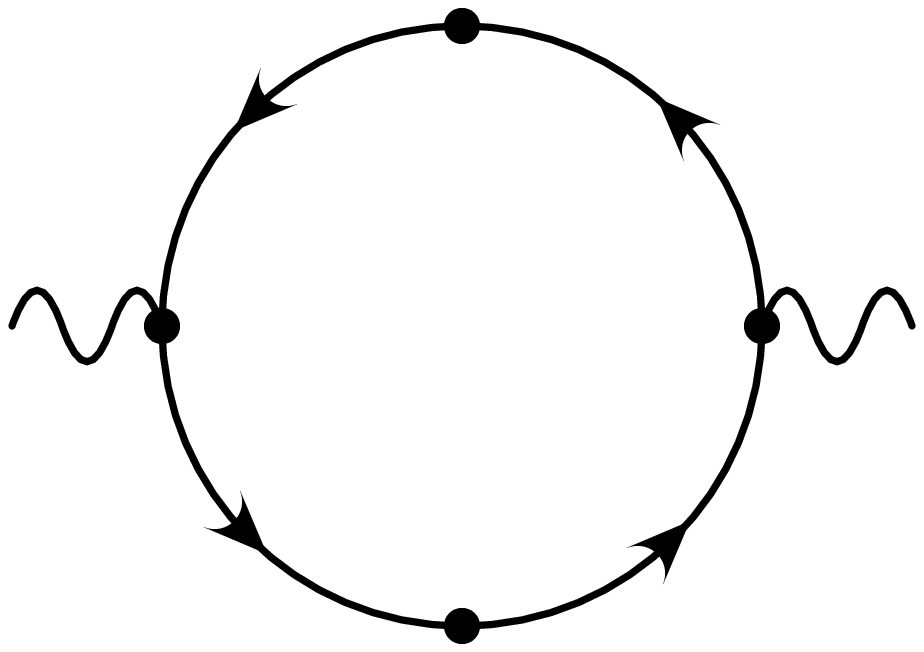}\hspace{0em}
   \hspace{.5em}\raisebox{2.8em}
   {\Large $+ \ \ 2\times \!\!\!\!$}
   \epsfxsize=3cm
   \raisebox{0em}{\epsffile[150 260 420 450]{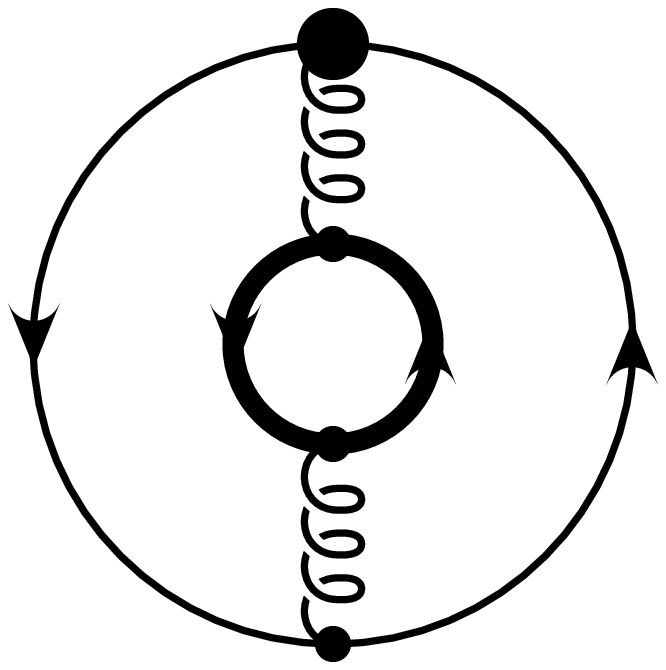}}
   \epsfxsize=3cm
   \raisebox{2.8em}{\Large $\star$}
   \epsffile[150 260 420 450]{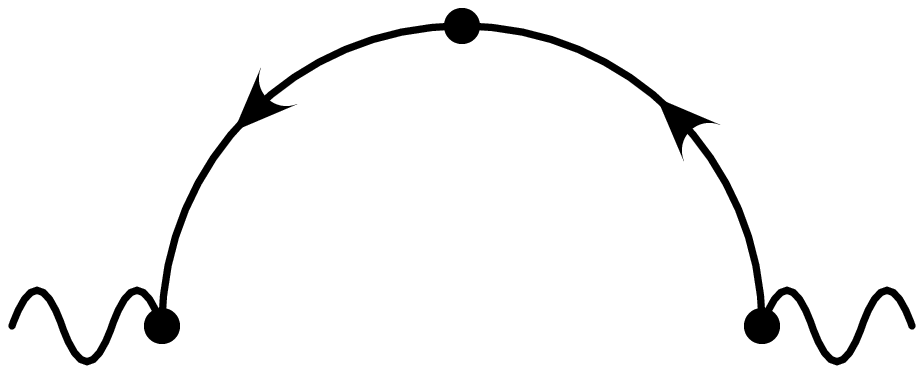}\hspace{0em}
   \parbox{\captionwidth}{
   \caption[]{\label{figdblmp} \sloppy
     Large-momentum procedure for the double-bubble diagram.
     The inner loop (thick lines)
     carries mass $m$, the outer one is massless, and so are the gluon
     lines. The square of the momentum $q$ flowing through
     the diagram is supposed to be much larger than $m^2$.  It is
     understood that the hard subgraphs (right of ``$\star$'') are to be
     expanded in $m$ and all external momenta except for $q$, and
     reinserted into the fat vertex dots of the co-subgraphs (left of
     ``$\star$'').  Contributions involving massless tadpoles are not
     displayed since they are zero in dimensional regularization.}}
 \end{center}
\end{figure}

Application of the large-momentum procedure to the diagram of
Fig.~\ref{figdb}(a) results in the subdiagrams displayed in
Fig.~\ref{figdblmp}. We have combined topologically identical diagrams
there. Similar terms arise from the graph of Fig.~\ref{figdb}(b). Since
the hard subgraphs have to be Taylor expanded in $m$ and any
``external'' momentum except $q$, they belong to the class of massless
two-point integrals, and can therefore be computed with the help of the
integration-by-parts technique.  The co-subgraphs, on the other hand,
are massive tadpole integrals, so that also here integration-by-parts
can be applied.  Note that the first of the four subdiagrams in
Fig.~\ref{figdblmp} represents the naive Taylor expansion of the
integrand with respect to $m$. It is clear that starting from a certain
order this term contains infra-red poles which are artificial as the
original diagram is infra-red finite.  The cancellation of these poles
in the sum of all terms on the r.h.s.\ of the equation in
Fig.~\ref{figdblmp} provides a non-trivial check of the
calculation.  The result of this expansion up to ${\cal O}(m^8/q^8)$
looks as follows:
\begin{eqnarray}
\bar{\Pi}_{gs}(q^2) &\stackrel{q^2\gg m^2}{=}&
  {3\over 16\pi^2}\,
  \left({\alpha_s\over \pi}\right)^2\,C_{\rm F}\,T\,\Bigg[
      -{3701\over 648} 
      + {38\over 9}\,\zeta_3 
                + \logqmums\,\bigg(
          {11\over 6} 
          - {4\over 3}\,\zeta_3
          \bigg) 
      - {1\over 6}\,\logqmums^2 
\nonumber\\&&\mbox{\hspace{1em}} 
      + {m^2\over q^2}\,\bigg(
          -{64\over 3} 
          + 16\,\zeta_3
          \bigg) 
      + \left({m^2\over q^2}\right)^2\,\bigg(
          -{67\over 3} 
          + 16\,\zeta_3 
          + \logqmms\,\big(
              -{26\over 3} 
              + 8\,\zeta_3
              \big)
          - \logqmms^2 
          \bigg)
\nonumber\\&&\mbox{\hspace{1em}} 
      + \left({m^2\over q^2}\right)^3\,\bigg(
          -{1552\over 243} 
          + {160\over 27}\,\zeta_3
          - {272\over 243}\,\logqmms 
          + {56\over 81}\,\logqmms^2 
          + {16\over 81}\,\logqmms^3 
          \bigg) 
\nonumber\\&&\mbox{\hspace{1em}} 
      + \left({m^2\over q^2}\right)^4\,\bigg(
          {1435\over 324} 
          - {10\over 3}\,\zeta_3
          + {113\over 54}\,\logqmms 
          - {1\over 18}\,\logqmms^2 
          - {1\over 9}\,\logqmms^3 
          \bigg) 
      \Bigg] + \cdots\,,
\label{eqdblmpres}
\end{eqnarray}
with $\logqmums=\ln(-q^2/\mu^2), \logqmms=\ln(-q^2/m^2)$ and
$\zeta_3=1.202056903\ldots$.  In Section~\ref{subsubexplmp} it is
shown how the computation of all subdiagrams can be done using program
packages designed to automate the large-momentum procedure.  The
logarithmic terms up to the fourth order can be found
in~\cite{CheKue94}, the others are new. We have adopted the
\msbar~scheme, i.e., after taking into account the counter-terms for
$\alpha_s$ induced by the (massless) one and two-loop diagrams, the
local poles are subtracted.  Note that because the mass $m$ is absent in
the lower order terms, it does not need to be renormalized.

As an application of the hard-mass procedure let us consider the hierarchy 
$m_1^2\ll q^2\ll m_2^2$. The imaginary part again leads to contributions
for $R(s)$. This time one may think of charm quark production ($m_1=M_c$)
in the presence of a virtual bottom quark ($m_2=M_b$). It turns out that
already the first term provides a very good approximation almost up
to the threshold $\sqrt{s}=2M_b$~\cite{Che93,HoaJezKueTeu94,Teudiss}.
For simplicity we set $m_1=0$ and $m_2=m$ in the following.

\begin{figure}[t]
  \begin{center}
  \leavevmode
   \epsfxsize=3cm
   \epsffile[150 260 420 450]{db.ps}\hspace{1em}
   \raisebox{2.8em}
   {\Large $=$}
   \epsfxsize=2.cm
   \raisebox{1.em}{\epsffile[150 260 420 450]{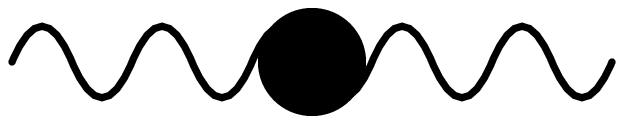}}\hspace{-1em}
   \raisebox{2.8em}{\Large $\ \ \star \!\!$}
   \epsfxsize=3.cm
   \epsffile[150 260 420 450]{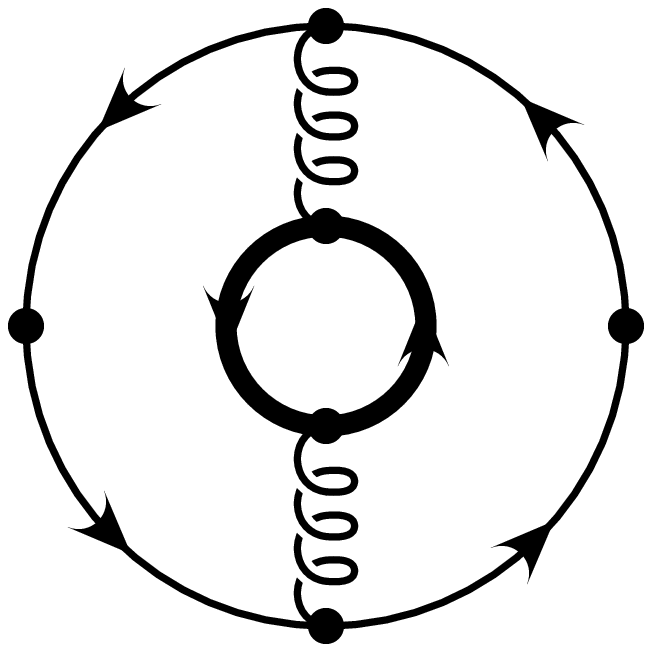}\hspace{0em}\\[1em]
   \mbox{\hspace{1em}}
   \raisebox{2.8em}
   {\Large $+ \ \ 2\times \ $}
   \epsfxsize=2.cm
   \raisebox{1.em}{\epsffile[150 260 420 450]{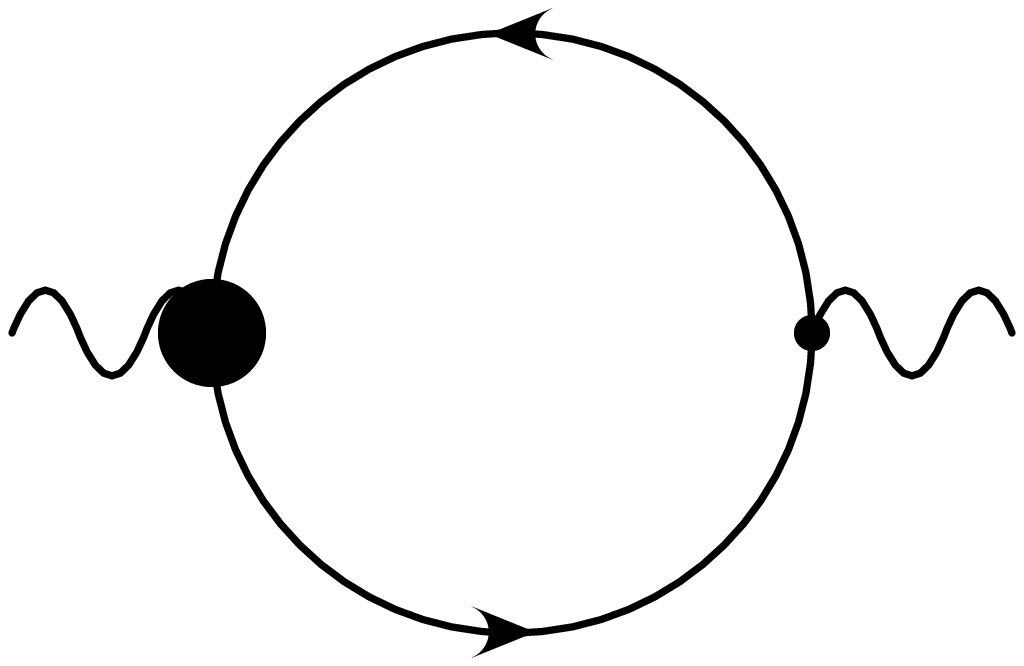}}\hspace{0em}
   \raisebox{2.8em}{\Large $\ \ \star \!\!$}
   \epsfxsize=3.cm
   \epsffile[150 260 420 450]{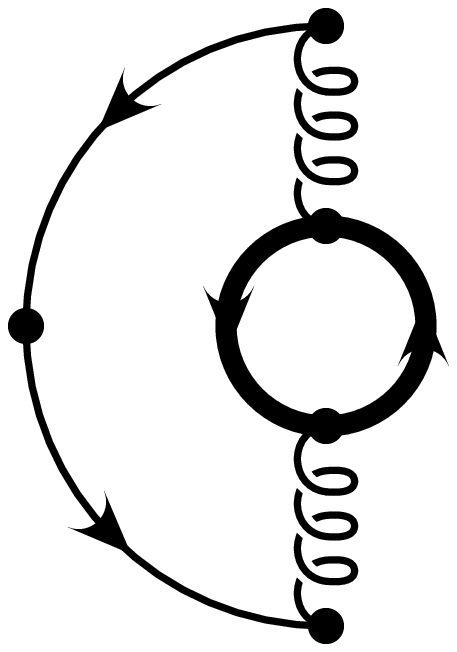}
   \raisebox{2.8em}
   {\Large $\!\!\!\!+\ \ $}
   \epsfxsize=3cm
   \raisebox{0em}{\epsffile[150 260 420 450]{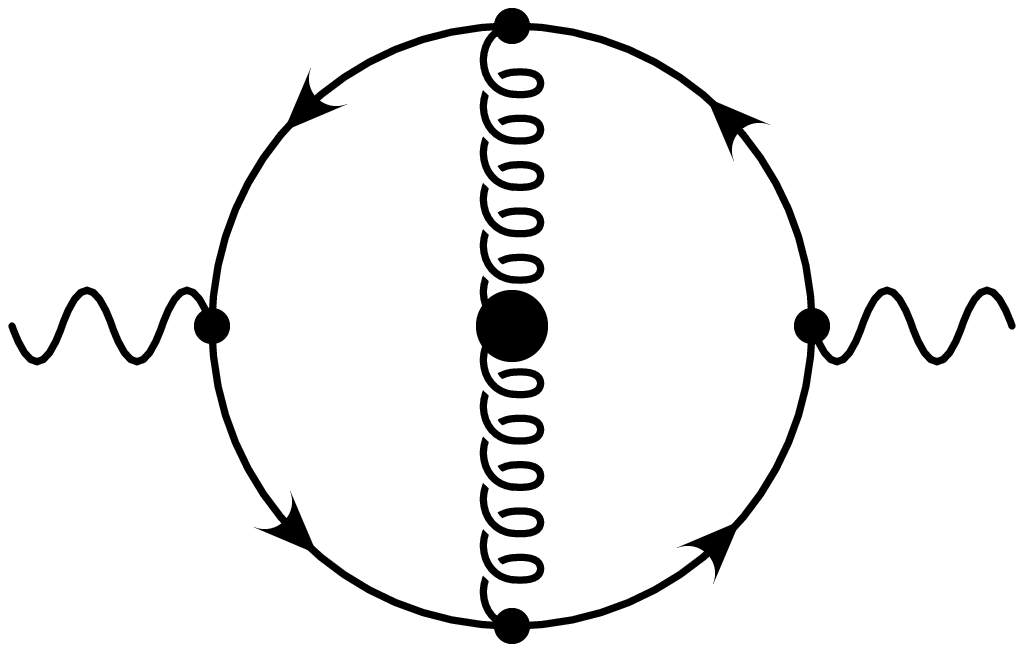}}
   \epsfxsize=3.cm
   \raisebox{2.8em}{\Large $\ \ \star \!\!\!\!\!\!$}
   \epsffile[150 260 420 450]{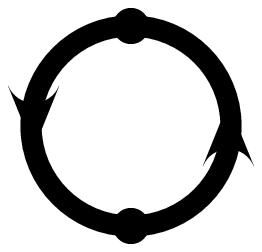}\hspace{0em}
    \parbox{\captionwidth}{
   \caption[]{\label{figdbhmp}\sloppy
     Hard-mass procedure for the double-bubble diagram. Now the
     hierarchy $q^2 \ll m^2$ is considered where $m$ is the mass of the
     inner line.  The hard subdiagrams (right of ``$\star$'') are to be
     expanded in all external momenta including $q$ and reinserted into
     the fat vertex dots of the co-subgraphs (left of~``$\star$'').}}
 \end{center}
\end{figure}

The corresponding diagrammatic representation
is shown in Fig.~\ref{figdbhmp}. There are three subdiagrams, one of
which again corresponds to the naive Taylor expansion of the integrand
in the external momentum $q$.
After Taylor expansion, the subdiagrams are reduced to tadpole
integrals with mass scale $m$. The scale of the co-subgraphs is given
by $q$ thus leading to massless propagator-type integrals.
The result for the first three terms reads~\cite{Stediss}:
\begin{eqnarray}
\bar{\Pi}_{gs}(q^2) &\stackrel{q^2\ll m^2}{=}&
   \frac{3}{16\pi^2}
  \left({\alpha_s\over \pi}\right)^2\,C_{\rm F}\,T\,\Bigg[
        \frac{295}{648} 
      + \frac{11}{6}\logqmums
      - \frac{1}{6}\logqmums^2
      - \frac{11}{6}\logqmms 
      + \frac{1}{6}\logqmms^2
\nonumber\\
&&\mbox{}
      - \frac{4}{3}\zeta_3\logqmums 
      + \frac{4}{3}\zeta_3\logqmms
      + \frac{q^2}{m^2}\left(
         \frac{3503}{10125} 
       - \frac{88}{675}\logqmms
       + \frac{2}{135}\logqmms^2
                       \right)
\nonumber\\
&&
       + \left(\frac{q^2}{m^2}\right)^2\left(
        - \frac{2047}{514500} 
        + \frac{1303}{529200}\logqmms
        - \frac{1}{2520}\logqmms^2
                                        \right)
\Bigg] + \cdots\,.
\label{eqdbhmpres}
\end{eqnarray}
If $m_1$ was different from zero, successive application of the
large-momentum procedure to the co-subgraphs would lead to a subsequent
expansion in $m_1^2/q^2$. An example for such a repeated use of
asymptotic expansions will be described in Section~\ref{sec::zbb}.
Alternatively, one may evaluate the integration analytically as it has
been done for the $q^2/m^2$ term in~\cite{Sei:dipl}, thus leading to the
full $m_1$ dependence of the power-suppressed result.



\subsection{Helicity-amplitude technique\label{sec::helamp}}
The standard way to obtain a cross section or a decay rate in
perturbative quantum field theories is to compute the squared amplitude
and integrate over the phase space for the final state particles. A
typical amplitude is a sum over terms of the form
\begin{equation}
  c\,\prod_i\epsilon_{\mu_i}(k_i,\chi_i)\,\prod_j \bar
  u(p_{j},\lambda_j)\,\Gamma_j\, u'(p'_j,\lambda'_j)\,,
  \label{eq::typamp}
\end{equation}
where $\epsilon$ are polarization vectors of vector particles with
momentum $k_i$ and polarization $\chi_i$, $u$ and $u'$ are spinors of
fermions (or anti-fermions) with momentum $p_j$, $p'_j$ and
helicity $\lambda_j$, $\lambda'_j$, respectively. $\Gamma_j$ are
matrix-valued objects in Dirac space, and $c$ is a scalar function of
momenta, masses, coupling constants etc.  Each term of the form
(\ref{eq::typamp}) corresponds to a certain Feynman diagram.  The
standard way to deal with the Dirac structure is to square the amplitude
{\it before} evaluating the expressions in (\ref{eq::typamp}) any
further. After summing over polarizations of final and initial states, one
employs the relations
\begin{equation}
\sum_{\lambda = \pm 1/2} u(p,\lambda)\bar u(p,\lambda) \sim p\!\!\!/ \pm m
\end{equation}
for fermions/anti-fermions and takes the trace in Dirac space which can
be evaluated with the help of the anti-commutator $\{\gamma_\mu,\gamma_\nu\}
= 2 g_{\mu\nu}$. Specific polarization configurations can be
investigated by introducing suitable projectors in Dirac space. 

However, the higher the order of perturbation theory, the more diagrams
contribute, and because the amplitude must be squared, the number of
terms to be evaluated in the above way increases even quadratically with
the number of diagrams.  As long as one stays with integrated quantities
(total rates), one way out is to apply the optical theorem, i.e., to
take the imaginary part of the forward scattering amplitude (see, e.g.,
Sections~\ref{sec::polfunc} and~\ref{sec::zbb}).
But as soon as one is interested in differential distributions,
this method is no longer applicable. A solution here is the so-called
helicity-amplitude technique. The basic idea is to rewrite
expressions of the form
\begin{equation}
u(p,\lambda)\bar u'(p',\lambda')
\end{equation}
for fixed helicities $\lambda,\lambda'$ in terms of Dirac matrices. This
allows the trace to be taken {\it before} squaring the amplitude. For
given four-momenta one arrives at a single complex number for the full
amplitude, one for each helicity configuration. If desired, the
summation over polarizations is done only {\it after} squaring the
amplitude.

This strategy goes back to \cite{BjoChe66} and was further developed in
\cite{ref::calkul}. Improved algorithms have been worked out in
\cite{ref::helamp,BalMai95}. For detailed discussions of the various
methods let us refer to these original works.


%
%
%

%% file: algebraic.tex
\subsection{\label{secalgprg}Algebraic Programs}
%
Automatic computation of Feynman diagrams would not be possible without
the development of powerful algebraic programming systems. While these
systems themselves are based on conventional languages like {\tt C} or
{\tt LISP}, in turn they constitute the breeding ground for all the
application software which is our main concern in this review.
Therefore, before discussing concrete packages that often are
exclusively designed for Feynman diagram calculations, it may be helpful
to briefly introduce four of the most frequently used algebraic
programming languages, with emphasis on their different strategies and
philosophy.

\subsubsection{{\tt Mathematica} and {\tt MAPLE}}
%
{\tt Mathematica} certainly is the program with the largest bandwidth of
possible applications. Apart from the algebraic operations, {\tt
  Mathematica} performs analytical and numerical integrations for quite
huge classes of integrals, it can do finite and infinite sums,
differentiation, matrix operations, it provides graphic routines etc.

{\tt MAPLE} is very similar to {\tt Mathematica} and often it is only a
question of taste which one to use. There certainly are fields where one
of them is superior to the other, and in extreme cases one may use them
in combination to find the result of a calculation.

The advantage of writing applications in {\tt Mathematica}
or {\tt MAPLE} is that a lot of algorithms are built in and, of course,
the whole environment --- the interactive platform constituting an
important piece in this respect --- is available for the user to operate
on the input and output. For example, one may perform numerical studies
by inserting numbers for the parameters in the symbolic result.
Meanwhile, a lot of users provide additional packages for
very different purposes.

One of the disadvantages of this generality is, of course, the rather
big requirements of disk and memory space. Furthermore, the improper use
of some very general commands ({\tt Integrate[]}, {\tt Simplify[]},
etc.) may lead to inadequately long operational times or even hang-ups, a
problem mainly occurring with unexperienced users.

\subsubsection{{\tt FORM}}
%
The power of {\tt FORM} lies in its capability of manipulating huge
algebraic expressions, the only limit in size being essentially the
amount of free disk space. While for {\tt Mathematica} or {\tt MAPLE}
operating on expressions of a few mega-bytes becomes rather
inconvenient, {\tt FORM} may easily deal with several hundred mega-bytes
and even giga-bytes. However, it is very restricted in its available
operations.  But after some experience one will be surprised how
powerful the so-called {\tt id}-statement is which replaces one
quantity by an in general more complicated expression.

The operational philosophy of {\tt FORM} is completely different to the
ones of {\tt Mathematica} and {\tt MAPLE}, and so far {\tt FORM} is
clearly optimized for algebraic operations in high energy physics.
The concept is mainly to bring any expression to a unique form by
fully expanding it into monomials (``terms''). Consider, for example, the
following screen-shot of a {\tt Mathematica}-session:
\begin{verbatim}
In[1]:= f1 = (a+b)^2

               2
Out[1]= (a + b)

In[2]:= f2 = a^2 + 2*a*b + b^2

         2            2
Out[2]= a  + 2 a b + b

In[3]:= diff = f1 - f2

          2            2          2
Out[3]= -a  - 2 a b - b  + (a + b)

In[4]:= Expand[diff]

Out[4]= 0
\end{verbatim}
This clearly demonstrates that the internal representations of {\tt f1}
and {\tt f2} inside {\tt Mathematica} are not the same.
{\tt FORM}, in contrast, will instantly expand the bracket in {\tt f1},
i.e., the input file
\begin{verbatim}
    Symbol a,b;
    NWrite statistics;

    L f1 = (a + b)^2;
    L f2 = a^2 + 2*a*b + b^2;

    L diff = f1 - f2;

    print;
    .end
\end{verbatim}
leads to the following output:
\begin{verbatim}
   f1 =
      2*a*b + a^2 + b^2;

   f2 =
      2*a*b + a^2 + b^2;

   diff = 0;
\end{verbatim}
All further operations on {\tt f1} or {\tt f2} will be {\it term-wise},
i.e., an {\tt id}-statement
\begin{verbatim}
id a = b + c;
.sort
\end{verbatim}
will first replace {\tt a} by {\tt b + c} in each of the three terms of
{\tt f1} and expand them again into monomials. The {\tt .sort} means to
collect all terms that only differ in their numerical coefficient and to
bring the new expression for {\tt f1} to a unique form again.  This
term-wise operation is the reason for {\tt FORM}'s capability of
dealing with such huge expressions: Only a single term is treated in
each operational step, all the others can be put on disk or wherever
{\tt FORM} considers it convenient.

In addition, identification of certain terms (``pattern matching'') is
more transparent
than in {\tt Mathematica}. For example, the user must be aware
of the fact that in order to nullify the mixed term in {\tt f1},
the function {\tt Expand[]} has to be applied before. 
To be concrete, in {\tt Mathematica}:
\begin{verbatim}
In[2]:= f1 /. a*b -> 0

               2
Out[2]= (a + b)

In[3]:= Expand[ f1 ] /. a*b -> 0

         2    2
Out[3]= a  + b
\end{verbatim}

It should be clear that the discussion above is not meant to judge if
{\tt FORM}'s or {\tt Mathematica}'s concept is the better one.  Each one
of them has its pros and cons, although the number of operations in {\tt
  FORM} is much more restricted than in {\tt Mathematica}.

\subsubsection{{\tt REDUCE}}
%
{\tt REDUCE}~\cite{reduce} is an algebra program that also was created
for particle physics. It is not so general like {\tt Mathematica}, but
for specific problems especially in high energy physics it may be more
efficient. In contrast to {\tt FORM} it has an interactive mode and many
more built-in functions, but again its capability of dealing with huge
expressions is more limited.

%
%

%% file: implement_intro.tex
%
\section{\label{secimpl}Implementation}
In Section~\ref{subsurvey} a brief survey of the program packages is
given which are very frequently used in high energy physics.
Section~\ref{subselpac} describes those packages in more detail which
were used to obtain the physical results presented in
Section~\ref{secapplications}.


%% file: survey.tex
%
\subsection{\label{subsurvey}Survey of the existing program packages}
%

This section gives a short overview of the existing program packages
written to automate the treatment of Feynman diagrams.  Meanwhile quite
a number of such packages exist, some of them published in journals and
available via anonymous ftp, others still under development and
therefore only accessible for a restricted group.  This review is not
supposed to serve as a catalogue to select the one package most suitable
for ones purposes. It shall simply provide an indication of what has
been done, what is doable, and what are the most urgent topics to
improve. Therefore we will not respect the question of availability.
Furthermore, the description of packages and applications unavoidably
will be biased, with emphasis on the topics the authors are and were
personally involved in.  However, since it allows us to go into details
as far as ever necessary, we hope the reader will benefit from this
strategy.  Nevertheless, we will try to be as objective as possible and
to fairly cover the class of most successful packages.

The first part will be devoted to programs concerned with the generation
of diagrams. For higher order corrections this becomes more and more
important, because the number of diagrams increases rapidly with the
order of perturbation theory.  The second part describes programs that
apply to the evaluation of the corresponding amplitudes on a
diagram-by-diagram basis. Some of them are optimized for the use in
combination with one of the generators mentioned before.  Part three
deals with two packages that automatically apply the rules of the
hard-mass and the large-momentum procedure.  Finally,
Section~\ref{subcompl} contains a collection of programs that mainly
combine some of the previously mentioned ones in order to treat full
processes from the very generation up to the summation of the results of
all diagrams.

The aim of this section is not to compare or judge the listed programs.
This is inadmissible anyway since each of them has its specific main
focuses.  We will just describe their needs, abilities and applications.
Each package is based on one (sometimes also several) algebraic
programs, the most important of which were introduced in
Section~\ref{secalgprg}.  A summary in table form will also include some
additional packages that could not be described in more detail.

\subsubsection{Generation of Feynman diagrams}
%
Two programs of quite different design will be discribed: {\tt FeynArts}
and {\tt QGRAF}.


\paragraph{{\tt FeynArts}}\mbox{}\\[1em]
%
{\tt FeynArts}~\cite{FeynArts} is written in {\tt
  Mathematica} and may be used interactively.  Given the
number of loops and external particles, {\tt FeynArts} first creates all
possible topologies, allowing for additional criteria to select some
subset of diagrams.  In the second step, fields must be attributed to
the lines which is most conveniently done by choosing a specific field
theoretical model.  The most popular models are predefined, but the user
may equally well provide his own ones.  {\tt FeynArts} has the nice
feature of drawing the generated diagrams which works to two loops with the
default setup but may be extended to higher loop order by the
user.  This is very helpful for debugging, for example, by figuring out
if the desired subset of diagrams was selected correctly.
Finally, a
mathematical expression is generated for each diagram which may then be
further evaluated.  This is preferably done using the program packages
{\tt FeynCalc}, {\tt FormCalc} or {\tt TwoCalc} (see below)
because the user's intervention remains minimal in that case.
It is very convenient that {\tt FeynArts} works within the {\tt
  Mathematica} environment since a lot of powerful commands are
available here, allowing the manipulation of intermediate and final
results.


\paragraph{{\tt QGRAF}}\mbox{}\\[1em]
%
The program {\tt QGRAF}~\cite{Nog93} is written in {\tt FORTRAN~77} and
is a rather efficient generator for Feynman diagrams.  It takes, for
example, only a few seconds to generate $10000$ diagrams.  The user has
to provide two files: the first one, called {\it model file}, contains
the vertices and propagators in a purely topological notation.  In the
second one, the {\it process file}, the initial and final states as well
as the number of loops must be defined.  Furthermore, as for {\tt
  FeynArts}, one may provide several options allowing the selection of
certain subclasses of diagrams.

Similar to the input, also the output of {\tt QGRAF} is very abstract.
It encodes the diagrams in a symbolic notation,
thereby reproducing all necessary combinatoric
factors as well as minus signs induced by fermion loops. While {\tt
  QGRAF} suggests a distribution for the loop- and external momenta
among the propagators, it is the users task to insert the Feynman rules,
i.e., the proper mathematical expressions for the vertices and
propagators.  Nevertheless, {\tt QGRAF} allows one to choose among different
output formats which increases the flexibility for further operations.



\subsubsection{\label{sec:surv:comp}Computation of diagrams}

%
This section lists programs to be used for the computation of Feynman
diagrams, all working on a diagram-by-diagram basis. Packages dealing
with whole processes will be described in Section~\ref{subcompl}.

We will start with three packages based on {\tt FORM}~\cite{form} and
dealing with single-scale integrals: massive integrals with zero
external momentum, massless integrals with one external momentum and
two-point functions on their mass shell.  Concerning their input
requirements and their principle structure they are quite comparable.

Furthermore we will describe certain {\tt Mathematica} packages
computing one- and two-loop diagrams and a {\tt MAPLE} program which
provides a graphical interface for such calculations.


\paragraph{{\tt MINCER}}\mbox{}\\[1em]
%
The package {\tt MINCER} computes one-, two- and three-loop integrals
where all lines are massless and only one external momentum is different
from zero.  The first version of {\tt MINCER}~\cite{MINCER1} was written
in {\tt SCHOONSCHIP}~\cite{VelSS}, but here we will only describe the
much more elaborate {\tt FORM}-version~\cite{mincer2}. {\tt MINCER} was
the first implementation of the integration-by-parts algorithm and
certainly is one of the most important programs for multi-loop
calculations. It is supposed to be highly efficient, in particular
because the author of {\tt FORM} was involved in the translation to this
system.

The user has to provide the diagrams that may all be listed in a single
file, separated with the help of the fold-option of {\tt FORM}.  The
input notation is based on the following scheme: The momenta carried by
the propagators are numerated by integer numbers fixed by the topology
of the diagram. The topology itself, uniquely classifying the diagram,
must be specified through a keyword.  The final result is given as an
expansion in $\varepsilon$ where one-loop results are expanded up to
${\cal O}(\varepsilon^2)$, two-loop ones up to ${\cal O}(\varepsilon)$
and three-loop ones up to the finite part.


\paragraph{{\tt MATAD}}\mbox{}\\[1em]
%
One-, two- and three-loop vacuum integrals can be evaluated with the
help of {\tt MATAD}~\cite{Stediss}, also written in {\tt FORM}. Each
propagator may either be massless or carry the common mass, $M$, and all
external momenta have to be zero.

The input notation as well as the whole concept is very similar to the
one of {\tt MINCER}. In particular, the core of the routines is again
formed by the integration-by-parts algorithm.  {\tt MATAD} provides an
interface for using {\tt MINCER} and, furthermore, it supports Taylor
expansion in small masses or momenta. For example, given a massive
diagram with one small external momentum, after Taylor expansion of the
integrand in the small momentum one is left with a vacuum integral
again.  At different stages of the computation there is the opportunity
to interact from outside in order to control the expansion, apply
projectors, or perform other operations.

The applicability of both {\tt MINCER} and {\tt MATAD} seems to be quite
restricted.  However, as we have seen in
Section~\ref{subasymp}, if a diagram has a certain hierarchy of mass scales
which happens to be the case for quite a lot of physical applications,
it can be reduced to products of single-scale integrals. Then {\tt
  MINCER} and {\tt MATAD} can be used in combination to arrive at an
analytical result.


\paragraph{{\tt SHELL2}}\mbox{}\\[1em]
%
The program {\tt SHELL2}~\cite{SHELL2} is also written in {\tt FORM} and
deals with one- and two-loop propagator-type on-shell integrals.  The
implemented topologies allow computations mainly in QED and QCD.  The
prototype examples are the two-loop contribution to $g-2$ of the
electron and the relation between the $\overline{\rm MS}$ and
on-shell mass up to order $\alpha_s^2$ in QCD.

The user has to provide the diagrams in terms of a polynomial
representing the numerator and abbreviated denominators. This must be
supplemented by a label indicating both the number of loops and the
topology. Except for the on-shell momentum no other external momentum
may be present because then the numerator can be completely decomposed
in terms of the denominators.  A small {\tt FORM} program ensures that an
input very similar to the one of {\tt MATAD} and {\tt MINCER} can be
applied so that one may run all three packages in parallel.


\paragraph{{\tt FeynCalc}, {\tt FormCalc} and {\tt TwoCalc}}\mbox{}\\[1em]
%
The {\tt Mathematica} package {\tt FeynCalc}~\cite{FeynCalc} is based on
a quite different philosophy than the ones described above.  Its main
applications are one-loop radiative corrections in the Standard Model and
its extensions. The diagrams may either be provided by hand or one uses
the output of the generator {\tt FeynArts} (see above).  {\tt FeynCalc}
performs the Dirac algebra and applies the tensor reduction algorithm of
Section~\ref{sectensdec} in order to express the result in terms of
scalar integrals. Special functions and the power of {\tt Mathematica}
allow to conveniently handle the intermediate and final expressions.

Since {\tt FeynCalc} is fully based on {\tt Mathematica}, its
performance is rather slow if the underlying expressions get large.
Combining the advantages of {\tt Mathematica} and {\tt FORM}, the
package {\tt FormCalc}~\cite{FormCalc} is a sped-up version of {\tt
  FeynCalc} well suited for huge problems at the one-loop level.

The {\tt Mathematica} package {\tt TwoCalc}~\cite{WeiSchBoe94} is to
some degree the extension of {\tt FeynCalc} to two loops. It applies,
however, only to two-point functions, since only for them may the tensor
reduction algorithm be generalized to two loops, as was already noted in
Section~\ref{sectensdec}.  Operations that are not specific for two-loop
calculations, like the evaluation of Dirac traces, are passed to {\tt
  FeynCalc}, so that mainly the reduction of the tensor integrals to a
basic set of one- and two-loop integrals is performed.

The numerical evaluation of the scalar one-loop integrals may
conveniently be performed with the help of {\tt LoopTools} to be
described below. The two-loop integrals resulting from the {\tt
  TwoCalc}-routines may be evaluated using the {\tt C}-programs {\tt
  s2lse} and {\tt master}~\cite{Bauetal}.

\pagebreak[4]

\paragraph{{\tt LoopTools}}\mbox{}\\[1em]
%
{\tt LoopTools}~\cite{FormCalc} is an integration of the {\tt FORTRAN}
program {\tt FF}~\cite{ffmanual} into {\tt Mathematica}. It allows a
convenient numerical evaluation of the results as obtained by {\tt
  FeynCalc} or {\tt FormCalc}. In addition, it extends the ability of
{\tt FF} from doing only scalar integrals to the coefficients of the
tensor decomposition $B_{ij}, C_{ijk}, D_{ijkl}$ as described in
Section~\ref{sectensdec}, so that it is not even necessary to fully
reduce the tensor integrals with the help of {\tt FeynCalc} in order to
arrive at numerical results.  Indeed, {\tt FormCalc} reduces the tensor
integrals only up to the point where {\tt LoopTools} is applicable.


\paragraph{{\tt ProcessDiagram}}\mbox{}\\[1em]
%
The package {\tt ProcessDiagram}~\cite{ProcDia} is written in {\tt
  Mathematica} and deals with one- and two-loop vacuum diagrams without
any restrictions concerning masses. One may apply Taylor expansions with
respect to external momenta before integration.  In addition, some
auxiliary functions to simplify the results are available. It is, for
example, possible to expand the result with respect to the masses.


\paragraph{{\tt XLOOPS}}\mbox{}\\[1em]
%
One- and two-loop diagrams can be processed in a convenient way with the
help of the program package {\tt XLOOPS}~\cite{XLOOPS}.  An {\tt
  Xwindows} interface based on {\tt Tcl/Tk} allows even unexperienced
users to compute loop diagrams.  After choosing the topology, the
particle type of each line must be specified.  {\tt XLOOPS} then
performs the Dirac algebra and expresses the result in terms of certain
one- and two-loop integrals.  Whereas the one-loop integrals are
evaluated analytically, for the two-loop ones the result is reduced to
an at most two-fold integral representation which is then evaluated
numerically.  The main underlying strategy in this reduction is based on
the parallel space technique~\cite{CzaKilKre95}.  For the analytical
part of the calculations the algebra program {\tt MAPLE~V}~\cite{MAPLE}
is used; the numerical integrations are done with the help of {\tt
  VEGAS}~\cite{VEGAS,pvegas}.



\subsubsection{\label{subgensd}Generation of Subdiagrams}
%

This section describes two packages, {\tt LMP} and {\tt
  EXP}, that automatically apply the hard-mass
and the large-momentum procedure to a given diagram. The basic
concept and the realization of these programs is quite similar.
However, whereas {\tt LMP} is written in {\tt PERL} and was specialized
for the large-momentum procedure, {\tt EXP} is written in {\tt
  FORTRAN~90} and applies to the case of large external momenta, large
internal masses, and a combination of both.


\paragraph{{\tt LMP} and {\tt EXP}}\mbox{}\\[1em]
%
{\tt LMP}~\cite{Har:diss} is written in {\tt PERL} and was developed to
fully exploit the power of the large-momentum procedure. It therefore
extends the range of analytically calculable three-loop diagrams from
single-scale ones to two-point functions carrying small masses. It found
its main application in the evaluation of quark mass corrections to
certain QCD processes.

Its basic concept was then carried over to the {\tt FORTRAN~90} program
{\tt EXP}~\cite{Sei:dipl} which not only evaluates the large-momentum as
well as the hard-mass procedure but also any successive application of
both. Although the first version of {\tt EXP} is only capable of dealing
with two-point functions, the next version will also incorporate the
case of an arbitrary number of external momenta and masses, so that any
three-loop integral will be calculable in terms of a (possibly nested)
series in ratios of the involved mass scales. So far, {\tt EXP} was
applied to the decay rate of the $Z$ boson into
quarks~\cite{HarSeiSte97}, but one can certainly think of a huge number
of further possible applications.

The usage of both {\tt LMP} and {\tt EXP} is very similar. Their input
and output is adapted to {\tt MINCER} and {\tt MATAD}, so that
the experienced user of these two packages will only have to
provide some additional input information.


\subsubsection{\label{subcompl}Complete packages}

%
In this section five program packages are described
which automate the evaluation of a given process from
the generation up to the computation of the corresponding amplitudes.
Both their methods and purposes are quite different.

{\tt GEFICOM} computes Feynman diagrams up to three loops in analytical
form by expanding them in terms of single-scale integrals; processes
containing different mass scales are evaluated in terms of expansions. A
different setup was used at NIKHEF (Amsterdam) to compute, for example,
four-loop tadpole diagrams up to their simple poles in $\varepsilon$.
Further on, {\tt CompHEP} is a multi-leg system calculating cross
sections at tree level, involving up to five particles in the final
state. The automation of {\tt GRACE} relies heavily on numerical methods
and has a similar field of application like {\tt CompHEP}.  Finally, an
approach to provide a rather general environment for the computation of
one- and two-loop diagrams is described with the programs {\tt TLAMM}
and {\tt DIANA}.


\paragraph{{\tt GEFICOM}}\mbox{}\\[1em]
%
The program package {\tt GEFICOM}~\cite{geficom} combines the generator
{\tt QGRAF}, the integration packages {\tt MATAD} and {\tt MINCER} and
the programs {\tt EXP} and {\tt LMP} concerned with asymptotic expansion
to compute Feynman diagrams up to three loops. The translation of the
{\tt QGRAF} output to {\tt MINCER}/{\tt MATAD} notation is done by a
collection of {\tt Mathematica} routines.  The links between the single
packages is done by script languages like {\tt AWK} and {\tt PERL}.

With the short descriptions of its components in the previous section it
is rather clear what the purpose of {\tt GEFICOM} is: Given the initial
and final states, {\tt GEFICOM} generates and computes all contributing
diagrams up to three loops in analytic form, provided that the involved
mass scales are subject to a certain hierarchy. The result is obtained in
terms of an in general multiple expansion in the ratios of the mass
scales. The input therefore is essentially the same as for {\tt QGRAF},
except that one additionally assigns masses to the different particles
and defines a hierarchy among them.


\paragraph{``NIKHEF setup''}\mbox{}\\[1em]
%
When the number of diagrams contributing to a single quantity increases
and gets of the order $10^4$ one should think carefully about the
organization of the calculation.  For example, storing the result of
each individual diagram to a separate file may push the operational
system over its limits.  In the ``NIKHEF setup'' a database-like tool
named {\tt Minos}~\cite{minos} is used to circumvent such book-keeping
problems.  It contains {\tt make}-like and lots of additional features.
For example, it helps to find bottlenecks of the setup by reporting on
the subproblem on which most of the CPU time was spent.

For the generation of the diagrams the program {\tt QGRAF} is used. The
output of {\tt QGRAF} is translated to the notation of
the {\tt FORM} routines {\tt MINCER} and {\tt BUBBLES}~\cite{bubbles}
which are concerned with the integration of the massless two-point
functions and the massive tadpole integrals, respectively.  Another {\tt
  FORM} program, named {\tt Color}~\cite{color}, determines the colour
factor. The resulting expressions are inserted into a database.  After
integration, the results of the single diagrams are written to another
database.  Finally, the diagrams are multiplied by their colour factor
and summed up.


\paragraph{{\tt CompHEP}}\mbox{}\\[1em]
%
{\tt CompHEP} \cite{comphep} is a program package which allows the
evaluation of scattering processes and decay rates at tree level.  A
menu-driven interface makes its use quite handy.  There are several
built-in models among which one finds the Standard Model both in
unitary and in 't~Hooft-Feynman gauge, or the Minimal Supersymmetric
Standard Model. Modification of these models and definition of new ones
can be done in a very convenient way. Using {\tt LanHEP} (see
Section~\ref{submisc}) which works out the Feynman rules in {\tt
  CompHEP} format, it even suffices to provide a Lagrangian density.

After a model is selected and the process is specified, the Feynman
diagrams are generated and graphically displayed on the screen.  The
user may now select the diagrams to be treated further. Then the squared
Feynman amplitudes are generated and displayed, and the corresponding
analytical expressions are computed.  They may be stored either in {\tt
  REDUCE} or {\tt Mathematica} format which simplifies further
symbolical manipulations.  For complicated processes {\tt FORTRAN} or
{\tt C} code is generated, allowing for numerical studies.  On the other
hand, if the number of diagrams is not too large, numerical integrations
may be performed immediately and plots showing angular distributions and
cross sections can be produced.

The numerical part of {\tt CompHEP} is based on the Monte-Carlo
integration routine {\tt VEGAS}.  It is possible to introduce
cuts on various kinematical variables.
Further on, distributions, cross sections
and particle widths can be evaluated.  For the incoming particles one
may define structure functions and then repeat the same integrations.
Finally, {\tt CompHEP} generates events and displays the corresponding
histograms.


\paragraph{{\tt GRACE}}\mbox{}\\[1em]
%
A different approach for the automatic computation of Feynman diagrams
is realized by the program package {\tt GRACE}~\cite{grace}.  It was
developed to compute cross sections and radiative corrections
to them. Also here several models are available, including the
Minimal Supersymmetric Standard Model~\cite{Jim95}.

{\tt GRACE} has its own generator for Feynman diagrams,
producing {\tt FORTRAN} source code for each of them and passing it
to the package {\tt CHANEL}~\cite{CHANEL} which performs the calculation
with purely numerical methods. The user is free to directly evaluate the
amplitudes rather than the squared ones. This significantly reduces
the size of the expressions (see Section~\ref{sec::helamp}).

In the final step the integration over the phase space for the
particular final state is performed with the help of the
multi-dimensional integration package {\tt BASES} and the event
generator {\tt SPRING} which allows to generate unweighted event
flow~\cite{BASES/SPRING}.


\paragraph{{\tt TLAMM} and {\tt DIANA}}\mbox{}\\[1em]
%
There are two projects, partly under development, the purpose of which
is to automate the evaluation of processes involving one- and two-loop
diagrams. The package {\tt TLAMM}~\cite{tlamm} is written in {\tt C} and was
mainly developed to compute the two-loop corrections to the anomalous
magnetic moment of the muon. It uses {\tt QGRAF} to generate the
diagrams, translates the output to {\tt FORM} code and executes the
integration routines.

The second package, called {\tt DIANA}~\cite{diana},
is supposed to work on a more
universal basis. Again, in a first step the output of {\tt QGRAF} is
read, the topologies are determined and internal representations for the
diagrams are created.  Subsequently, an interpreter executes a ``special
text manipulating language'' ({\tt TM}) which allows one to select the
algebra language to be adopted for the calculation, to pass the
expressions to {\tt FORTRAN} in order to perform a numerical calculation
or to generate, for example, PostScript files of the diagrams.



\subsubsection{\label{submisc}Miscellaneous}
%
There are many more programs developed by several different groups.
Many of them are neither published nor documented in the literature,
especially if they were designed for a very special task only.  This
section is supposed to touch on some of them in a more or less encyclopedic
form\footnote{ The links to the corresponding {\tt www} and {\tt ftp}
  sites can also be found at\\ {\tt
    http://www-ttp.physik.uni-karlsruhe.de/Links/algprog.html } }.  For
completeness also the programs already discussed are included in the
list below.  Note that when we did not include information on the
availability of a program this means that we could not find an
apropriate {\tt ftp} or {\tt http} site.  Interested readers should
contact the authors of the corresponding programs in this case.  Some
programs are also available from the CPC program library ({\tt
  http://www.cpc.cs.qub.ac.uk}) upon request, even if it is not
indicated below.

\newlength{\savebaselineskip}
\setlength{\savebaselineskip}{\baselineskip}
\begin{itemize}
\setlength{\baselineskip}{.5em}


\item{\tt BASES} \cite{BASES/SPRING}:
  \begin{itemize}
  \item{\it Availability:} CPC program library
  \item{\it Purpose:} Monte Carlo integration
  \item{\it Source:} {\tt FORTRAN}
  \end{itemize}
  

\item{\tt BUBBLES} \cite{bubbles}:
  \begin{itemize}
  \item{\it Purpose:} analytical computation of purely massive
    four-loop tadpole integrals up to ${\cal O}(1/\varepsilon)$
  \item{\it Algorithms:} integration-by-parts
  \item{\it Source:} {\tt FORM}
  \end{itemize}
  

\item{\tt CHANEL} \cite{CHANEL}:
  \begin{itemize}
  \item{\it Availability:} CPC program library
  \item{\it Purpose:} library for the calculation of helicity amplitudes
  \item{\it Source:} {\tt FORTRAN}
  \end{itemize}
  

\item {\tt Color} \cite{color}:
  \begin{itemize}
  \item {\it Availability:}
    {\tt http://norma.nikhef.nl/\~\/t68/FORMapplications/Color}
  \item{\it Purpose:} computation of colour factors
  \item {\it Source:} {\tt FORM}
  \end{itemize}
  

\item {\tt CompHEP} \cite{comphep}:
  \begin{itemize} 
  \item {\it Availability:} {\tt http://theory.npi.msu.su/\~\/comphep}
    or {\tt http://www.ifh.de/\~\/pukhov}
  \item {\it Purpose:} symbolic and numerical computation of tree level
    processes with up to six external legs
  \item {\it Algorithms:} symbolic evaluation of squared diagrams,
    recursive representation of kinematics, Monte Carlo integration
  \item{\it Source:} {\tt FORTRAN}, {\tt C}
  \item{\it Uses:} {\tt VEGAS}
  \item {\it Preferably combined with:} {\tt LanHEP}
  \end{itemize}


\item {\tt DIANA} \cite{diana}:
  \begin{itemize}
  \item{\it Availability:} upon request from the author
  \item{\it Purpose:} general environment for generating and evaluating
    Feynman diagrams
  \item{\it Source:} {\tt C}
  \item{\it Uses:} {\tt QGRAF}
  \end{itemize}


\item {\tt EXP} \cite{Sei:dipl}:
  \begin{itemize}
  \item{\it Purpose:} reduce two-point functions to single-scale integrals
  \item{\it Algorithms:} large-momentum procedure, hard-mass procedure
    (see Section~\ref{subasymp})
  \item {\it Preferably combined with:} {\tt MATAD}, {\tt MINCER}
  \item{\it Source:} {\tt FORTRAN~90}
  \end{itemize}


\item {\tt FeynArts} \cite{FeynArts}:
  \begin{itemize}
  \item{\it Availability:} {\tt
      http://www-itp.physik.uni-karlsruhe.de/feynarts}
  \item{\it Purpose:} diagram generator with main focus on one- and
    two-loop cases
  \item{\it Source:} {\tt Mathematica}
  \item{\it Preferably combined with:} {\tt FeynCalc}, {\tt FormCalc}, 
    {\tt TwoCalc}
  \end{itemize}


\item {\tt FeynCalc} \cite{FeynCalc}:
  \begin{itemize}
  \item{\it Availability:} {\tt
      http://www.mertig.com} (commercial);\\ the latest free version is
      available at {\tt http://www.mertig.com/oldfc}
  \item{\it Purpose:} reduction of arbitrary one-loop to a set of basis
    integrals
  \item{\it Algorithms:} tensor decomposition and tensor reduction by
    means of \cite{PasVel79} (see Section~\ref{sectensdec})
  \item{\it Source:} {\tt Mathematica}
  \item{\it Preferably combined with:} {\tt FeynArts}, {\tt LoopTools}
  \end{itemize}


\item {\tt FF} \cite{OldVer89,ffmanual}:
  \begin{itemize}
  \item{\it Availability:} {\tt http://www.xs4all.nl/\~\/gjvo/FF.html}
  \item{\it Purpose:} numerical computation of scalar and vector one-loop
    integrals up to six-point functions
  \item{\it Source:} {\tt FORTRAN}
  \end{itemize}


\item {\tt FormCalc} \cite{FormCalc}:
  \begin{itemize}
  \item{\it Availability:} {\tt
      http://www-itp.physik.uni-karlsruhe.de/formcalc}
  \item{\it Purpose:} slimmed, high speed version of {\tt FeynCalc}
  \item{\it Algorithms:} tensor decomposition
  \item{\it Source:} {\tt Mathematica}, {\tt FORM}
  \item{\it Preferably combined with:} {\tt FeynArts}, {\tt LoopTools} 
  \end{itemize}


\item {\tt GEFICOM} \cite{geficom}:
  \begin{itemize}
  \item{\it Purpose:} automatic generation and computation of three-loop
    Feynman diagrams in terms of expansions
  \item {\it Uses:} {\tt QGRAF}, {\tt MATAD}, {\tt MINCER},
    {\tt EXP} or {\tt LMP}
  \item {\it Source:} {\tt Mathematica}, {\tt FORM}, {\tt AWK}
  \end{itemize}


\item {\tt GRACE} \cite{grace}:
  \begin{itemize}
  \item {\it Availability:} {\tt ftp://ftp.kek.jp/kek/minami/grace}
  \item {\it Purpose:} numerical computation of $2\to 2$ scattering
    processes to one-loop and multi-particle scattering processes at
    tree level
  \item {\it Algorithms:} helicity-amplitude method (see
    Section~\ref{sec::helamp}), Monte Carlo integration
  \item {\it Uses:} {\tt CHANEL}, {\tt BASES}, {\tt SPRING}
  \item{\it Source:} {\tt C}, {\tt FORTRAN}
  \end{itemize}


\item {\tt HELAS} \cite{HELAS}:
  \begin{itemize}
  \item{\it Purpose:} helicity-amplitude subroutines for Feynman diagram
    evaluations 
  \item{\it Source:} {\tt FORTRAN}
  \end{itemize}


\item {\tt HEPLoops} \cite{HEPLoops}:
 \begin{itemize}
  \item {\it Availability:} upon request from the author
  \item{\it Purpose:} analytical computation of massless propagator-type
    diagrams up to three loops
  \item {\it Algorithms:} integration-by-parts (see Section~\ref{sec::IP})
  \item {\it Source:} {\tt FORM}
  \end{itemize}


\item {\tt LanHEP} \cite{lanhep}:
  \begin{itemize}
  \item {\it Availability:}
    {\tt http://theory.npi.msu.su/\~\/semenov/lanhep.html}
  \item {\it Purpose:} generate Feynman rules from Lagrangian
  \item {\it Preferably combined with:} {\tt CompHEP}
  \item{\it Source:} {\tt C}
  \end{itemize}


\item {\tt LMP} \cite{Har:diss}:
  \begin{itemize}
  \item{\it Purpose:} factorize large external momentum in two-point
    functions
  \item{\it Algorithms:} large-momentum procedure (see
    Section~\ref{subasymp})
  \item {\it Source:} {\tt PERL}
  \item {\it Preferably combined with:} {\tt MATAD}, {\tt MINCER}
  \end{itemize}


\item {\tt LOOPS} \cite{LOOPS}:
  \begin{itemize}
  \item {\it Availability:} CPC Program Library
  \item{\it Purpose:} computation of one- and two-loop propagator type
                      integrals
  \item {\it Algorithms:} integration-by-parts (see
    Section~\ref{sec::IP})
  \item {\it Source:} {\tt REDUCE}
  \end{itemize}


\item {\tt LoopTools} \cite{FormCalc}:
  \begin{itemize}
  \item{\it Availability:} {\tt
      http://www-itp.physik.uni-karlsruhe.de/looptools}
  \item{\it Purpose:} implementation and extension of {\tt FF} in {\tt
      Mathematica}
  \item{\it Uses:} {\tt FF}
  \item{\it Source:} {\tt Mathematica}, {\tt FORTRAN}
  \item{\it Preferably combined with:} {\tt FeynCalc}, {\tt FormCalc}
  \end{itemize}


\item {\tt MadGraph} \cite{MadGraph}:
  \begin{itemize}
  \item {\it Availability:} {\tt
      http://pheno.physics.wisc.edu/Software/MadGraph/}
  \item{\it Purpose:} automatic generation of Feynman diagrams; calculation of
    helicity amplitudes
  \item {\it Uses:} {\tt HELAS}
  \item{\it Source:} {\tt FORTRAN}
  \end{itemize}


\item{\tt master} \cite{Bauetal}: supplement to {\tt s2lse}
    

\item {\tt MATAD} \cite{Stediss}:
  \begin{itemize}
  \item {\it Purpose:} analytical computation of massive three-loop tadpole
    integrals
  \item {\it Algorithms:} integration-by-parts (see Section~\ref{sec::IP})
  \item {\it Source:} {\tt FORM}
  \end{itemize}


\item {\tt MINCER} ({\tt FORM} version) \cite{mincer2}:
  \begin{itemize}
  \item {\it Availability:} {\tt
      ftp://nikhefh.nikhef.nl/pub/theory/form/libraries/form2/mincer}
  \item{\it Purpose:} analytical computation of massless propagator-type
    diagrams up to three loops
  \item {\it Algorithms:} integration-by-parts (see Section~\ref{sec::IP})
  \item {\it Source:} {\tt FORM}
  \end{itemize}


\item {\tt MINCER} ({\tt SCHOONSCHIP} version) \cite{MINCER1}:
  \begin{itemize}
  \item {\it Availability:} CPC program library
  \item{\it Purpose:} analytical computation of massless propagator-type
    diagrams up to three loops
  \item {\it Algorithms:} integration-by-parts (see Section~\ref{sec::IP})
  \item {\it Source:} {\tt SCHOONSCHIP}
  \end{itemize}


\item {\tt MINOS} \cite{minos}:
  \begin{itemize}
  \item{\it Purpose:} controlling facility for the calculation of
    processes with a huge number of diagrams
  \item{\it Source:} {\tt C}
  \end{itemize}


\item {\tt oneloop} \cite{oneloop}:
  \begin{itemize}
  \item{\it Availability:} {\tt http://wwwthep.physik.uni-mainz.de/\~\/xloops}
  \item{\it Purpose:} algebraic and numerical calculation of one-loop diagrams
  \item{\it Source:} {\tt MAPLE}
  \end{itemize}


\item{\tt PHACT} \cite{phact}:
  \begin{itemize}
  \item{\it Purpose:} numerical computation of tree processes up to four
    particles in the final state
  \item {\it Algorithms:} helicity amplitudes by means of
    \cite{BalMai95}
  \item{\it Source:} {\tt FORTRAN}
  \end{itemize}
  

\item {\tt ProcessDiagram} \cite{ProcDia}:
  \begin{itemize}
  \item{\it Purpose:} computation of one- and two-loop bubble diagrams
    allowing for several different masses
  \item {\it Source:} {\tt Mathematica}
  \end{itemize}


\item{\tt PVEGAS} \cite{pvegas}: parallel version of {\tt VEGAS}
  \begin{itemize}
  \item{\it Availability:} {\tt
      ftp://ftpthep.physik.uni-mainz.de/pub/pvegas/}
    \item{\it Source:} {\tt C}
  \end{itemize}
  

\item {\tt RECURSOR} \cite{recursor}:
  \begin{itemize}
  \item{\it Purpose:} analytical computation of massive three-loop tadpole
    integrals
  \item {\it Algorithms:} integration-by-parts (see Section~\ref{sec::IP})
  \item {\it Source:} {\tt REDUCE}
  \end{itemize}


\item{\tt s2lse} \cite{Bauetal}:
  \begin{itemize}
  \item{\it Availability:} {\tt
      ftp://ftp.physik.uni-wuerzburg.de/pub/hep/index.html}
  \item{\it Purpose:} numerically evaluate integral
    representations for scalar two-loop self-energy integrals
  \item{\it Source:} {\tt C}
  \item{\it Preferably combined with:} {\tt TwoCalc}
  \end{itemize}


\item {\tt SHELL2} \cite{SHELL2}:
  \begin{itemize}
  \item {\it Availability:} CPC Program Library
  \item{\it Purpose:} computation of propagator-type on-shell integrals up to
    two loops
  \item {\it Algorithms:} integration-by-parts (see Section~\ref{sec::IP})
  \item {\it Source:} {\tt FORM}
  \end{itemize}


\item {\tt SIXPHACT} \cite{sixphact}:
  extension of {\tt PHACT} to six particles in the final state


\item {\tt SLICER} \cite{slicer}:
  \begin{itemize}
  \item{\it Purpose:} analytical computation of massless propagator-type
    diagrams up to three loops
  \item {\it Algorithms:} integration-by-parts (see Section~\ref{sec::IP}) 
  \item {\it Source:} {\tt REDUCE}
  \end{itemize}


\item{\tt SPRING} \cite{BASES/SPRING}:
  \begin{itemize}
  \item{\it Purpose:} event generation
  \item{\it Source:} {\tt FORTRAN}
  \end{itemize}


\item {\tt TARCER} \cite{MerSch98}:
  \begin{itemize}
  \item {\it Availability:} {\tt http://www.mertig.com/tarcer/}
  \item{\it Purpose:} tensor reduction of two-loop propagator-type
    integrals and reduction to a set of basic integrals
  \item {\it Algorithms:} tensor reduction by shifting the space-time
      dimension (see Section~\ref{sec::tredtar})
  \item {\it Source:} {\tt Mathematica}
  \end{itemize}


\item {\tt TLAMM} \cite{tlamm}:
  \begin{itemize}
  \item{\it Purpose:} automatic evaluation of two-loop vertex diagrams
  \item{\it Source:} {\tt C}
  \end{itemize}


\item {\tt TRACER} \cite{tracer}:
  \begin{itemize}
  \item{\it Purpose:} Dirac trace calculations in {\tt Mathematica},
    optionally with 't~Hooft-Veltman or naive anti-commuting $\gamma_5$
  \item {\it Source:} {\tt Mathematica}
  \end{itemize}


\item {\tt TwoCalc} \cite{WeiSchBoe94}:
  \begin{itemize}
  \item{\it Availability:} upon request from the author
  \item{\it Purpose:} reduction of two-loop propagator diagrams to 
    a set of basis integrals
  \item{\it Algorithms:} two-loop tensor reduction by means of
    \cite{WeiSchBoe94}
  \item{\it Source:} {\tt Mathematica}
  \item{\it Uses:} {\tt FeynCalc}
  \item{\it Preferably combined with:}  {\tt FeynArts}, {\tt s2lse},
    {\tt master}
  \end{itemize}


\item {\tt QGRAF} \cite{Nog93}:
  \begin{itemize}
  \item{\it Availability:} {\tt ftp://gtae2.ist.utl.pt/pub/qgraf/}
  \item{\it Purpose:} efficiently generate multi-loop Feynman diagrams in
    symbolic notation
  \item{\it Source:} {\tt FORTRAN}
  \end{itemize}


\item{\tt VEGAS} \cite{VEGAS}:
  \begin{itemize}
  \item{Availability:} see, e.g.~\cite{NumRecBook}
  \item{\it Purpose:} adaptive multi-dimensional integration
  \item{\it Source:} {\tt FORTRAN}
  \end{itemize}
  

\item {\tt WPHACT} \cite{wphact}: extension of {\tt PHACT} to $W$ and
  Higgs physics
  \begin{itemize}
  \item {\it Availability:}
    {\tt http://www.to.infn.it/\~\/ballestr/wphact/}
  \end{itemize}


\item {\tt XLOOPS} \cite{XLOOPS}:
  \begin{itemize}
  \item{\it Availability:} {\tt
      http://wino.physik.uni-mainz.de/\~\/xloops/}
  \item{\it Purpose:} compute one- and two-loop diagrams using
    X-interface
  \item{\it Algorithms:} parallel space technique \cite{CzaKilKre95}
  \item{\it Uses:} {\tt PVEGAS}, {\tt oneloop}
  \item{\it Source:} {\tt MAPLE}, {\tt Tcl/Tk}
  \end{itemize}
\end{itemize}



\setlength{\baselineskip}{\savebaselineskip}

%% file: selected_intro.tex
\subsection{\label{subselpac}Selected packages}
We will start with the description of the {\tt FORM} programs {\tt
  MATAD} and {\tt MINCER}.  Details on the automatic application of
asymptotic expansions using {\tt LMP} and {\tt EXP} are given in
Section~\ref{subsubexplmp}.  In Section~\ref{subgeficom} the package
{\tt GEFICOM} will be explained. The way the generator {\tt QGRAF} works
should also become clear then, which is why it will not be described
separately.  The subsequent section is devoted to the {\tt Mathematica}
packages {\tt FeynArts}, {\tt FeynCalc}, {\tt FormCalc}, {\tt TwoCalc}
and {\tt LoopTools}.  In the concluding section we discuss the package
{\tt CompHEP} as an application of automated multi-leg computations.

%% file: matmin.tex
\subsubsection{\label{sec::matmin}{\tt MATAD} and {\tt MINCER}}
\begin{figure}[hb]
  \begin{center}
    \begin{tabular}{c}
      \epsfxsize=4cm
      \epsffile[131 278 481 514]{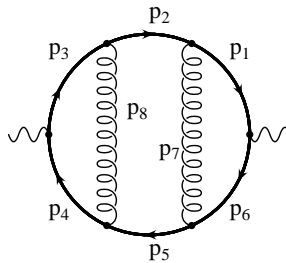}
      \smallskip
    \end{tabular}
    \parbox{\captionwidth}{
      \caption[]{\label{figladder}\sloppy 
        Ladder-type diagram contributing to the
        photon propagator. The solid, wavy and curly lines represent
        quarks, photons and gluons, respectively.  }}
  \end{center}
\end{figure}

Let us examine the three-loop ladder diagram (see Fig.~\ref{figladder})
in order to see in detail how the programs {\tt MATAD} and {\tt MINCER}
work.  We will focus on the case where the external momentum, $q_1$, is
much smaller than the mass of the fermions, $m$.  Application of the
rules for the hard-mass procedure (Section~\ref{subasymp}) shows that it
suffices to perform a Taylor expansion of the integrand in order to
arrive at an expansion in $q_1^2/m^2$.  The resulting integrals to be
evaluated are of the tadpole type, and the calculation can completely be
passed to {\tt MATAD}.  In the opposite limit, $q_1^2\gg m^2$, in
addition to the naive Taylor expansion of the integrand in $m^2/q_1^2$,
the large-momentum procedure generates non-trivial subgraphs.
In general their evaluation requires to use {\tt MATAD} and
{\tt MINCER} in combination.

The input for the Feynman graph is written to a file using the fold
construct of {\tt FORM}~\cite{form}:
\begin{verbatim}
*--#[ ladder:
        ((-1)
        *Dg(nu1,nu2,p7)
        *Dg(nu3,nu4,p8)
        *S(mu1,q1,p3m,nu3,q1,p2m,nu1,q1,p1m,mu2,p6m,nu2,p5m,nu4,p4m)
        );

        #define TOPOLOGY "LA"
*--#] ladder:
\end{verbatim}
Some words concerning the notation are in order: The function
\verb/S(...)/ represents the fermion string and contains momenta
(\verb/p1/ $\widehat = \,p_1$, \ldots) and Lorentz indices (\verb/mu1/
$\widehat = \,\mu_1$, \ldots) as its arguments. The momenta are labeled
in accordance with the {\tt MINCER} conventions for the specific
topology which is \verb/LA/ here, meaning ``ladder''. Note that it
is more convenient to express the input in terms of eight momenta and to
use momentum conservation at a later stage than to impose it already in
the input.  The ending \verb/m/ indicates that massive fermions should be
considered.  A small {\tt FORM} program transforms each Lorentz index to
the corresponding $\gamma$ matrix and the momenta to fermion
propagators.  The expansion in the external momentum, \verb/q1/, is
indicated by the consecutive appearance of \verb/q1/ together with an
integration momentum. The depth of the expansion, which is encoded
as a pre-processor variable in {\tt FORM}, must be defined at program
call. \verb/Dg(...)/ represents the gluon propagator.  After reading the
diagram these functions are replaced according to the Feynman rules.
Then {\tt MATAD} evaluates the fermion trace and expands the
denominators with respect to \verb/q1/.  At this point the user may
influence the procedure by providing an additional file, the so-called
``special {\tt treat}-file''. For the matter in hand, this file contains
a projector on the transverse part
of the diagram:
\begin{verbatim}
multiply, (-d_(mu1,mu2)+q1(mu1)*q1(mu2)/q1.q1)*deno(3,-2);
\end{verbatim}
where \verb/deno(3,-2)/ means $1/(3-2\varepsilon)$.  In this way one
obtains a scalar expression where \verb/q1/ is absent in the
denominators of the propagators. However, it may still appear in scalar
products with integration momenta.  Application of d'Alembertian
operators performs an averaging in the numerator, i.e. terms like
$(q_1\cdot p)^2$ are replaced by $p^2 q_1^2/D$. This projects out the
coefficients of $(q_1^2)^n$ which have no dependence on the external
momenta any more.  After that it is possible to decompose the numerators
in terms of the denominators.  At this stage the expression may already
have split into millions of different integrals.  Recurrence relations
derived with the integration-by-parts technique reduce them to a small
set of tabulated master integrals which have been evaluated once and for
all.  In the case of the photon propagator, for example, all three-loop
integrals reduce to only three massive three-loop master integrals.  The
same is true for the massless case with non-vanishing external momentum,
which is where {\tt MINCER} applies.

In intermediate steps, when the recurrence relations are applied, it may
happen that the expression blows up and disk space up to several
Gigabytes is needed. But the final result is rather compact and fills
only of the order of one output screen (depending, of course, on the
depth of the expansion).

%% file: explmp.tex
\subsubsection{\label{subsubexplmp}{\tt EXP} and {\tt LMP}}
The programs {\tt EXP} and {\tt LMP} are computer-technical realizations
of the prescriptions for the asymptotic expansion of Feynman
diagrams described in Section~\ref{subasymp}.
They both work at the graphical level, i.e.~they are not
dealing with the analytic expressions corresponding to the Feynman
diagrams but rather with the diagrams themselves. 

Given an arbitrary diagram with a single external momentum (not counting
momenta that can be gotten rid of by a simple Taylor expansion), {\tt EXP}
and {\tt LMP} generate all possible subdiagrams by successively removing
any combination of lines (propagators). For those subdiagrams matching
the conditions of the large-momentum, respectively, the hard-mass
procedure, the programs determine the momentum distribution. In
particular, they determine which lines should be Taylor expanded in
which momenta, and how the momenta of the hard subgraph are related to
the ones of the co-subgraph. Thus, every diagram is split into
several subdiagrams, all of them being products of single-scale
integrals as discussed in Section~\ref{subasymp}.

As it was already mentioned in Section~\ref{subgensd}, {\tt LMP} is
written in {\tt PERL} and is only concerned with the large-momentum
procedure. {\tt EXP}, being in some sense its successor, is a {\tt
  FORTRAN~90} program\footnote{A forthcoming version will be written in
  {\tt C}.} and deals also with the hard-mass procedure. In addition, if
several scales are involved, it repeatedly applies both methods in
combination, thus reducing an arbitrary three-loop self-energy diagram
to single-scale integrals (see Section~\ref{sec::zbb}).

To be specific, let us consider the example for the double-bubble diagram
of Section~\ref{subsubexa} where the outer line is massless and the inner one
carries mass $m$. The input file for {\tt LMP} looks as follows:
\begin{verbatim}
*--#[ GLOBAL :
#define POWER "12"
*--#] GLOBAL :

*--#[ TREAT :
multiply, (-d_(mu,nu) + q(mu)*q(nu)/q.q)*deno(3,-2);
*--#] TREAT :

*--#[ db :
    *S(mu,p1,ro,p2,nu,p3,si,p4)
    *SS(tau,p6m,al,-p7m)
    *Dg(tau,ro,p8)
    *Dg(si,al,p5)
    ;
    #define TOPOLOGY "O1"
*--#] db :
\end{verbatim}
where the fold construct of {\tt FORM} is used not only to encode the
diagram itself but also the projectors ({\tt TREAT}-fold) and possible
options ({\tt GLOBAL}-fold). The projector is again on the transverse
part of the diagram (see Section~\ref{sec::matmin}).  The external
momentum is denoted by $q$ this time.  The only option here is {\tt
  POWER}, denoting the required depth for the expansion in the mass $m$.

The diagram is contained in the fold named {\tt db} (for ``double
bubble''). The encoding is based on {\tt MINCER/MATAD} notation (see
Section~\ref{sec::matmin}). In particular, the momentum distribution
is in accordance with the corrsponding topology
defined in these programs.

{\tt LMP} reads the content of the {\tt db}-fold, translates it to an
internal graph-based notation and applies the large-momentum procedure.
There are two main output files; the first one is called {\tt db.dia}
and contains the relevant subdiagrams:
\begin{verbatim}
*--#[ db_1 :
    *S(mu,+p1,ro,+p2,nu,+p3,si,+p4)
    *SS(tau,+p6mexp,al,-p7mexp)
    *Dg(tau,ro,+p5)
    *Dg(si,al,+p5)
    ;
#define DIANUM "1"
#define TOPOLOGY "arb"
#define INT1 "inpo1"
#define MASS1 "0"
*--#] db_1 :

[...]
\end{verbatim}
\begin{verbatim}
*--#[ db_5 :
    *S(mu,+p11,ro,+p12,nu,+p13,si,+p14)
    *SS(tau,p1m,al,+aexp,-p15mexp)
    *Dg(tau,ro,+p15)
    *Dg(si,al,+p15)
    ;
#define DIANUM "5"
#define TOPOLOGY "arb"
#define INT1 "topL1"
#define MASS1 "M"
#define INT2 "inpt1"
#define MASS2 "0"
*--#] db_5 :

[...]
\end{verbatim}
Only two out of six contributing subdiagrams are displayed here, the
remaining ones being represented by {\tt [...]}.  The information on the
topology of the diagrams is now split according to the factorization of
loops. It is contained in the pre-processor variables {\tt INT1}, {\tt
  INT2}, etc. The variable {\tt TOPOLOGY} is set to the dummy value {\tt
  arb}.

The first subdiagram, {\tt db\_1}, corresponds to the naive
Taylor expansion of the integrand with respect to $m$. This can be seen
from the fact that the momentum distribution is the same as in the
original diagram, and that the massive lines are denoted by {\tt
  p}$i${\tt mexp} instead of {\tt p}$i${\tt m}. In this way {\tt MATAD}
is advised to perform a Taylor expansion in $m$.  The second subdiagram
displayed above, {\tt db\_5}, corresponds to the second one on the
right-hand side of Fig.~\ref{figdblmp} (note that there is a
topologically identical one, indicated by the factor 2 in
Fig.~\ref{figdblmp}; {\tt LMP}, however, generates both of them as
separate diagrams).  The co-subgraph is a one-loop tadpole-diagram
carrying the momentum $p_1$. It must not be expanded in the mass,
therefore its propagator is simply denoted by {\tt p1m} above. The hard
subgraph is a two-loop diagram of the topology shown in
Fig.~\ref{figtriangle}.  Its momenta $p_i$ ($i=1,\ldots,5$) are denoted
by {\tt p1}$i$, where the additional ``{\tt 1}'' is introduced to
distinguish them from the momenta (here it is only one momentum,
actually) of the co-subgraph. One propagator carries the momentum $p_1 -
p_{15}$ which is indicated by the notation {\tt +aexp,-p15mexp}. Note
that this propagator also has to be expanded w.r.t. $m$. Here, $p_1$ is
denoted by {\tt aexp} to let {\tt MATAD} know that it has to expand with
respect to $p_1$. After expansion, {\tt aexp} is identified with {\tt
  p1}. This is one of the main contents of the second output file of
{\tt LMP}, called {\tt treat.db}. It again splits into folds, one for
each subdiagram.  The fold corresponding to the subdiagram {\tt db\_5}
above, for example, contains the line
\begin{verbatim}
id aexp = + p1;
\end{verbatim}
which does the identification of momenta mentioned above.  Most of the
remaining statements of the {\tt treat.db}-file are concerned with the
proper ``power-counting'', i.e., to take care that the Taylor expansions
are performed such that not too few and not too many terms are
generated.  (Expanding up to unnecessarily high power may drastically
slow down the computation.)  

To compute the second example of Section~\ref{subsubexa},
the one concerned with
the hard-mass procedure, one has to use {\tt EXP}. The input file,
however, is almost identical to the one shown above. The only difference
is the {\tt GLOBAL}-fold, since in the case of {\tt EXP} also the hierarchy
of scales has to be fixed (for {\tt LMP}, only $q^2\gg
m_1^2,m_2^2,\ldots$ is allowed):
\begin{verbatim}
*--#[ GLOBAL :
#define POWERM "6"
#define SCALES "M,q"
*--#] GLOBAL :
\end{verbatim}
This means to assume $m^2\gg q^2$ and to expand up to $(q^2/m^2)^3$. For
multi-scale problems, this fold may look like
\begin{verbatim}
*--#[ GLOBAL :
#define POWERMa "4"
#define POWERMc "2"
#define POWERQ "8"
#define SCALES "Ma,q,Mc,Mb"
*--#] GLOBAL :
\end{verbatim}
indicating the hierarchy $m_a^2\gg q^2\gg m_c^2\gg m_b^2$ and an
expansion up to the corresponding degrees.  The output files of {\tt
  EXP} are very similar to the ones above, although the power-counting
statements are obviously much more involved than for {\tt LMP}.

In addition, {\tt LMP} and {\tt EXP} produce the whole set of
administrative files, namely {\tt FORM} files which control the running of
{\tt MINCER}/{\tt MATAD}, and {\tt make} files concerned with program
calls and the reduction of loss caused by system crashes.

The only task to be done by the user is to feed in the input file shown
above, call {\tt LMP} or {\tt EXP}, and finally to type ``{\tt make}''
in order to start the very computation of the diagrams. The output is
exactly a result like the one quoted in Eq.~(\ref{eqdblmpres}) and
(\ref{eqdbhmpres}), respectively.

%% file: geficom.tex
\subsubsection{\label{subgeficom}{\tt GEFICOM}}
\begin{figure}[ht]
  \begin{center}
    \leavevmode
    \epsfxsize=8cm
    \epsffile[77 200 420 780]{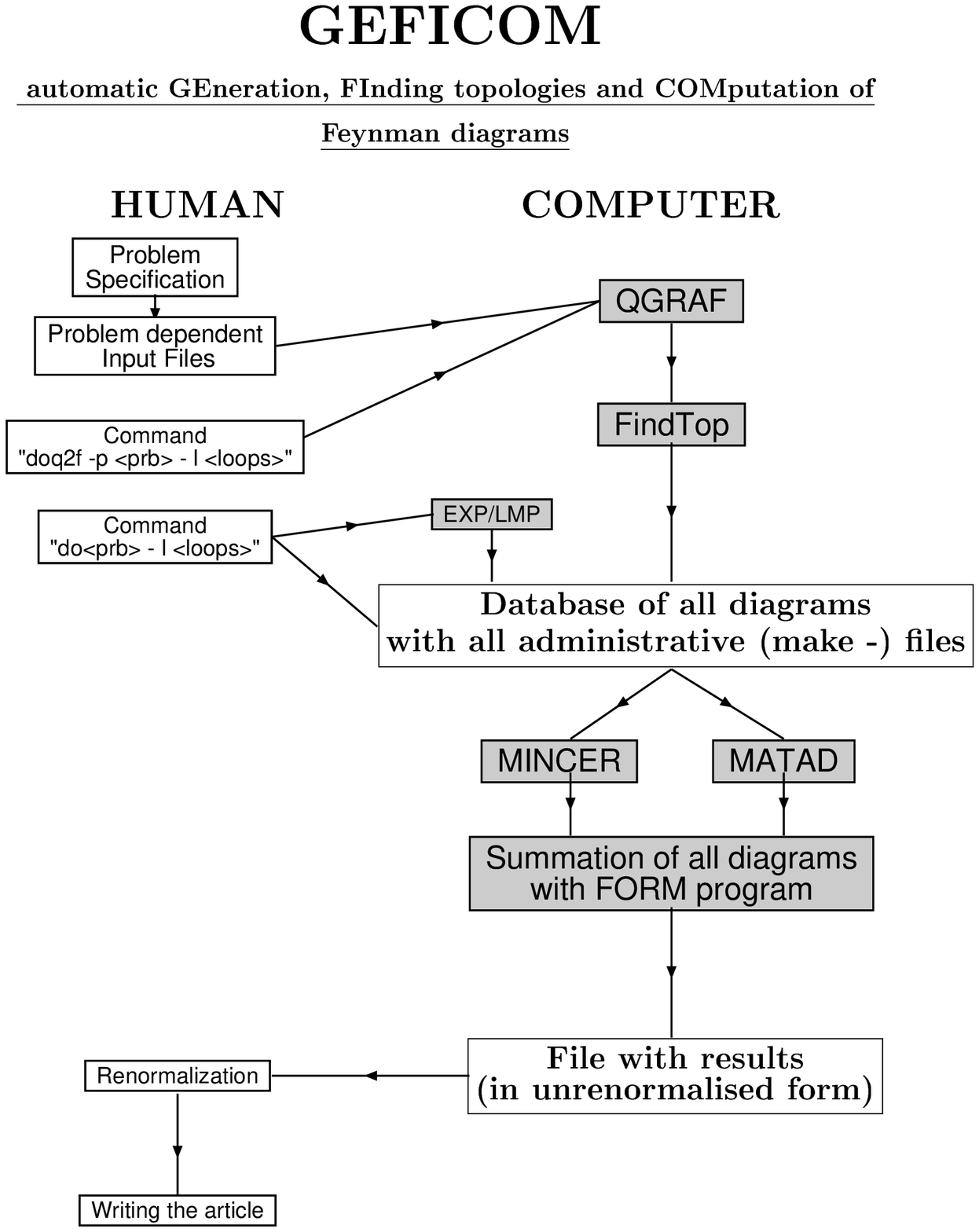}\\
    \parbox{\captionwidth}{
      \caption[]{\label{figfchart}The structure of {\tt GEFICOM}.
        }}
  \end{center}
\end{figure}

This section investigates in more detail how to use {\tt GEFICOM} to
evaluate a whole set of Feynman diagrams.  Fig.~\ref{figfchart} gives a
schematic overview of {\tt GEFICOM} and the possible interactions from
outside. It unifies several packages described above and thus allows to
automatically compute a specified class of diagrams in analytic form as
long as a certain hierarchy among the involved mass scales is justified.
The strategy then is to apply asymptotic expansions in order to split
all integrals into massless propagator-type diagrams and massive
tadpoles.

As an example let us consider corrections to the gluon propagator from
which follows, for example, the gluon wave function renormalization constant.
{\tt GEFICOM} needs a name for each problem, and we will choose ``{\tt
  gg}'' for it.  A more sophisticated example involving up to four
different dimensional parameters is considered in Section~\ref{sec::zbb}.

In a first step the user sets up a file containing the
Lagrangian, or rather its topological content, i.e., the plain
vertices and propagators without the corresponding analytical expressions.
In our example it is called {\tt gg.lag} and looks as follows:
\begin{verbatim}
*  propagators 
  [g,g;3+t]
  [c,C;1-]
  [q,Q;2-]
  [sigma,sigma;1+t]
*  vertices
  [Q,q,g]
  [C,c,g]
  [g,g,g]
  [g,g,sigma]
\end{verbatim}
This file serves as input for {\tt QGRAF}\footnote{The notation of the
  {\tt QGRAF} version 1.3 has been adopted.}  and the notation is
adopted from this program.  In the first part the propagators are
defined for gluons (\verb/g/), ghosts (\verb/c/) and quarks (\verb/q/).
The auxiliary particle $\sigma$ (\verb/sigma/) is introduced to split
the four-gluon vertex into three-linear vertices which are much easier
to deal with.  In addition to the spin multiplicity one also has to
specify if the particle is a boson (``$+$'') or fermion (``$-$''). The
optional parameter \verb/t/ guarantees that no diagrams with tadpoles of
the corresponding particle are generated.  The notation for the vertices
in the second part is rather obvious.

A second file, called {\tt gg.def}, contains the underlying process,
accompanied by additional options:
\begin{verbatim}
*** MINCER

* gauge 1
* scheme 2
* power 0

      list = symbolic ;
      lagfile = 'q.lag' ;
      in = g[q];
      out = g[q]; 
      nloop = ;
      options = ;
      true = bridge[ g,c,q, 0,0 ]; 
\end{verbatim}
The first line determines that the output of the generation procedure
should be transformed into {\tt MINCER} notation.  The next two
statements fix the gauge and the regularization scheme, and the variable
\verb/power/ defines the depth of the expansion in small masses and
momenta.  Since in our case all particles are taken to be massless, only
one scale is left, namely the external momentum. Thus, an expansion is
not needed and the variable \verb/power/ may be assigned the value zero.
The remaining part again uses {\tt QGRAF} notation and specifies the
diagrams to be generated.  For example, ``\verb/in/'' and ``\verb/out/''
denote in- and outgoing states, and the last line excludes diagrams that
are one-particle reducible w.r.t.\ gluons, ghosts or quarks.  {\tt
  options} is left empty, and {\tt nloop}, the number of loops, will be
defined via a command line option later.

Like in Section~\ref{sec::matmin}, the user again may provide ``special
{\tt treat}-files'' --- in the case of the gluon propagator two of them.
The first one, called
{\tt treat.gg.1}, again contains the projector on the transverse
structure of the correlator:
\begin{eqnarray}
\frac{1}{D-1}\left(-g^{\mu_1\mu_2}+\frac{q^{\mu_1}q^{\mu_2}}{q^2}\right)
\,.
\end{eqnarray}

Besides the consideration in momentum space 
{\tt GEFICOM} also takes care about the colour factor which is why in
general one should supply a projector also in colour space.
For the gluon propagator with the colour indices $a(1)$ and $a(2)$ it is
given by
\begin{eqnarray}
\frac{\delta^{a(1)a(2)}}{n_g}
\,,
\end{eqnarray}
where $n_g=N_c^2-1$ is the number of gluons. The corresponding {\tt
  FORM} file, called {\tt treatcol.gg.1}, looks as follows:
\begin{verbatim}
multiply, prop(a(1),a(2))/ng;
\end{verbatim}

After providing these simple input files,
{\tt GEFICOM} may be called by executing the shell-script
command:
\begin{verbatim}
> doq2f -p <prb> -l <loops>
\end{verbatim}
where \verb/<prb>/ gives the name of the problem ({\tt gg} here) and
\verb/<loops>/ is the number of loops. {\tt GEFICOM} calls {\tt QGRAF}
and transforms the output to {\tt Mathematica} format. Unfortunately {\tt
  QGRAF} does not provide the topology of the generated diagrams.
While for a human being it is very simple
to figure out this information by just looking at the diagram, for a
computer this is a non-trivial task, especially if the number of
loops exceeds two.  One of the core parts of {\tt GEFICOM} indeed is a
{\tt Mathematica} program precisely concerned with this problem. 

\begin{figure}[ht]
  \begin{center}
    \leavevmode
    \epsfxsize=6.cm
    \epsffile[130 260 450 460]{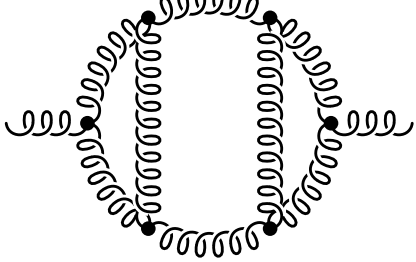}
    \hfill
    \parbox{\captionwidth}{
    \caption[]{\label{fig::gg}\sloppy
      Sample diagram contributing to the gluon propagator at three loops.
      }}
  \end{center}
\end{figure}

Once the topology is available, the notation for the diagrams is
translated to a format which is suitable for the {\tt FORM} packages
{\tt MINCER} and {\tt MATAD}. In addition, a set of administrative files
controlling the calculation and ensuring minimal loss in the case of
computer breakdowns is generated.  For the case of the gluon propagator
considered in this section, at three-loop level 494~diagrams are
generated, and it takes roughly 12 hours of CPU time on a DEC-Alpha
machine with 600~MHz for the complete evaluation. A sample diagram for
the three-loop gluon propagator is shown in Fig.~\ref{fig::gg}.

In analogy to the momentum-space diagram also a {\tt FORM} version of the
``colour diagram'' is generated which is evaluated with the help of
a small {\tt FORM} program.

The computation of the diagrams is not initiated automatically.
Instead, {\tt GEFICOM} generates a shell-script, \verb/do<prb>/
(\verb/dogg/ in our case) that allows, for example, only a subset of
diagrams, or even only a single diagram, to be computed via command line
options.  The advantage of this strategy is that different computers may
be used in parallel for different subsets, or that a part of the
diagrams may even be evaluated before the generation is finished.  After
the results of all diagrams are available they are summed up, taking
into account the correct colour (and other) factors and the
(unrenormalized) result is at hand.

%% file: feynarts.tex
\subsubsection{{\tt FeynArts}, {\tt FeynCalc}, {\tt FormCalc}, 
  {\tt TwoCalc} and {\tt LoopTools}}
\paragraph{{\tt FeynArts}:}
{\tt FeynArts} is a {\tt Mathematica}-package generating Feynman
diagrams and the corresponding amplitudes. Its great flexibility has
allowed it to become an enormously useful tool not only for high energy
particle physicists. It has found its applications also in effective
field theories, and by knowing its power one could imagine that it may
be useful even for at first sight completely unrelated subjects.

There exists a manual describing its features in a rather
clear and pedagogical way. The four following functions constitute the
heart of {\tt FeynArts}: 
{\tt CreateTopologies}, {\tt InsertFields}, {\tt Paint} and {\tt
  CreateFeynAmp}.

\begin{figure}[ht]
  \begin{center}
    \leavevmode
    \epsfxsize=6.cm
    \epsffile[110 265 465 560]{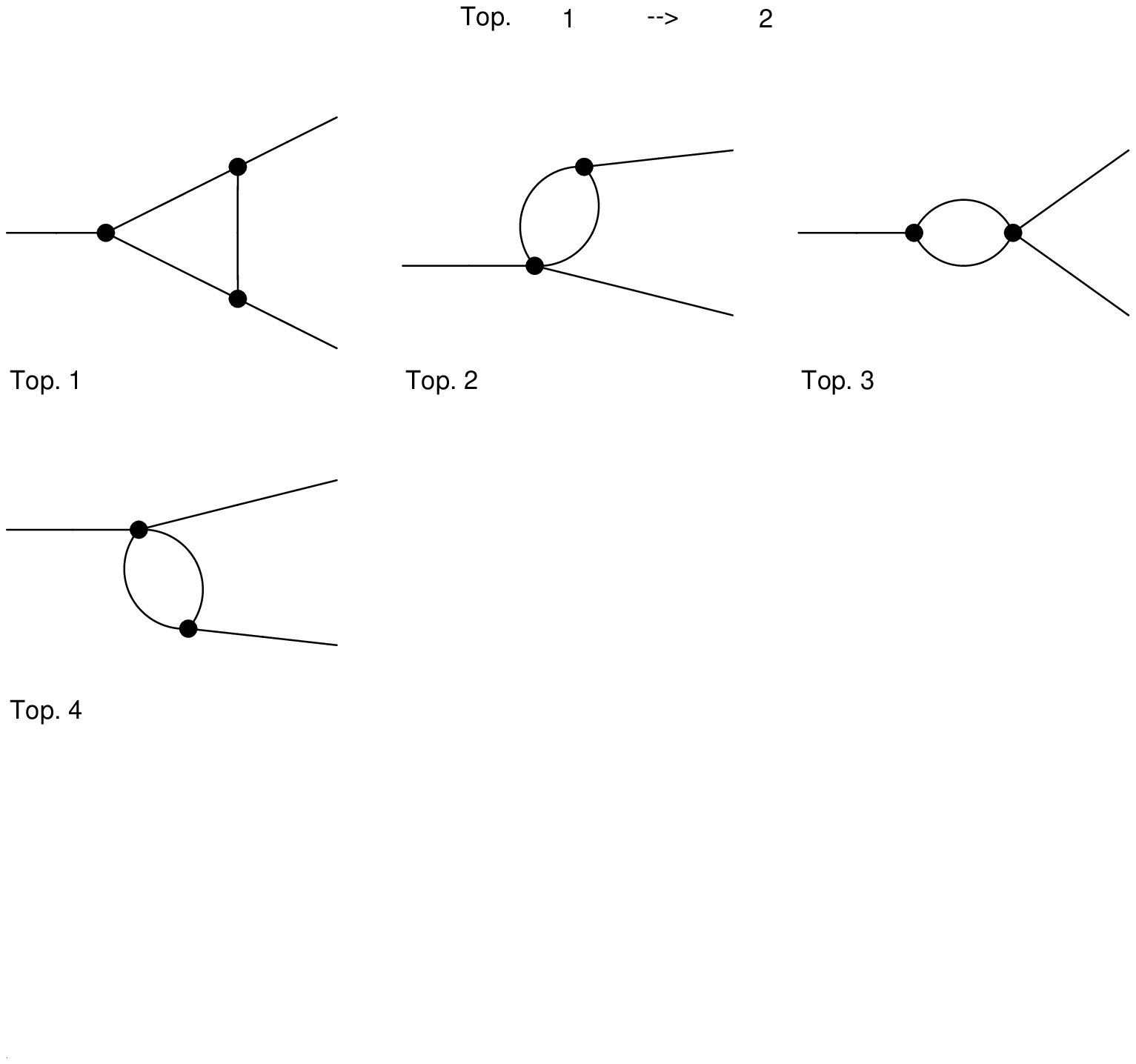}
    \hfill
    \parbox{\captionwidth}{
    \caption[]{\label{FAtops1l.ps}\sloppy
      Topologies created by {\tt FeynArts}.
      }}
  \end{center}
\end{figure}

As an example let us consider the one-loop corrections to the
triple Higgs vertex in the Standard Model.
The use of {\tt CreateTopologies} in this case is the following\footnote{In
  what follows we will use the notation of the {\tt FeynArts} version 1.2}:
\begin{verbatim}
tops = CreateTopologies[1,3, Tadpoles -> False, SelfEnergies -> False];
\end{verbatim}
The first two arguments of {\tt CreateTopologies} determine the number
of loops and external legs, respectively. The remaining arguments are
options, where the first one prevents creation of tadpole insertions and
the second one states that no diagrams with self-energy insertions on
external legs are generated\footnote{ 
  There is a single option
  comprising both of the latter, named {\tt WFCorrections} (for ``wave
  function corrections'').}.  To view the topologies, one types
\begin{verbatim}
Paint[tops,1,2];
\end{verbatim}
where the first argument specifies the list of topologies and the second
and third one refer to the number of incoming and outgoing fields,
respectively. The outcome is the graphic shown in Fig.~\ref{FAtops1l.ps}.

\begin{figure}[ht]
  \begin{center}
    \leavevmode
    \epsfxsize=8.cm
    \epsffile[110 160 520 590]{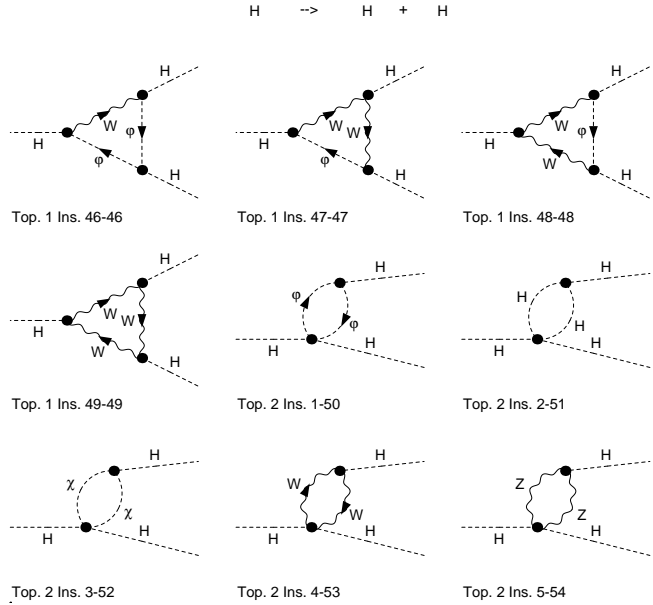}
    \hfill
    \parbox{\captionwidth}{
    \caption[]{\label{FAHHH.ps}\sloppy
      Sample of diagrams generated by {\tt FeynArts} for the triple Higgs
      vertex to one-loop order.
      }}
  \end{center}
\end{figure}

Next the fields are attributed to the lines. This is done by typing
\begin{verbatim}
ins = InsertFields[tops, {S[1]} -> {S[1],S[1]}, Model -> {SM}];
\end{verbatim}
where the first argument again refers to the list of topologies created
above, the third one fixes the model ({\tt SM} $\widehat =$ Standard
Model), and the second one defines the in- and outgoing particles ({\tt
  S[1]} for the physical Higgs boson).  This produces a list of 64
diagrams that can be viewed using {\tt Paint} again, this time, however,
with only one argument, since in- and outgoing fields are already
specified in ``{\tt ins}'':
\begin{verbatim}
Paint[ins];
\end{verbatim}
An extract of the list of diagrams is shown in Fig.~\ref{FAHHH.ps}.
\begin{figure}
  \begin{center}
    \leavevmode
    %
    %
    %
    \epsfxsize=5.cm
    \epsffile[80 180 520 600]{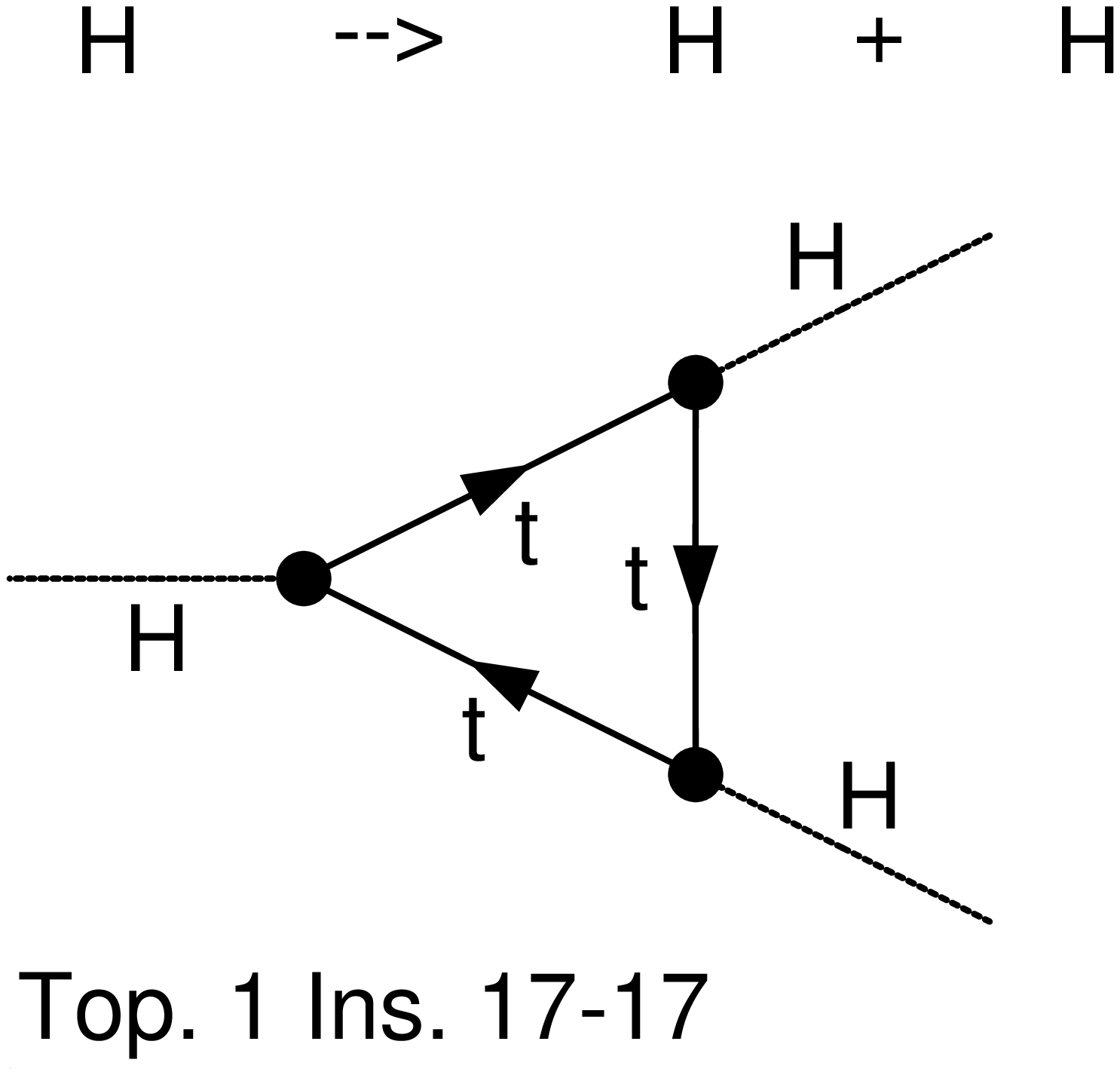}
    \hfill
    \parbox{\captionwidth}{
    \caption[]{\label{fig:fa:17}\sloppy
      $17^{\rm th}$ diagram produced by {\tt FeynArts} for the
      triple Higgs vertex.
      }}
  \end{center}
\end{figure}

The last step to be done by {\tt FeynArts} is to generate analytic
expressions which are understood by {\tt FeynCalc}, {\tt FormCalc} and
{\tt TwoCalc}:
\begin{verbatim}
amps = CreateFeynAmp[ins];
\end{verbatim}
This produces a list containing the amplitudes of the individual diagrams. The
$17^{\rm th}$ element corresponds to the diagram with a top-quark
triangle (Fig.~\ref{fig:fa:17}) and reads as follows:
\begin{verbatim}

In[12]:= amps[[17]]

                                        3 I   3   3
Out[12]= FeynAmp[S1S1S1, T1, I17, N17][(--- EL  MT  Integral[q1] 
                                        128
 
>       tr[(MT + gs[k1 - p1 + q1]) . (MT + gs[q1]) . (MT + gs[k1 + q1])]) / 
 
         3   4     2     2      2            2      2                 2    3
>     (MW  Pi  (-MT  + q1 ) (-MT  + (k1 + q1) ) (-MT  + (k1 - p1 + q1) ) SW )]
\end{verbatim}
{\tt p}$i$ are the ingoing, {\tt k}$i$ the outgoing, and {\tt q}$i$ the
loop momenta. {\tt EL} is the electric charge, {\tt gs[p]} means
$p\!\!\!/$ and {\tt tr} the Dirac trace.

\paragraph{{\tt FeynCalc}, {\tt FormCalc} and {\tt TwoCalc}:}
The list of expressions named {\tt amps} above may now be directly fed
into {\tt FeynCalc}\footnote{ We use version $2.2\beta$ here which is
  the latest version which is freely available (see
  Section~\ref{submisc}). } or {\tt FormCalc}.  Its entries will
automatically be transformed to the internal notation (see below).  The
key function of these programs is {\tt OneLoop}, respectively derivatives of
it, in particular {\tt OneLoopSum}. As a simple example consider the
integral
\begin{equation}
\int{\dd^D q\over (2\pi)^D}{1\over (m_1^2 - q^2)(m_2^2-(k+q)^2)} =
{i\over 16\pi^2}\, B_0(k^2,m_1^2,m_2^2)\,.
\label{eq::B0def}
\end{equation}
In the notation of {\tt FeynCalc} this reads
\begin{verbatim}
In[3]:= OneLoop[q,1/(2*Pi)^4 FeynAmpDenominator[
               PropagatorDenominator[q,m1],
               PropagatorDenominator[q+k,m2]]]

        I            2    2
        -- B0[k.k, m1 , m2 ]
        16
Out[3]= --------------------
                  2
                Pi
\end{verbatim}
Note that the pre-factor in {\tt In[3]} above is $D$-independent.  In
principle, this could be assigned to a definition of {\tt B0} in {\tt
  Out[3]} that differs from the one in (\ref{eq::B0def}) by a
$D$-dependent factor; however, another interpretation is to implicitely
assume \msbar-regularization of all integrals. In fact, as will become
clear in a moment, this is what is done in the numerical routine {\tt
  LoopTools}.  

In this way {\tt FeynCalc} rewrites any one-loop diagram to the standard
integrals defined in Eq.~(\ref{eq::tmunuN}).  For example, the diagram
corresponding to {\tt amps[[17]]} defined above produces the output
\begin{verbatim}
In[4]:= OneLoopSum[amps,SelectGraphs -> {17}]

Out[4]= K[7]

In[5]:= FixedPoint[ReleaseHold[#]&,%]
\end{verbatim}
\pagebreak[4]
\begin{verbatim}
              3   4             2    2        3   4             2    2
         -3 EL  MT  B0[k1.k1, MT , MT ]   3 EL  MT  B0[p1.p1, MT , MT ]
Out[5]= ------------------------------ - ----------------------------- - 
                      3   2   3                       3   2   3
                 32 MW  Pi  SW                   32 MW  Pi  SW
 
      3   4                               2    2
  3 EL  MT  B0[k1.k1 - 2 k1.p1 + p1.p1, MT , MT ]
  ----------------------------------------------- - 
                       3   2   3
                  32 MW  Pi  SW
 
       3   4                                             2    2    2
  (3 EL  MT  C0[k1.k1, p1.p1, k1.k1 - 2 k1.p1 + p1.p1, MT , MT , MT ] 
 
          2                                   3   2   3
     (4 MT  - k1.k1 + k1.p1 - p1.p1)) / (32 MW  Pi  SW )
\end{verbatim}
The action of {\tt FormCalc} is quite similar to the one of {\tt
  FeynCalc}, except that is does not perform the full reduction to the
scalar integrals of Eq.~(\ref{eq::scalint}), but rather only decomposes
the expressions into covariants, keeping the corresponding coefficients
as they are. In general this reduces the length of the expressions and
still can be evaluated numerically with the help of {\tt LoopTools}, to
be described below. The same strategy can also be followed by {\tt
  FeynCalc}, simply by setting the option {\tt ReduceToScalars} in {\tt
  OneLoop} to {\tt False}.  The main feature of {\tt FormCalc} in
comparison to {\tt FeynCalc} is its speed which is mainly due to the
fact that it passes large expressions to {\tt FORM}.

{\tt TwoCalc} is the extension of {\tt FeynCalc} to two-loop propagator
type diagrams, and its usage is very similar.

\paragraph{{\tt FF} and {\tt LoopTools}:}
{\tt FF} is a {\tt FORTRAN} program which numerically evaluates the
scalar one-loop integrals defined in (\ref{eq::scalint}). The extension
to the coefficients appearing in the tensor decomposition of tensor
integrals and its integration into {\tt Mathematica} is done in {\tt
  LoopTools}.  E.g., the numerical evaluation of the $B_0$-integral
defined above may be simply done in {\tt Mathematica} by saying
\begin{verbatim}
In[2]:= B0[k.k,m1^2,m2^2] //. {k.k -> m1^2,m1 -> 100,m2 -> 10}

Out[2]= -7.49109
\end{verbatim}
This example shows that the $1/\varepsilon$ poles (together with the
constants $\gamma_{\rm E}$ and $\ln 4\pi$) are set to zero as it corresponds
to an \msbar-renormalized quantity; this was already pointed out in the
discussion below Eq.~(\ref{eq::B0def}). Both the ultra-violet and the
infra-red divergent parts are controlled with the help of parameters
which specify the square of the renormalization scale $\mu$.  In a
similar way, it is possible to obtain numerical values for the diagram
{\tt amps[[17]]} above by providing numbers for the scalar products and
masses.

%% file: comphep.tex
\subsubsection{\label{subcomphep}{\tt CompHEP}}
\begin{figure}
  \begin{center}
    \leavevmode
    \epsfxsize=14.cm
    \epsffile{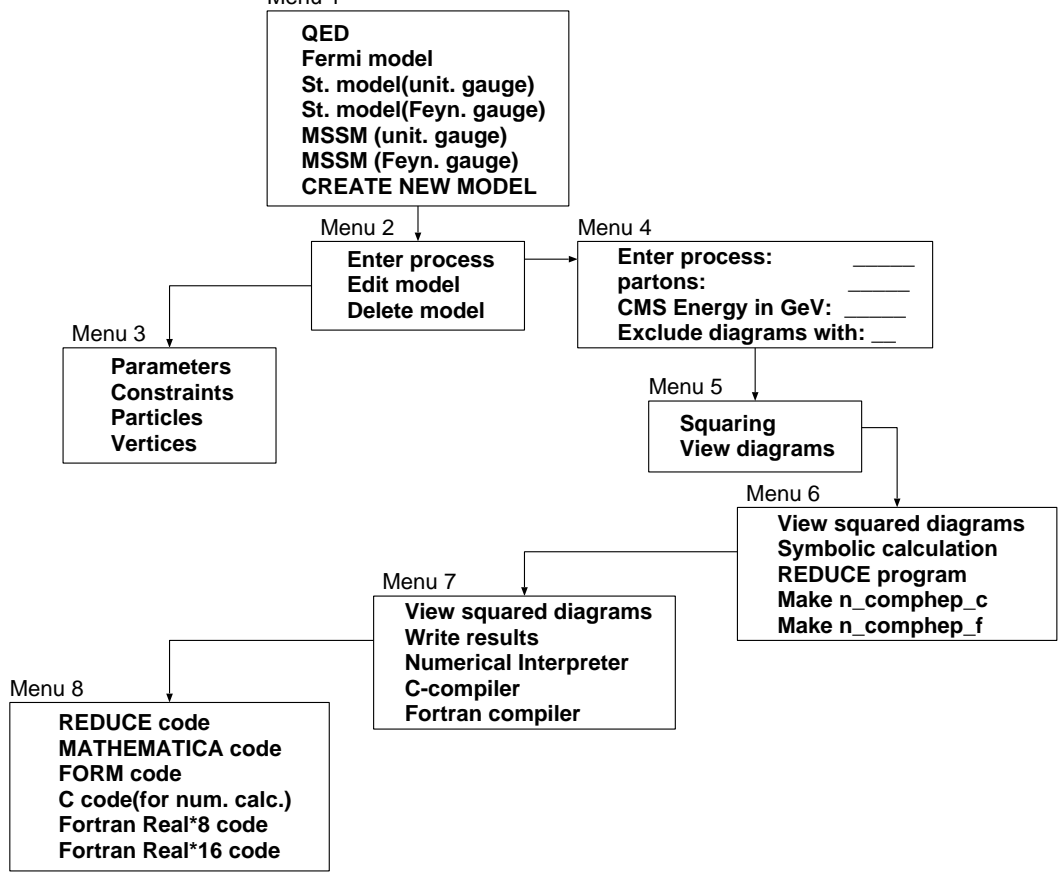}
    \hfill
    \parbox{\captionwidth}{
    \caption[]{\label{figcomphep1}\sloppy
        Menus of the symbolic part of {\tt CompHEP}~\cite{pukhov}.}}
  \end{center}
\end{figure}

Let us now discuss in more detail how the package {\tt CompHEP} works.
As already mentioned in Section~\ref{subcompl}, {\tt CompHEP}
essentially consists of two parts: the symbolic and the numerical one.
A flowchart representing the symbolic part is shown in
Fig.~\ref{figcomphep1}.

After choosing the underlying model in {\it menu~1}, one gets to {\it
  menu~2} where the process must be defined and certain diagrams may be
discarded from further considerations with the help of {\it menu 4}.
One may also modify the parameters of the selected model (see {\it
  menu~3}).  {\it menu~5} is concerned with the graphical output and the
algebraic squaring of the matrix elements for the generated diagrams.

In {\it menu~6}, analytical expressions for the squared matrix elements
are computed. They can be stored to disk either in {\tt C}, {\tt
  FORTRAN}, {\tt REDUCE} or {\tt Mathematica} format (see {\it menu~8}).
Instead of applying the helicity-amplitude technique, {\tt CompHEP} uses
the conventional method of squaring the amplitudes before calculating
traces. Whether this is an advantage or a disadvantage depends on the
specific problem under consideration.  For simple processes a built-in
{\it numerical interpreter} (see {\it menu~7}) is available to perform
the phase space integration. For more complicated problems one has to
initiate the actual numerical part of {\tt CompHEP}. It deals with
processes that exceed the capability of the built-in calculator.

\begin{figure}
  \begin{center}
{\small
\vbox{
\centerline{
\begin {tabular}{|ll|}
 \multicolumn{2}{c}{\it menu 1}\\ \hline
 Subprocess & IN state\\
 Model parameters & QCD scale\\
 Breit-Wigner & Cuts\\
 Kinematics & Regularization\\
 Vegas & Simpson\\
 Batch & \\
 \hline
\end{tabular}
}
  }
}
\parbox{\captionwidth}{
\caption[]{\label{figcomphep2}Main menu of the numerical part of {\tt
  CompHEP}~\cite{pukhov}.}}
\end{center}
\end{figure}

The main options of the numerical part of {\tt CompHEP} are displayed in
Fig.~\ref{figcomphep2}.  Before starting the computation, the
environment for the Monte Carlo integration (using {\tt VEGAS}) must be
set.  This includes, for example, the modification of model parameters
and the choice of the center-of-mass energy. Furthermore, the user
decides whether structure functions for the incoming particles should be
used or not.  A central point is represented by the item {\it
  Regularization}, where the integration variables, previously defined
in the item {\it Kinematics}, are mapped in such a way that the
integrand becomes a smooth function.  This is one of the main features
of {\tt CompHEP}: The users task is just to provide possible
singularities in the propagators that are close to the edges of the
phase space region; the mappings themselves are done completely
automatically.  Other items are concerned with cuts over energy, angles,
transverse momenta, squared momentum transfers, and invariant masses or
the rapidity for a set of outgoing particles.  Note that all the
numerical calculations may equally well be performed in batch mode which
is useful as soon as the run-time becomes large.

In order to reduce the CPU time, {\tt CompHEP} saves the matrix elements
computed during a {\tt VEGAS} run into a file. This file is short
because it contains only one number for each event.  Later-on, a special
program called {\tt genEvents} repeats the Monte Carlo evaluations,
reading the previously calculated matrix elements from the file instead
of performing a re-calculation. Thus the CPU time is drastically reduced
for this second run, being typically of the order of only a few minutes
as compared to a few hours for the initial run.  Therefore, it is
possible to study the influence of various additional cuts on the
cross-section and to fill histograms rather quickly by performing
several subsequent runs of {\tt genEvents}.  Each working session
produces two output files: one containing the results of the calculation
together with a list of model parameters and a copy of the screen
report, and a second one containing the computed matrix elements.


%% file: applications_intro.tex
\section{Applications\label{secapplications}}
This section reports on recent results that would have been impossible
to obtain without the use of computer algebra. The selected examples are
supposed to underline the fields of applications for some of the
programs described above.

The first example is concerned with one of the most classical subjects
for multi-loop calculations, the photon polarization function. We will
describe its evaluation, using the program {\tt MATAD} in the limit
$q^2\ll m^2$, and additionally {\tt MINCER} and {\tt LMP} in the
opposite case, $q^2\gg m^2$.
The technique of repeatedly applying the hard-mass procedure was
successfully applied to the decay rate of the $Z$ boson into $b$ quarks
using {\tt EXP} in the {\tt GEFICOM} environment.
The calculation of renormalization group functions was the
driving item for {\tt BUBBLES} to be developed.
As part of the ``NIKHEF-setup'' it was used to compute four-loop
tadpole diagrams up to their simple poles in $\varepsilon$.
Another example for the calculation of a four-loop quantity is the decay
rate of the Higgs boson into gluons. Applying the approach of an effective
Hamiltonian, {\tt GEFICOM} could be used to compute the corresponding
diagrams.
As an application of the multi-leg program {\tt CompHEP} the
scenario of a strong interaction among electroweak gauge bosons will be
discussed. It is traded as an alternative to the Higgs mechanism to
restore unitarity at high energies.
Calculations in the full electroweak theory usually involve a large
number of different scales. For important contributions to the parameter
$\Delta r$ it was possible to reduce the contributing diagrams
to two-point functions which then were accessible with the help of the
programs {\tt FeynArts} and {\tt TwoCalc}.

%% file: polfunc.tex
\subsection{QCD corrections to the photon polarization
  function\label{sec::polfunc}}
\vspace{2ex}
{\bf Notation and methods}\\[2ex]
This section is concerned with the computation of QCD corrections
to the photon propagator. We define the polarization function in the
following way:
\begin{eqnarray}
\left(-g_{\mu\nu}q^2+q_\mu q_\nu\right)\,\Pi(q^2)
&=&i\int \dd^4 x\,e^{iqx}\langle 0 |Tj_\mu(x) j_\nu(0)|0 \rangle
\,,
\end{eqnarray}
where $j^\mu=\bar{\psi}\gamma^\mu\psi$ is the diagonal vector current of
two quarks with mass $m$.  The main motivation is the simple connection
of $\Pi(q^2)$ to the physical quantity $R(s)$ which is defined as the
normalized total cross section to the production of heavy quarks:
\begin{eqnarray}
R(s)&\equiv&\frac{\sigma\left(e^+e^-\to\mbox{hadrons}\right)}
            {\sigma\left(e^+e^-\to\mu^+\mu^-\right)}
    \,\,=\,\,12\pi\mbox{Im}\Pi(q^2=s+i\epsilon)
\,.
\end{eqnarray}
The advantage of making a detour to $\Pi(q^2)$ and not considering
$R(s)$ from the very beginning is the implicit summation of real
radiation and infra-red singularities. The fact that the number of loops
to be evaluated for $\Pi(q^2)$ is larger by one is made up for by having
to deal with two-point functions only, for which a huge amount of
technology is available.

\begin{figure}
  \begin{center}
    \leavevmode
    \epsfxsize=10.cm
    \epsffile[90 430 510 730]{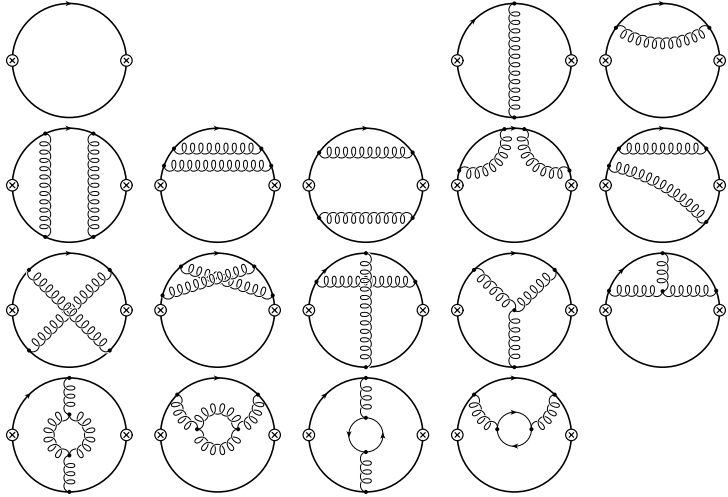}
    \hfill
    \parbox{\captionwidth}{
    \caption[]{\label{figpol}\sloppy
      Diagrams contributing to the one-, two- and three-loop polarization
      function. Solid lines represent quarks, loopy ones are gluons.
      }}
  \end{center}
\end{figure}

It is convenient to separately consider the following contributions:
\begin{eqnarray}
\Pi(q^2) &=& \Pi^{(0)}(q^2) 
         + \frac{\alpha_s(\mu^2)}{\pi} C_F \Pi^{(1)}(q^2)
         + \left(\frac{\alpha_s(\mu^2)}{\pi}\right)^2\Pi^{(2)}(q^2)
         + \ldots\,\,,
\\
\Pi^{(2)} &=&
                C_F^2       \Pi_A^{(2)}
              + C_A C_F     \Pi_{\it NA}^{(2)}
              + C_F T   n_l \Pi_l^{(2)}
              + C_F T       \Pi_F^{(2)},
\label{eqpi2}
\end{eqnarray}
and similarly for $R(s)$.  The colour factors $C_F=(N_c^2-1)/(2N_c)$
and $C_A=N_c$ correspond to the Casimir operators of the fundamental
and adjoint representations of $SU(3)$, respectively.  The case of QCD
corresponds to $N_c=3$, the trace normalization of the fundamental
representation is $T=1/2$, and the number of light (massless) quark
flavours is denoted by $n_l$.  In Eq.~(\ref{eqpi2}), $\Pi_A^{(2)}$ is
the Abelian contribution (corresponding to quenched QED) and 
$\Pi_{\it NA}^{(2)}$ is
the non-Abelian part specific for QCD.  There are two fermionic
contributions arising from diagrams with two closed fermion lines,
so-called double-bubble diagrams: For $\Pi_l^{(2)}$ the quark in the
inner loop is massless, the one in the outer loop massive, whereas for
$\Pi_F^{(2)}$ both fermions carry mass $m$.  The case where the external
current couples to massless quarks and these via gluons to massive ones
will not be addressed here.  Its imaginary part was considered in
\cite{HoaJezKueTeu94}.  The Feynman diagrams contributing to one-, two-
and three-loop order are shown in Fig.~\ref{figpol}.

The two-loop corrections of ${\cal O}(\alpha_s)$ were computed in
analytic form in the context of QED quite some time ago~\cite{KalSab55}.
In subsequent works the calculation was redone and more convenient
representations were found~\cite{BarRem73,Kni90,BroFleTar93}.  It is, at
least with the currently available technology, out of range to compute
the three-loop diagrams in complete analytic form for arbitrary $q$ and
$m$.  Only for a subclass of diagrams, namely the ones with two closed
fermion lines where one of them is massless, was it possible to compute
the corresponding contribution to $R(s)$
analytically~\cite{HoaJezKueTeu94,HoaKueTeu95,Teudiss,Hoadiss}.
Nevertheless, there are essentially two approaches which provide very
good approximations for $\Pi^{(2)}(q^2)$, both of them relying heavily
on the use of computer algebra.

One method is to consider the polarization function $\Pi(q^2)$ in the
limit $q^2\gg m^2$. Application of the large-momentum procedure (see
Section~\ref{subasymp}) leads to an expansion in $m^2/q^2$ with the
coefficients still depending on logarithms $\ln(-q^2/m^2)$.  The idea is
to evaluate as many terms as possible (the limitations essentially
coming from the CPU time and the size of the intermediate expressions)
and thus to approximate the true function $\Pi(q^2)$ even for rather
small values of $q^2$.

Whereas the above method results in analytical expressions, an
alternative way is to construct a semi-numerical result for $\Pi(q^2)$
which will, however, be valid for all values of $q$ and $m$. In addition
to the high energy expansion discussed above, this method also requires
the expansion of $\Pi(q^2)$ in the limit $q^2\ll m^2$.  Together
with some information about the (two-particle) threshold $q^2=4m^2$, a
suitable conformal mapping and the use of Pad\'e approximation, one
obtains an approximation for $\Pi(q^2)$ over the full kinematical region
\cite{CheKueSte96,CheHarSte98}.

Referring to the literature for physical applications (e.g.,
\cite{ckk96,chkst98}), in the following we will concentrate on the
technical aspect of the calculation in the limits $q^2\ll m^2$ and
$q^2\gg m^2$. While the latter case directly corresponds to the first
method above, the former one is needed for the second method.

As can be seen from Fig.~\ref{figpol} only a small number of diagrams
contribute at three-loop level. This makes the use of a diagram generating
program like {\tt QGRAF} unnecessary.  The main challenge instead is to
compute as many terms in the expansions as possible.  Apart from fast
computers this requires an efficient algebraic language and optimized
programs to deal with a large number of terms. The choice for this
problem was {\tt MATAD} and {\tt MINCER}, both written in {\tt FORM}
(see Section~\ref{sec::matmin}).

\vspace{4ex}
{\bf Large mass limit}\\[2ex]
In the limit $q^2\ll m^2$ application of the hard-mass procedure (see
Section~\ref{subasymp}) shows that it suffices to keep the naive Taylor expansion
of the diagrams in their external momentum. Thus one stays with the
calculation of massive tadpole integrals. Let us consider the moments $C_n$ of
the polarization function, defined through the Taylor series
\begin{eqnarray}
\Pi^{(2)}(q^2) &=& 
              \frac{3}{16\pi^2}
              \sum_{n>0} C_{n}^{(2)} \left(\frac{q^2}{4m^2}\right)^n
\,,
\label{eqpolfunpi}
\end{eqnarray}
where $m$ is the on-shell mass.
Although this series does not develop an imaginary part, one may gain
information on the rate from it by exploiting the analyticity of
$\Pi(q^2)$. We do not want to discuss this approach here and refer the
interested reader to~\cite{BaiBro95,CheKueSte96}.  At three-loop level
the evaluation of the coefficients up to $C_8$ was performed in
\cite{CheKueSte96}. The calculation required disk space of the order of
several GB for {\tt FORM} (so-called ``formswap'') to store the
intermediate expressions.  The total CPU time on a 256~MHz DEC-Alpha
workstation with 128~MB main memory was roughly two weeks.  With the
input for the ladder-type diagram already shown in
Section~\ref{sec::matmin}, let us now have a look at the output of {\tt
  MATAD}. The expansion up to terms of order $(q^2/m^2)^4$ reads:
\begin{verbatim}
   ladder =
       + ep^-3 * (  - 8/9*Q.Q )

       + ep^-2 * (  - 1/693*Q.Q^5*M^-8 + 8/945*Q.Q^4*M^-6 - 2/35*Q.Q^3*M^-4 + 
         8/15*Q.Q^2*M^-2 - 134/27*Q.Q )

       + ep^-1 * (  - 20*M^2 - 4042699/108056025*Q.Q^5*M^-8 + 273004/1488375*
         Q.Q^4*M^-6 - 63029/66150*Q.Q^3*M^-4 + 10658/2025*Q.Q^2*M^-2 - 473/81*
         Q.Q - 4/3*Q.Q*z2 )

       + 96*z3*M^2 - 482/3*M^2 - 19508/225*Q.Q^5*z3*M^-8 - 1/462*Q.Q^5*M^-8*z2
          + 129728592122581/1248047088750*Q.Q^5*M^-8 + 42496/675*Q.Q^4*z3*M^-6
          + 4/315*Q.Q^4*M^-6*z2 - 559517166977/7501410000*Q.Q^4*M^-6 - 44*
         Q.Q^3*z3*M^-4 - 3/35*Q.Q^3*M^-4*z2 + 6119442979/125023500*Q.Q^3*M^-4
          + 328/27*Q.Q^2*z3*M^-2 + 4/5*Q.Q^2*M^-2*z2 - 880963/121500*Q.Q^2*
         M^-2 + 11909/486*Q.Q - 392/9*Q.Q*z3 - 67/9*Q.Q*z2;
\end{verbatim}
Note that this expression still contains an overall factor $q^2$ (the
results above are in Euclidean space, indicated by the capital $Q$,
i.e.\ $Q^2 = -q^2$).  After dividing by $q^2$ there are terms of ${\cal
  O}(m^2/q^2)$ which cancel in the sum of all three-loop diagrams.  The
constant terms disappear after requiring the QED-like on-shell condition
$\Pi(0)=0$, and finally the structure of Eq.~(\ref{eqpolfunpi}) is
obtained.

After all diagrams up to three loops are computed and added up (taking
into account the correct colour factors), the parameters $\alpha_s$ and
$m$ are renormalized. To give an impression of the structure of
the final result, we list the first two and the eighth term of the
Abelian part, $\Pi_A^{(2)}$:
\begin{eqnarray}
   C^{(2)}_{A,1} &=&
          - {8687\over 864}
          - {32\over 5}\,\zeta_2\,\ln 2
          + 4\,\zeta_2
          + {22781\over 1728}\,\zeta_3\,,
\nonumber\\
   C^{(2)}_{A,2} &=&
          - {223404289\over 1866240}
          - {192\over 35}\,\zeta_2\,\ln 2
          + {24\over 7}\,\zeta_2
          + {4857587\over 46080}\,\zeta_3\,,
\label{eq::pfcn}
\\
&\cdots&
\nonumber\\
   C^{(2)}_{A,8} &=&
          - {190302182417255312898886115648452691\over 
             63063833203636585050931200000}
\nonumber\\
&&\mbox{}
          - {786432\over 230945}\,\zeta_2\,\ln 2
          + {98304\over 46189}\,\zeta_2
          + {31209476560803609727258477\over 12432176686773043200}\,\zeta_3
\nonumber
\,,
\end{eqnarray}
where $\zeta_n \equiv \zeta(n)$, with Riemann's $\zeta$-function. For
the specific values above it is $\zeta_2 = \pi^2/6$ and $\zeta_3 =
1.20206\ldots$.  The left-out coefficients and the ones for the other
colour structures of (\ref{eqpi2}) can be found in~\cite{CheKueSte96}.

\vspace{4ex}
{\bf Large momentum limit}\\[2ex]
The calculation for $q^2\gg m^2$ is more involved in the sense that
here the asymptotic expansion does not reduce to a naive Taylor
expansion.  Instead, the large-momentum procedure to be applied in this
case generates a large number of subgraphs for each of the diagrams of
Fig.~\ref{figpol}. Whereas for the small-momentum expansion a computer
only had to be used for the computation of the diagrams, in this
case even the generation of the subdiagrams is a non-trivial task and
program packages like {\tt EXP} or {\tt LMP} (see
Section~\ref{subsubexplmp}) are indispensable.
For the three-loop case the 18 initial diagrams produce 240 subgraphs
and their manual selection would be very tedious and error-prone if not
impossible.  The corresponding numbers in the two- (one-) loop case are
2 (1) initial and 14 (3) subdiagrams which would still allow a manual
treatment.  These numbers also show that there is a big step in
complexity when going from two to three loops.

\begin{figure}[t]
  \begin{center}
  \leavevmode
   \epsfxsize=8em
   \raisebox{-2.7em}{\epsffile[150 260 420 450]{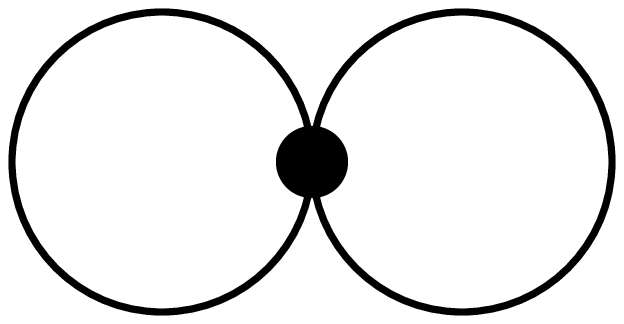}}\hspace{0em}
   \raisebox{0em}{\Large $\star$}\hspace{1em}
   \epsfxsize=8em
   \raisebox{-2.7em}{\epsffile[150 260 420 450]{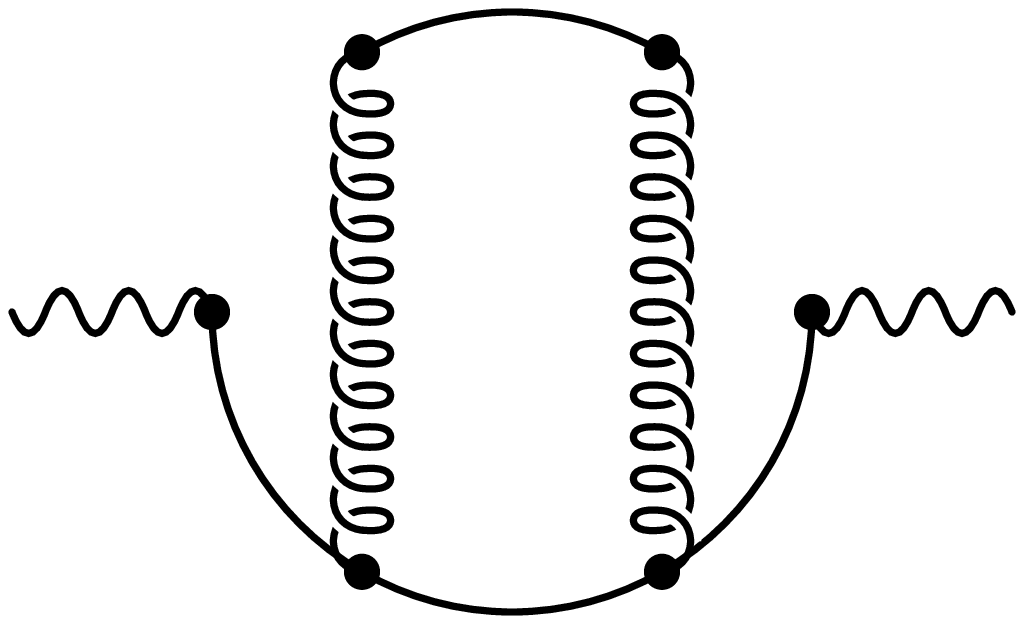}}
   \parbox{\captionwidth}{
   \caption[]{\label{figsub}
     Sixth subdiagram as generated by {\tt LMP} for the large-momentum
     procedure of the ladder diagram (cf.\ Fig.~\ref{figladder}).  }}
 \end{center}
\end{figure}

For the calculation the package {\tt LMP} (see
Section~\ref{subsubexplmp}) was used.  As an example let us again
consider the ladder diagram of Fig.~\ref{figladder}.  The input file,
looking similar to the one of the low-energy case above, produces 27
subdiagrams.  The one of Fig.~\ref{figsub} (the sixth subgraph generated
by {\tt LMP}), for example, corresponds to the following code:
\begin{verbatim}
*--#[ d3l3_6 :
        ((-1)
        *Dg(nu1,nu2,+aexp,-p2)
        *Dg(nu3,nu4,-bexp,+p2)
        *S(nu2,+p1mexp,nu4,-bexp,+qmexp,mu2,-p21m,nu3,-p2mexp,nu1,
           -p11m,mu1,-aexp,+qmexp)
        *1);
#define DIANUM "6"
#define TOPOLOGY "arb"
#define INT1 "inpl1"
#define MASS1 "0"
#define INT2 "topL1"
#define MASS2 "M"
#define INT3 "topL1"
#define MASS3 "M"
*--#] d3l3_6 :
\end{verbatim}
After Taylor expansion, the integral factorizes into a massless and two
massive one-loop integrals. The topology of each of these one-loop
integrals is encoded in the pre-processor variables {\tt INT1}, {\tt
  INT2}, and {\tt INT3}, where {\tt inp..} denotes a massless, {\tt
  top..} a massive topology (see also Section~\ref{subsubexplmp}). The
corresponding mass is given by {\tt MASS1}, {\tt MASS2}, and {\tt MASS3}.

Again the first two and the highest available (in this case the seventh)
term in the expansion for $\Pi_A^{(2)}$ will be displayed:
\begin{eqnarray}
   \bar{\Pi}^{(2)}_{A} &=& {3 \over 16 \pi^2}\, \bigg\{
       - {143\over 72}
          - {37\over 6}\,\zeta_3
          + 10\,\zeta_5
          + {1\over 8}\,\logqmums
\nonumber\\&&\mbox{}
       + {\bar{m}^2\over q^2} \, \bigg[
            {1667\over 24}
          - {5\over 3}\,\zeta_3
          - {70\over 3}\,\zeta_5
          - {51\over 2}\,\logqmums
          + 9\,\logqmums^2
          \bigg] 
\nonumber\\[2ex]&&\mbox{}
+ \,\,\cdots
\nonumber\\[1ex]&&\mbox{}
       + \left({\bar{m}^2\over q^2}\right)^{6} \, \bigg[
          - {420607059143\over 19440000}
          - {13059229\over 2700}\,\zeta_3
          - 1120\,\zeta_4
          + 4000\,\zeta_5
          + {560\over 3}\,B_4
\nonumber\\&&\mbox{\hspace{1.0cm}}
          + \left( {133099291\over 972000}
          + {16842\over 5}\,\zeta_3 \right)\,\logqmms
          + {54076013\over 7200}\,\logqmms^2
          + {8575579\over 2700}\,\logqmms^3
\nonumber\\&&\mbox{\hspace{1.0cm}}
          + \left( {13274779\over 450}
          - {142256\over 15}\,\logqmms
          - 8992\,\logqmms^2 \right)\,\logqmums
\nonumber\\&&\mbox{\hspace{1.0cm}}
          + \left( - 10854
          + 7560\,\logqmms \right)\,\logqmums^2
          \bigg]
\bigg\} + \ldots\,,
\label{pi2abar}
\end{eqnarray}
with $\logqmms = \ln(-q^2/\bar m^2)$ and $\logqmums = \ln(-q^2/\mu^2)$,
$\zeta_n$ as in (\ref{eq::pfcn}) and the additional values
$\zeta_4 = \pi^4/90$, $\zeta_5 = 1.03693\ldots$.  $B_4$ was
analytically calculated in \cite{Bro92} and evaluates numerically to
$B_4 = -1.76280\ldots$.  $\bar m$ is the \msbar\ mass which can easily be
transformed to the on-shell scheme~\cite{GraBroGraSch90}. For the
left-out terms in (\ref{pi2abar}) we refer to
\cite{CheHarKueSte96,CheHarKueSte97}.

\begin{figure}[ht]
\begin{center}
\begin{tabular}{cc}
    \leavevmode
    \epsfxsize=5.5cm
    \epsffile[110 265 465 560]{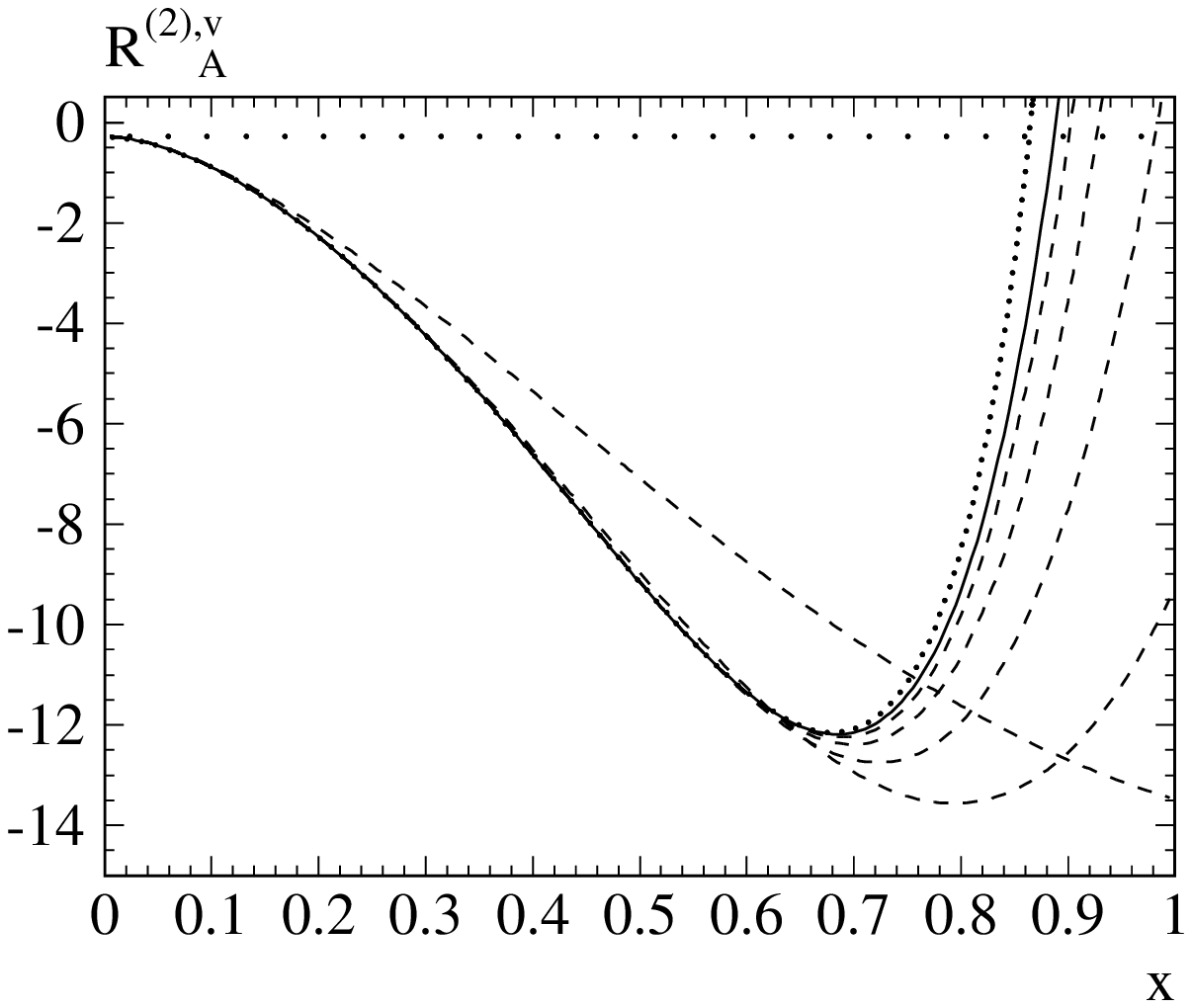}
&
    \epsfxsize=5.5cm
    \epsffile[110 265 465 560]{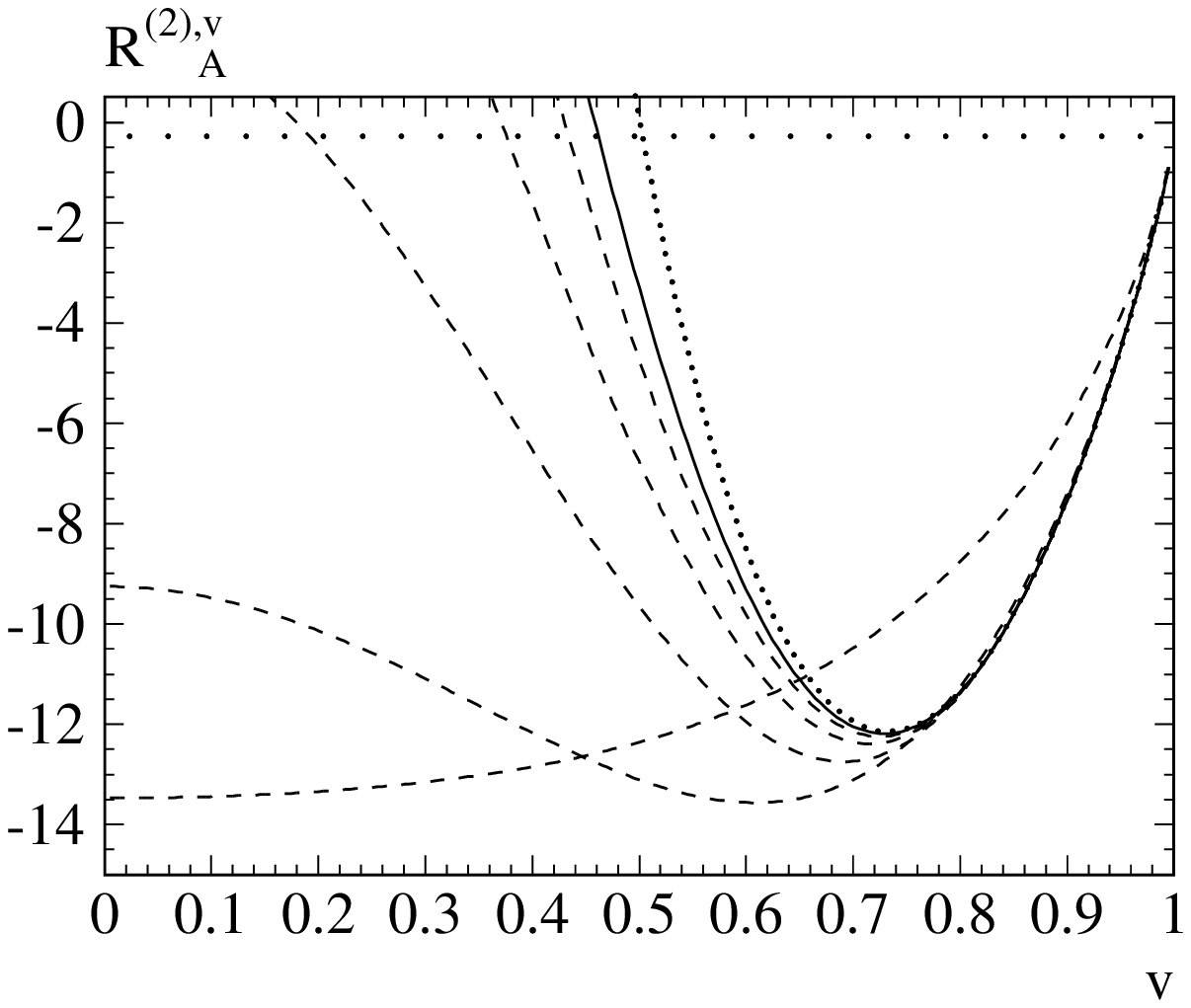}
\\
$(a)$ & $(b)$
\end{tabular}
\parbox{\captionwidth}{
    \caption[]{\label{figrAv}\sloppy
      The Abelian contribution $R_A^{(2)}$ as functions of $(a)$ $x =
      2m/\sqrt{s}$ and $(b)$ $v=\sqrt{1-4m^2/s}$.  Wide dots: no mass
      terms; dashed lines: including mass terms $(m^2/s)^n$ up to $n=5$;
      solid line: including mass terms up to $(m^2/s)^6$; narrow dots:
      semi-analytical result obtained via Pad\'e
      approximation~\cite{CheKueSte96}.  }}
\end{center}
\end{figure}

In Fig.~\ref{figrAv}~(a) the imaginary part of the Abelian contribution
is plotted as a function of $x=2m/\sqrt{s}$ including successively
higher orders in $x$. Up to $x\approx 0.7$ very quick convergence is
observed. For larger values of $x$, however, the inclusion of higher
terms does not significantly improve the approximation.  This becomes
manifest in an even more drastic way in Fig.~\ref{figrAv}~(b), showing
the same curves as functions of $v=\sqrt{1-4m^2/s}$.  This choice of
variable enlarges the threshold region and thus demonstrates the
breakdown of the high-energy expansion close to threshold.

%
%

%% file: zbb.tex
\subsection[Corrections of ${\cal O}(\alpha\alpha_s)$ to the decay of
the $Z$ boson into bottom quarks]{Corrections of \bld{{\cal
      O}(\alpha\alpha_s)} to the decay of the \bld{Z} boson into bottom
  quarks\label{sec::zbb}}
In the examples considered so far essentially only the QCD part of the
Standard Model is involved, which is, as already noted, the most
important field for multi-loop calculations. The electroweak sector, on
the contrary, resides on a broken gauge symmetry which is the reason why
the diagrams in general carry a lot more scales. Higher order
calculations are thus much more involved.

As an in some sense intermediate case, one may consider mixed
electroweak/QCD corrections. An important example in this class are the
${\cal O}(\alpha\alpha_s)$ corrections to the decay rate of the $Z$
boson into quarks.  One may distinguish two cases, namely the decay into
the light $u,d,s$ or $c$ quarks and the one into $b\bar{b}$.  The latter
case is of special interest as the top quark enters the vertex diagrams
already at one-loop order and leads to corrections proportional to
$M_t^2$.  The complete ${\cal O}(\alpha)$ corrections were considered
in~\cite{AkhBarRie86,BeeHol88}.  Since the mixed ${\cal
  O}(\alpha\alpha_s)$ correction to the $Z$ decay into $u,d,s$ and $c$
turned out to be quite sizable~\cite{CzaKue96}, it was tempting to
consider also the decay into bottom quarks at this order.  The leading
terms of ${\cal O}(G_F M_t^2)$ and ${\cal O}(G_F \ln(M_t^2/M_W^2))$ were
computed in~\cite{FleJegRacTar92,CheKwiSte93,KwiSte95,Per95}.

However, the full $M_t$ dependence to
${\cal O}(\alpha\alpha_s)$ is currently out of reach.  On the other
hand, the top quark is certainly the heaviest particle entering the $Z b
\bar b$ vertex at this order, which makes asymptotic expansions a
well suited tool to obtain a very good approximation to the full answer.

Let us describe the technical aspects of the calculation for $Z\to b
\bar b$ in more detail. The ${\cal O}(\alpha\alpha_s)$ corrections are
computed by applying the optical theorem, i.e., by evaluating the
imaginary part of the $Z$ boson propagator.
\begin{figure}[t]
  \begin{center}
    \leavevmode 
    \epsffile[105 558 503 717]{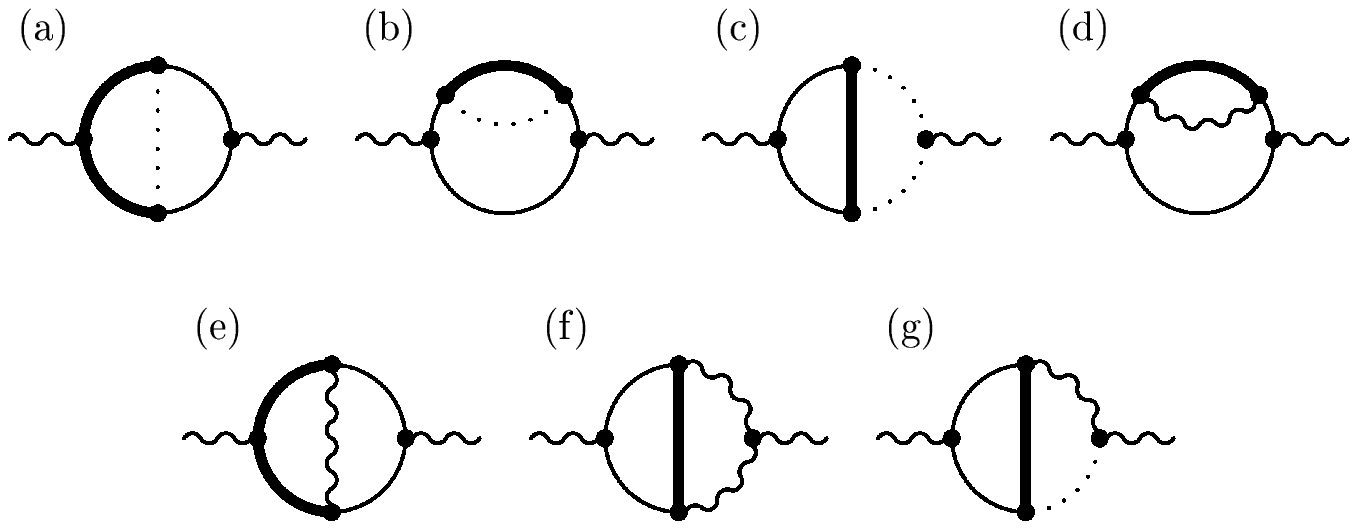}\\
    \parbox{\captionwidth}{
      \caption[]{\label{fig::zbbdias}Diagrams containing a top quark that 
        contribute to $Z\to b\bar b$. Thin lines correspond to bottom
        quarks, thick lines to top quarks, dotted lines to Goldstone
        bosons and inner wavy lines represent $W$ bosons.}}
  \end{center}
\end{figure}
Our concerns are only the diagrams involving the top quark.  The ones
contributing to ${\cal O}(\alpha)$ are shown in Fig.~\ref{fig::zbbdias}.
The diagrams to be evaluated to ${\cal O}(\alpha\alpha_s)$ can be
obtained by attaching a gluon in all possible ways, thereby increasing
the number of loops by one.

The different scales in these diagrams are given by $M_t,M_W,M_Z$ and
$M_\Phi$. The latter one is the mass of the charged Goldstone boson. It
is related to $M_W$ via $M_\Phi^2=\xi_W M_W^2$, where $\xi_W$ is the
corresponding gauge parameter which is
arbitrary and should drop out in the final result.  In a first step we
will consider it such that $M_t^2 \gg M_\Phi^2$. For the diagrams (a),
(b), (d) and (e) the application of the hard-mass procedure with respect
to $M_t$ immediately leads to a complete factorization of the
different scales, i.e., only single-scale integrals with at most three
loops are left. For the diagrams (c), (f) and (g), however, some
co-subgraphs still carry more than one scale which makes an evaluation
quite painful, especially as their ${\cal O}(\varepsilon)$ part is also
needed.  It is very suggestive to apply the hard-mass procedure again to
these co-subgraphs.  In order to avoid unwanted imaginary parts arising
from $W$ and $\Phi$ cuts the hierarchy to be chosen among the physical
masses is $M_t^2 \gg M_W^2 \gg M_Z^2$.  The last inequality is seemingly
inadequate, but a closer look at the corresponding diagrams shows that
in this context it is actually equivalent to $4M_W^2 \gg M_Z^2$,
respectively, $(M_W+M_t)^2 \gg M_Z^2$. The mass of the Goldstone boson
may be incorporated using the possible hierarchies $\xi_WM_W^2\gg M_W^2
\gg M_Z^2$ or $M_W^2 \gg \xi_W M_W^2 \gg M_Z^2$.  This procedure leads
to a nested series in $M_W^2/M_t^2$ and $M_Z^2/M_W^2$.
Diagrammatically, an example for this procedure of successively applying
asymptotic expansions looks as follows:
\begin{eqnarray*}
&& \epsfxsize=7em
\raisebox{-2em}{\epsffile[120 260 460 450]{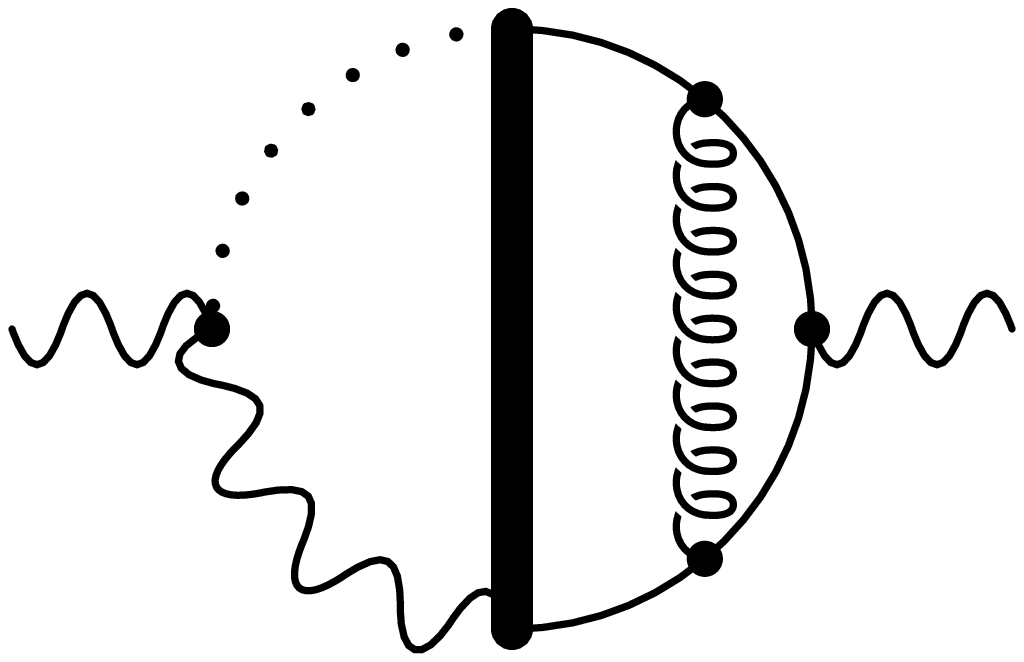}}
\stackrel{M_t^2\to \infty}{\longrightarrow}
\,\,\,\,\,
\epsfxsize=8em \raisebox{-1.7em}{\epsffile[120 260 560 450]{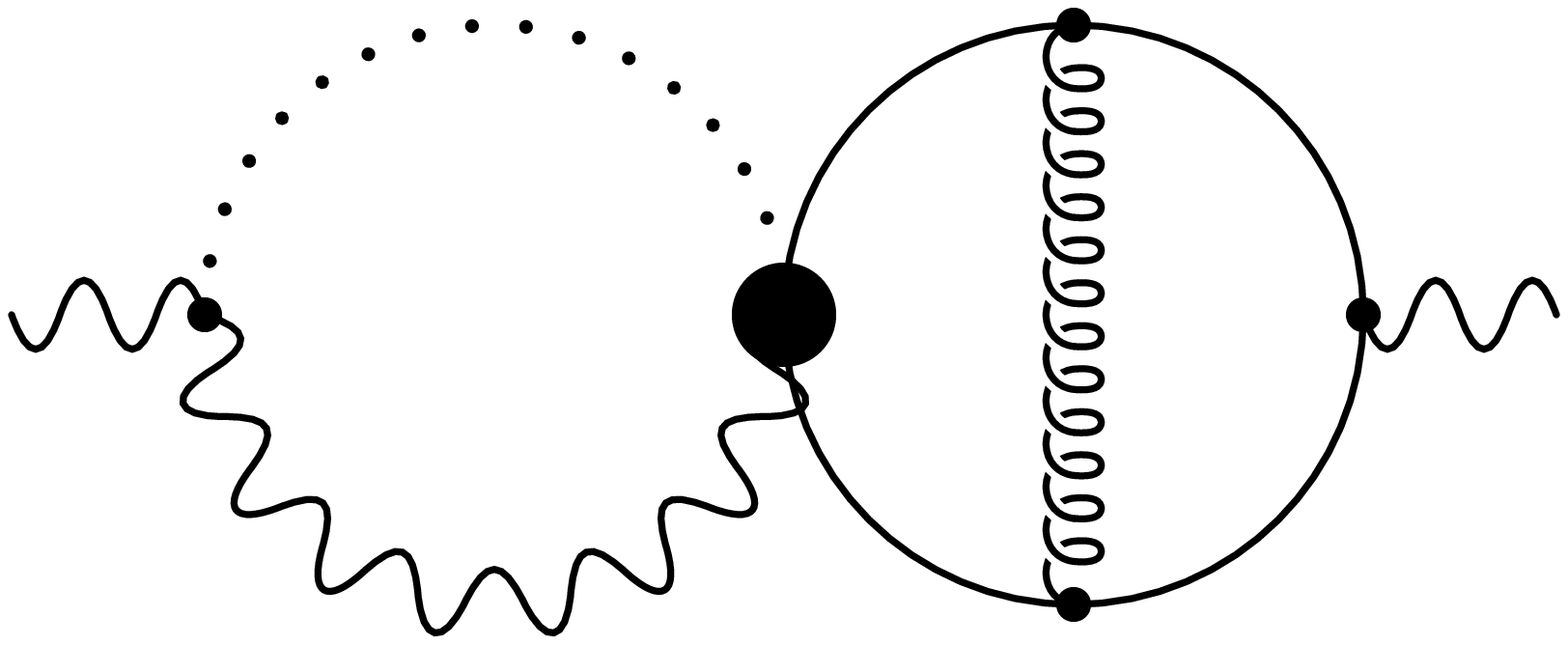}}
  \star
\epsfxsize=.9em
\raisebox{-1.7em}{\epsffile[260 260 310 450]{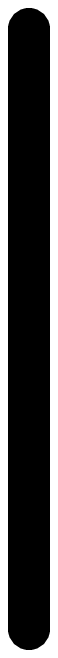}}
\,\,+ \,\,\cdots \\[.5em]&&\hspace{8em}
\stackrel{\xi M_W^2\to \infty}{\longrightarrow}
  \bigg(\!\!\!\!
  \epsfxsize=7em
  \raisebox{-2em}{\epsffile[120 260 460 450]{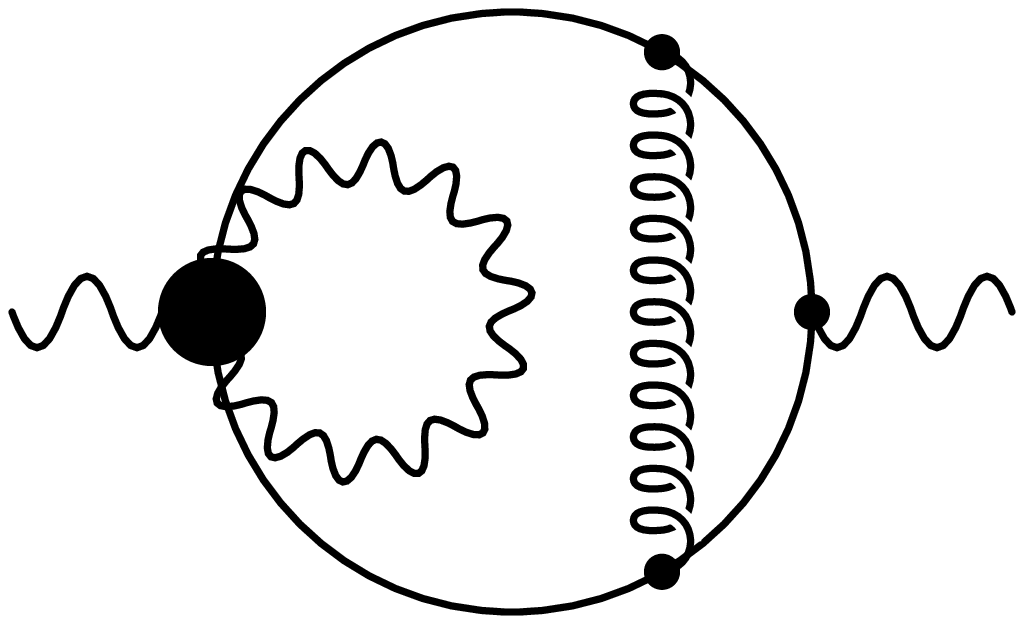}}
  \star\!\!\!\!\!
  \epsfxsize=7em \raisebox{-2em}{\epsffile[120 260 460
  450]{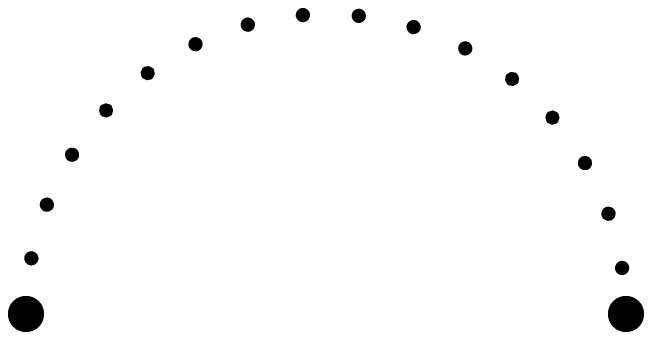}} \hspace{-1em}\bigg)
\star
\epsfxsize=.9em
\raisebox{-1.7em}{\epsffile[260 260 310 450]{zbbrhst.ps}} + \cdots\,,
\end{eqnarray*}
where only those terms are displayed which are relevant in the
discussion above and all others contributing to the hard-mass procedure
are merged into the three dots. The mass hierarchy is assumed to be
$M_t^2\gg \xi_W M_W^2 \gg M_W^2 \gg M_Z^2$.

The technical realization of the calculation was performed with the help
of {\tt GEFICOM} (see Section~\ref{subgeficom}). The main new ingredient
in comparison to the examples quoted so far is the use of the automated
version of the hard-mass procedure and its successive application,
implemented in the program {\tt EXP}, in combination with the generator
{\tt QGRAF} and the {\tt FORM} packages {\tt MINCER} and {\tt MATAD}.
This automation not only avoids human errors, but also allows several
checks of the results. For example, since the gauge parameter $\xi_W$ is
arbitrary, it is possible to choose it such that no longer the top
quark, but rather the Goldstone boson is the heaviest particle involved
in the process. Then one is left with the hierarchy $\xi_W M_W^2\gg
M_t^2 \gg M_W^2\gg M_Z^2$.  Note that for this choice the hard-mass
procedure produces completely different subdiagrams than for the cases
above, where $M_t$ was considered to be the largest scale.  Furthermore,
while a manual treatment of both cases would almost double the effort,
the automated version allows one to go from one hierarchy to the other by
simply interchanging two input parameters, as we will see below.

To be concrete, let us consider the input file needed for {\tt GEFICOM}.
It is certainly more sophisticated than the examples mentioned so far,
but this is mainly due to the enriched
particle spectrum.
\begin{verbatim}
*** MINCER

* scheme 2
* gauge 1
* exp y ma mc mb q
* powerma 2
* powermb 4
* powermc 4
* mass t Ma 
* mass Wp Mb
* mass pp Mc
* loops 1 true=iprop[t,0,0];
* loops 2 true=iprop[g,0,0];
* loops 2 false=iprop[t,0,0];
* loops 3 true=iprop[g,1,1];
* loops 3 false=iprop[t,0,0];

      list = symbolic ;
      lagfile = 'q.lag' ;
      in = Z[q];
      out = Z[q];
      nloop = ;
      options = onepi;
      true = iprop[Wm,pm,0,2];
      false = iprop[bq,0,1];
\end{verbatim}
The meaning of most of the lines can be deduced by comparison with the
example of Section~\ref{subgeficom}.
The mass of the top quark ({\tt t} $\widehat =
\,t$), the $W$ ({\tt Wp} $\widehat = \,W^+$), and the Goldstone boson
({\tt pp} $\widehat =\, \Phi^+$) are denoted by {\tt Ma}, {\tt Mb} and
{\tt Mc}, respectively.  The change in mass hierarchy mentioned before is
achieved by rewriting the line
\begin{verbatim}
* exp y ma mc mb q
\end{verbatim}
to
\begin{verbatim}
* exp y mc ma mb q
\end{verbatim}
This input file is combined together with the one containing
the vertices and propagators, which shall not be displayed here,
in order to generate in a first step the relevant diagrams
at three-loop level. After inserting the Feynman
rules, {\tt EXP} applies
the hard-mass procedure and reduces all diagrams to single-scale
integrals, if necessary by applying the procedure twice.
{\tt EXP} also produces the relevant administrative files which
then call {\tt MINCER} and {\tt MATAD}. 
The runtime for {\tt QGRAF} is of the order of a few seconds
and it takes a few minutes for {\tt EXP} to generate the
subdiagrams.
The time spent in the integration routines strongly depends on the
required depth of the expansion, of course. The computation
in~\cite{HarSeiSte97} took about three weeks.

The contribution of the vertex corrections to the decay rate of the $Z$
boson to bottom quarks, induced by the exchange of a $W$ or a Goldstone
boson, may be written in the following way:
\begin{equation}
\delta\Gamma_b^W = 
\delta\Gamma_d^W + (\delta\Gamma_b^W - \delta\Gamma_d^W) = 
\delta\Gamma_d^W + (\delta\Gamma_b^{0,W} - \delta\Gamma_d^{0,W}) = 
\delta\Gamma_d^W + \delta\Gamma^W_{b-d}\,,
\label{eq::delgamb}
\end{equation}
where $\delta\Gamma_q^{W}$ and $\delta\Gamma_q^{0,W}$ denote
renormalized and unrenormalized quantities. In the difference between
the $b$ and $d$ contributions the relevant counter-terms drop out.  This
means that $\delta\Gamma^W_{b-d}$ is independent of the renormalization
scheme and is therefore well suited for installation in data analyzing
programs like the ones described, e.g., in~\cite{YelRep95}. In addition,
$\delta\Gamma^W_{b-d}$ is also gauge independent. $\delta\Gamma_d^W$ has
been computed in~\cite{CzaKue96}.  
For convenience we list the result for $\delta\Gamma^W_{b-d}$ in
numerical form~\cite{HarSeiSte97}:
\begin{eqnarray}
\delta\Gamma^W_{b-d} &=&
    \Gamma^0 {1\over s^2_\theta}
    {\alpha\over \pi}
  \bigg\{ - 0.50
  + (0.71 -0.48)+ (0.08 - 0.29) + (-0.01 - 0.07) + (-0.007 - 0.006)
  \nonumber\\[0em]&&\mbox{}
  + {\alpha_s\over \pi} \bigg[ 1.16 + (1.21 - 0.49) + (0.30 - 0.65) +
  (0.02 - 0.21 + 0.01)
%
%
+ (-0.01 - 0.04
  + 0.004) \bigg] \bigg\} =
\nonumber\\
&=&\Gamma^0 {1\over s^2_\theta} {\alpha\over \pi} \bigg\{- 0.50 - 0.07 +
{\alpha_s\over \pi} \bigg[ 1.16 + 0.13 \bigg]\bigg\}\,,
\label{eq::zbb1l2l}
\end{eqnarray}
with $\Gamma^0 = N_cM_Z\alpha/(12s_\theta^2c_\theta^2$), $s_\theta =
\sin\theta_W$, $\theta_W$ being the weak mixing angle, $c_\theta^2 =
1-s_\theta^2$, and $M_t$ the on-shell top mass.  For
Eq.~(\ref{eq::zbb1l2l}) we used the values $M_t = 175$~GeV,
$M_Z=91.19$~GeV and $s_\theta^2 = 0.223$.  The numbers after the first
equality sign correspond to successively increasing orders in $1/M_t^2$,
where the brackets collect the corresponding constant, $\ln
(M_t^2/M_W^2)$ and, if present, $\ln^2(M_t^2/M_W^2)$ terms. The numbers
after the second equality sign represent the leading $M_t^2$ term and
the sum of the subleading ones.  The ${\cal O}(\alpha)$ and ${\cal
  O}(\alpha\alpha_s)$ results are displayed separately. Comparison of
this expansion of the one-loop terms to the exact result of
\cite{BeeHol88} shows agreement up to $0.01\%$ which gives quite some
confidence in the $\alpha\alpha_s$ contribution.  One can see that
although the $M_t^2$, $M_t^0$ and $M_t^0\ln(M_t^2/M_W^2)$ terms are of the same
order of magnitude, the final result is surprisingly well represented by
the $M_t^2$ term, since the subleading terms largely cancel among each
other.

%% file: beta.tex
\subsection[Four-loop $\beta$ function]{Four-loop \bld{\beta} 
  function\label{sec::beta}}
The $\beta$ function is an object of common interest in any field
theory, especially in non-Abelian ones. It describes the dependence of
the corresponding coupling constant with respect to the energy scale.
For QCD it is convenient to define
\begin{eqnarray}
\mu^2{\dd\over \dd\mu^2}\frac{\alpha_s(\mu^2)}{\pi} &=& 
\beta(\alpha_s)
\,\,=\,\,
-\left(\frac{\alpha_s(\mu^2)}{\pi}\right)^2
\sum_{i\ge0}\beta_i\left(\frac{\alpha_s(\mu^2)}{\pi}\right)^i
\,.
\end{eqnarray}
The $\beta$ function is directly related to the renormalization factor
$Z_g$ defined through $\alpha_s^0
= Z_g^2\alpha_s $, where $\alpha_s^0$ and $\alpha_s$ are
the bare and renormalized coupling constant of QCD, respectively:
\begin{eqnarray}
  \beta(\alpha_s)= {\alpha_s^2\over \pi}
  {\partial\over \partial\alpha_s} Z_g^{2,(1)}\,,
\label{eqbeta}
\end{eqnarray}
where $Z_g^{2,(1)}$ is the residue of $Z_g^2$ with respect to its
Laurent expansion in $\varepsilon$.
In Eq.~(\ref{eqbeta}) it is already indicated that in mass independent
renormalization schemes, like the \msbar\ scheme, the renormalization
constants and thus also the $\beta$ function only depend on $\alpha_s$.

There are several ways to compute $Z_g$, all of them related through
Slavnov-Taylor identities.  One choice would be to combine the
renormalization constants of the four-gluon vertex and the gluon
propagator.  At four-loop level, however, this requires the computation
of about half a million diagrams.  On the other hand, the approach
of~\cite{RitVerLar97} was based on the relation
\begin{eqnarray}
Z_g &=& \frac{\tilde{Z}_1}{\tilde{Z}_3\sqrt{Z_3}}
\,,
\end{eqnarray}
where $\tilde{Z}_1$ is the renormalization constant of the
ghost-ghost-gluon vertex, and $\tilde{Z}_3$ and $Z_3$ are the ones of
the ghost and gluon propagators, respectively.  Thus the pole parts of
the corresponding Green functions had to be computed up to four loops.

The roughly 50,000 contributing diagrams were generated by means of {\tt
  QGRAF}.  Introducing an artificial mass $M$ in each propagator as
infra-red regulator allowed a Taylor expansion to be performed with
respect to all external momenta, leading to four-loop massive tadpole
integrals.  Recurrence relations based on the integration-by-parts
algorithm were implemented in the program {\tt BUBBLES}~\cite{bubbles}
to reduce the integrals to a minimal set of master integrals. At one-,
two- and three-loop level only one, at four-loop level two master
integrals are required.  The colour factors of the individual diagrams
were determined with the help of a {\tt FORM} program~\cite{color}.  To
cope with the huge number of diagrams a special database-like tool
called {\tt MINOS} was developed. It controls the calculation and allows
to conveniently access the results of single diagrams.  The total CPU
time for this calculation was of the order of a few months.
 
The final result which we will only quote for the physically most
interesting case of QCD reads
\begin{eqnarray}
\renewcommand{\arraystretch}{ 1.3}
 \beta_0 & = & {1\over 4}\left(11 - \frac{2}{3} n_f \right)\,,  \nonumber\\
\beta_1 & = & {1\over 16}\left( 102 - \frac{38}{3} n_f \right)\,, \nonumber \\
 \beta_2 & = & {1\over 64}\left( \frac{2857}{2} - \frac{5033}{18} n_f +
 \frac{325}{54} n_f^2 \right)\,, \nonumber \\
 \beta_3 & = &  {1\over 256} \bigg[
 \left( \frac{149753}{6} + 3564 \zeta_3 \right)
        - \left( \frac{1078361}{162} + \frac{6508}{27} \zeta_3 \right) n_f
  \nonumber \\ & &
       + \left( \frac{50065}{162} + \frac{6472}{81} \zeta_3 \right) n_f^2
       +  \frac{1093}{729}  n_f^3
       \bigg]\,,
\end{eqnarray}
where $n_f$ is the number of active flavours.


%% file: hgg.tex
\subsection{Hadronic Higgs decay\label{sec:applic:hgg}}
%
%

\subsubsection{Introduction}
As was already pointed out in the introduction, the only particle of the
Standard Model not yet discovered is the Higgs boson.  Thus it is
necessary to learn as much as possible about potential production and
decay mechanisms of such a particle in order to correctly interpret the
signals in the detector.  An important decay channel of the Higgs boson
is the one into gluons. Although it is a loop-induced process it is
numerically not negligible mainly due to the fact that the ${\cal
  O}(\alpha_s)$ corrections are very
large~\cite{InaKubOka83,DjoSpiZer91}. In this section it is
shown how to compute the ${\cal O}(\alpha_s^2)$ corrections exploiting
the powerful tools of automatic Feynman diagram computation.  For
completeness also the decay into quarks will be discussed.

The Higgs boson will be assumed to be of intermediate-mass range, i.e.,
$M_H\lsim 2M_W$, which suggests the use of the approach of an effective
Lagrangian where the top quark is integrated out. For convenience
of the reader let us, in a first step, collect the relevant formulae.
Detailed derivations can be found
in~\cite{Klu75,Spi84,InaKubOka83,CheKniSte97hbb}.

The starting point is the Yukawa Lagrange density describing the
coupling of the $H$ boson to quarks.  In the limit $m_t\to\infty$ it can
be written as a sum over five operators~\cite{Klu75,Spi84} formed by the
light degrees of freedom, with the corresponding coefficient functions
containing the residual dependence on the top quark:
\begin{eqnarray}
{\cal L}_Y\,\,=\,\,-\frac{H^0}{v^0}
\sum_{q\in\{u,d,s,c,b,t\}}m_{q}^0\bar\psi_{q}^0\psi_{q}^0
&\longrightarrow&
{\cal L}_{\rm eff}\,\,=\,\,
-\frac{H^0}{v^0}\sum_{i=1}^5C_i^0{\cal O}_i^\prime
\,.
\label{eqlagr}
\end{eqnarray}
It turns out that only two of the operators, in the following called
${\cal O}_1^\prime$ and ${\cal O}_2^\prime$, contribute to physical processes.
Expressed in terms of bare fields they read
\begin{eqnarray}
{\cal O}^\prime_1
\,\,=\,\,
\left(G^{0\prime,a}_{\mu\nu}\right)^2\,,
&\quad&
{\cal O}^\prime_2
\,\,=\,\,
\sum_{q\in\{u,d,s,c,b\}}m_{q}^{0\prime}
  \bar\psi_{q}^{0\prime}\psi_{q}^{0\prime}
\,,
\end{eqnarray}
where $G^{0\prime,a}_{\mu\nu}$ is the gluonic field strength tensor.
The renormalization matrix for the coefficient functions 
can be expressed in terms of the well-known renormalization
constants of QCD and may be found in the papers cited above.

To evaluate the ingredients entering the calculation of
$\Gamma(H\to\mbox{hadrons})$ at three-loop level, namely $C_i^0$ and the
imaginary part of the correlators $\langle{\cal O}'_i{\cal O}'_j\rangle$,
a large amount of diagrams must be calculated. Thus it is necessary to
use both a generator for the Feynman diagrams and an efficient interface
transforming the output to a format readable by the integration packages.
In the matter at hand one is either confronted with
propagator-type diagrams or with tadpole integrals where the scales are
given by $M_H$ and $M_t$, respectively.  Therefore the package {\tt
  GEFICOM} in association with {\tt QGRAF} (for the generation) and {\tt
  MATAD/MINCER} (for the computation) is an ideal candidate for the
evaluation of the Feynman diagrams contributing to
$\Gamma(H\to\mbox{hadrons})$.

In the next section we will give an example of the calculation of three-loop
objects in the case of $C_1^0$ in some detail and concentrate in
Section~\ref{subhggcol} on the evaluation of the massless correlators.
{\tt GEFICOM} was even used to compute the ${\cal O}(\alpha_s^3)$
contribution to $\langle {\cal O}'_2 {\cal O}'_2\rangle$ which actually
is a four-loop calculation. It is shown how such a calculation is
nevertheless possible using the three-loop tools provided by {\tt
  GEFICOM}.  Finally, in Section~\ref{subhggres} the physical results
are briefly presented.


\subsubsection{Coefficient functions}
\begin{figure}[t]
\begin{center}
\leavevmode
\epsfxsize=15cm
\epsffile[76 635 552 724]{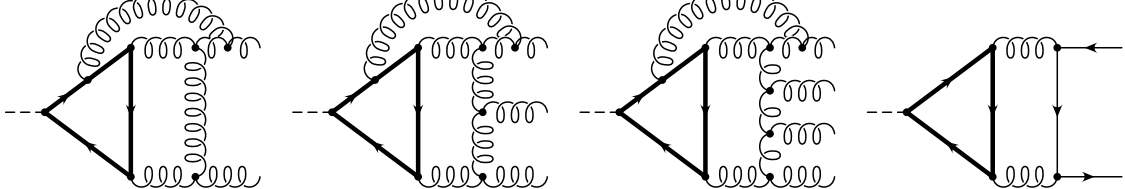}\\
\parbox{\captionwidth}{
\caption[]{\label{fighggc1}
  Typical Feynman diagrams contributing to the coefficients
  $C_1^0$ and $C_2^0$.  Looped, bold-faced, and dashed lines represent
  gluons, $t$ quarks, and $H$ bosons, respectively.}}
\end{center}
\end{figure}

The top quark dependence of the effective Lagrangian is contained in the
coefficient functions $C_i^0$. Thus it is not surprising that their
evaluation can be reduced to massive tadpole integrals with the scale
given by $M_t$.  As an example let us consider $C_1^0$.  There are three
independent ways to obtain $C_1^0$, namely from the sets of $ggH$
three-point, $gggH$ four-point, or $ggggH$ five-point diagrams.  At
three-loop level, these sets contain 657, 7362, and 95004 diagrams,
respectively. A sample diagram of each class is displayed in
Fig.~\ref{fighggc1} (together with a diagram to be evaluated for
$C_2^0$). The first set requires an expansion up to second order in the
external momenta, whereas for the second one it suffices to keep only
linear terms. For the diagrams of the third class all external momenta
may be set to zero from the very beginning. However, this advantage is
counterbalanced by the huge number of contributing diagrams.  Note that the
depth of the expansion is given by the individual terms of the operator
${\cal O}_1$ which dictate the
structure in the external momenta.

Ref.~\cite{CheKniSte97hgg} deals with the $ggH$ channel, using the package {\tt
  GEFICOM} to evaluate the 657 diagrams.  The file defining the process
is rather short because only the package {\tt MATAD} has to be
used:
\begin{verbatim}
*** MATAD
* power 2
* gauge 1
* dala12

      list = symbolic ;
      lagfile = 'q.lag' ;
      in = h[q1p2];
      out = g[q1], g[q2]; 
      nloop = ;
      options =;
      true = bridge[ g,q,c, 0,0 ];  
\end{verbatim}
The second line indicates that an expansion in the external momenta up
to second order should be performed, and {\tt dala12} prompts the
application of projectors for the $\sprod{q_1}{q_2}$ terms since we are dealing
with on-shell gluons ($q_1^2=q_2^2=0$).  The file containing the
propagators and vertices is essentially identical to the pure QCD case
(cf.~Section~\ref{subgeficom}) except for the additional coupling of the Higgs
boson to top quarks.

If one chooses a covariant gauge with arbitrary gauge parameter, the CPU
time for the calculation is of the order of a few days.  This is
drastically reduced after going, for example, to Feynman gauge.
Recently also the decay of a pseudo-scalar Higgs boson into gluons was
considered~\cite{CheKniSteBar98}. The number of diagrams being the same
as in the scalar case, there is a slight complication according to the
treatment of $\gamma^5$ in this case. Therefore the 7362 diagrams of the
three-gluon channel were evaluated in Feynman gauge to have a stringent
check of the calculation at one's disposal. The CPU time amounts to
roughly two weeks.  Let us refrain from listing the
analytical results and refer to \cite{CheKniSte97hgg,CheKniSte98dec}
instead.  Nevertheless, they are incorporated in the numerical formula
given in Section~\ref{subhggres}.


\subsubsection{\label{subhggcol}Correlators}

\begin{figure}[t]
  \begin{center}
    \leavevmode
    \epsfxsize=14.0cm
    \epsffile[125 640 500 730]{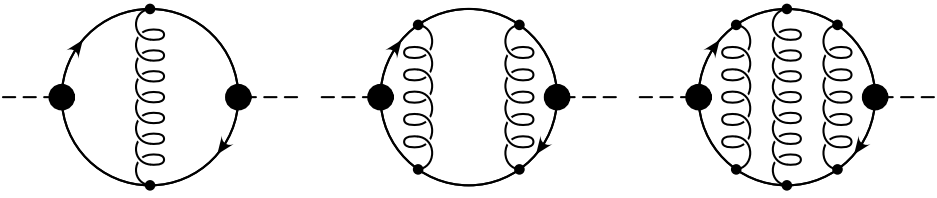}\\
    \parbox{\captionwidth}{
      \caption{\label{figo2o2}
        Typical Feynman diagrams contributing to
        $\langle{\cal O}_2^\prime{\cal O}_2^\prime\rangle$.
        The solid circles represent the operator
        ${\cal O}_2^\prime$.}}
  \end{center}
\end{figure}

The computation of the one-, two- and three-loop correlators
$\langle{\cal O}_i^\prime{\cal O}_j^\prime\rangle$ proceeds in close
analogy to the coefficient functions described in the previous section.
Therefore, let us in this subsection focus on the computation of the
imaginary part of the four-loop contribution to $\langle{\cal
  O}_2^\prime{\cal O}_2^\prime\rangle$ which was originally performed
in~\cite{Che97higgs}. Some sample diagrams are shown in
Fig.~\ref{figo2o2}. Although at first sight this seems to require the
evaluation of four-loop diagrams, the calculation can be reduced to
massless propagator-type and massive tadpole integrals at three-loop level.
The underlying idea is quite simple: since the imaginary part of
massless propagator-type diagrams only arises from $\ln(-q^2)$ terms and
the logarithms are in one-to-one correspondence to the poles in
$\varepsilon$ it suffices to compute the divergent parts of the
integrals. It is a feature of dimensional regularization accompanied
with MS-like schemes that the ultra-violet poles and thus the
renormalization constants are polynomials in the masses and external
momenta.  Consequently, for logarithmically divergent diagrams the poles
are independent of any mass scale.  This allows us to assign additional
masses to individual lines of the diagrams and to nullify the external
momentum.  If we take one of the lines at the left vertex to be massive
and set the external momentum to zero, the four-loop integrals can be
expanded in terms of three-loop propagator-type diagrams and one-loop
tadpole integrals for which the technology was described in some detail
above.  The drawback of the described method is that artificial
infra-red singularities are introduced which, of course, influence the
pole structure. They have to be removed using the so-called Infra-red
Rearrangement~\cite{Vla80CheKatTka80} which heavily relies on the
$R^*$ operation.  It was developed on a diagram-by-diagram basis
in~\cite{CheSmi84}.  In order to be able to perform the calculation
automatically a global version of the $R^*$ operation is
mandatory~\cite{Che97R,Che97higgs,Che97radcor96}.

In~\cite{Che97higgs} the computation of
$\Im\langle{\cal O}_2^\prime{\cal O}_2^\prime\rangle$
is presented and explicit formulae are given, demonstrating
that the combination of one-, two- and three-loop
massless propagator-type and massive vacuum integrals are
adequate to get the ${\cal O}(\alpha_s^3)$ corrections.
The computation was performed with the help of {\tt GEFICOM}
which had to be used at four-loop order.
A proper flag in the file defining the process tells
{\tt GEFICOM} to introduce the auxiliary mass and to rename
the momenta in accordance with the factorization taking place.
Again we refrain from listing explicit expressions and refer
to Section~\ref{subhggres} for numerical results.


\subsubsection{\label{subhggres}Results}

This section summarizes the current knowledge of hadronic Higgs decay by
presenting numerical results for the decay rates $\Gamma(H\to gg)$ and
$\Gamma(H\to b\bar{b})$.

Note that $C_1$ starts at order $\alpha_s$. Hence the combination
$C_1^2\mbox{Im}\langle {\cal O}_1^\prime {\cal O}_1^\prime\rangle$
governing the gluonic decay rate of the Higgs boson is available up to
${\cal O}(\alpha_s^4)$. Normalized to the Born rate it reads 
in numerical form ($\mu=M_H$):
\begin{eqnarray}
\frac{\Gamma(H\to gg)}{\Gamma^{\rm Born}(H\to gg)}
&\approx&
1+17.917\,\frac{\alpha_s^{(5)}(M_H)}{\pi}
+\left(\frac{\alpha_s^{(5)}(M_H)}{\pi}\right)^2
\left(156.808-5.708\,\ln\frac{M_t^2}{M_H^2}\right)
\nonumber\\
&\approx&
1+0.66+0.21
\,,
\label{eqhggfin}
\end{eqnarray}
with $\Gamma^{\rm Born}(H\to gg)
      =G_FM_H^3/(36\pi\sqrt2) \times (\alpha_s^{(5)}(M_H)/\pi)^2$.
In the last line $M_t=175$~GeV and $M_H=100$~GeV is inserted.
The analytical expressions can be found in~\cite{CheKniSte97hgg}.
We observe that the new ${\cal O}(\alpha_s^2)$ term further increases the
well-known ${\cal O}(\alpha_s)$ 
enhancement~\cite{InaKubOka83,DjoSpiZer91} by about one third.
If we assume that this trend continues to ${\cal O}(\alpha_s^3)$ and beyond,
then Eq.~(\ref{eqhggfin})
may already be regarded as a useful approximation to the
full result.
Inclusion of the new ${\cal O}(\alpha_s^2)$ correction leads to an increase of 
the Higgs-boson hadronic width by an amount of order 1\%.

Concerning the decay rate into quarks we restrict ourselves
to the case of bottom quarks. Inserting numerical values into the 
coefficient functions $C_1$ and $C_2$ and the correlators
$\mbox{Im}\langle {\cal O}_1^\prime {\cal O}_2^\prime\rangle$~\cite{CheSte97}
and
$\mbox{Im}\langle {\cal O}_2^\prime {\cal O}_2^\prime\rangle
$~\cite{Che97higgs,CheSte97}
leads to:
\begin{eqnarray}
\Gamma(H\to b\bar{b}) 
&=& 
A_{b\bar{b}}\Bigg\{
1
+ 5.667 \,  a_H^{(5)}
+ 29.147 \, \left(a_H^{(5)}\right)^2
+ 41.758 \left(a_H^{(5)}\right)^3
\label{eqbb}
\\&&\mbox{}
+\left(a_H^{(5)}\right)^2
\left[
3.111
-0.667\,L_t
\right] 
+\left(a_H^{(5)}\right)^3
\left[
50.474
-8.167\,L_t
-1.278\,L_t^2
\right]
\Bigg\},
\nonumber
\end{eqnarray}
with $A_{b\bar b}=3G_FM_Hm_b^2/(4\pi\sqrt{2})$,
$L_t=\ln M_H^2/M_t^2$ and $a_H^{(5)}=\alpha_s^{(5)}(M_H)/\pi$.
In Eq.~(\ref{eqbb})
electromagnetic~\cite{Kat97} and 
electroweak~\cite{KniSte95,CheKniSte97hbb}
corrections have been neglected. Also 
mass effects~\cite{HarSte97}
and second order QCD corrections which are suppressed by the 
top quark mass~\cite{LarRitVer95CheKwi96}
are not displayed.
One observes from Eq.~(\ref{eqbb}) that the top-induced corrections
at ${\cal O}(\alpha_s^3)$ (second line) are of the same order of magnitude
as the ``massless'' corrections (first line).

%% file: strongW.tex
\subsection{Strongly interacting vector bosons\label{sec:applic:strongW}}

As an application for the package {\tt CompHEP} let us consider a
scenario where the violation of unitarity in elastic scattering of
massive vector bosons at high energies is avoided by letting them become
strongly interacting.  This is an alternative approach to the Standard
Model solution of introducing a scalar Higgs boson. In~\cite{BHKPYZ98} a
new fundamental strong interaction among the vector bosons is assumed
that occurs at a scale of order 1~TeV.  The model is described by a
Lagrangian with global chiral symmetry which is spontaneously broken in
order to generate the masses of the vector bosons.  The corresponding
interaction terms have been worked out and implemented in {\tt CompHEP}.
Although this would have been much simpler in unitary gauge due to the
absence of ghost particles, the authors of~\cite{BHKPYZ98} decided to
use 't~Hooft-Feynman gauge as then the size of the final expressions is
much smaller.  One possibility to test the new interaction terms are
elastic and quasi-elastic $2\to 2$ scattering experiments with $W^\pm$
and $Z$ bosons which requires colliders with a center-of-mass energy in
the TeV range. Indeed, a future $e^+e^-$ linear collider is expected to
operate at energies of 1.5 to 2~TeV in a second stage~\cite{Acc97}.

\begin{figure}[h]
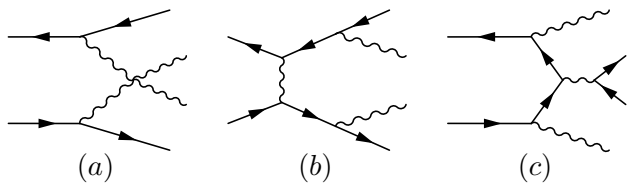

  \leavevmode
  \begin{center}
    \begin{tabular}{ccc}
      \epsfxsize=2.5cm
      \parbox{1cm}{\epsffile[0 0 114 86]{wwgraphs.7}}
      &
      \epsfxsize=2.5cm
      \parbox{1cm}{\epsffile[0 0 114 86]{wwgraphs.9}}
      &
      \epsfxsize=2.5cm
      \parbox{1cm}{\epsffile[0 0 114 86]{wwgraphs.13}}
      \\
      $(a)$&$(b)$&$(c)$
    \end{tabular}
    \parbox{\captionwidth}{
      \caption[]{\label{figstrongwdia}
        Sample diagrams contributing to processes specified in
        Eq.~(\ref{eqstrongw1}). Diagram $(a)$ is part of 
        the signal of the vector boson scattering process whereas $(b)$ and
        $(c)$ belong to the background. 
        }}
  \end{center}
\end{figure}

The following processes are considered
in~\cite{BHKPYZ98}:
\begin{eqnarray}
e^+e^- \to \nu_e \bar{\nu}_e W^+ W^-
&\quad:&
W^+ W^- \to W^+ W^-
\nonumber\\
e^+e^- \to \nu_e \bar{\nu}_e Z Z
&\quad:&
Z Z \to Z Z
\nonumber\\
e^-e^- \to \nu_e \bar{\nu}_e W^- W^-
&\quad:&
W^- W^- \to W^- W^-
\nonumber\\
e^+e^- \to e^- \bar{\nu}_e W^+ Z
&\quad:&
W^+ Z \to W^+ Z
\nonumber\\
e^+e^- \to e^+ e^- Z Z
&\quad:&
Z Z \to Z Z
\label{eqstrongw1}
\end{eqnarray}
where after the colon the subprocess
characterizing the vector boson scattering is given. However, not only
those diagrams that contain the desired subprocess contribute.
Instead, also background diagrams must be computed where, for example, the
final state vector bosons are radiated off the fermion lines or one of
the massive vector bosons is replaced by a photon.  Furthermore,
diagrams are involved where the neutrinos result from the decay of a $Z$
boson which is in turn generated from a $W^\pm$ pair.  In
Fig.~\ref{figstrongwdia} three sample diagrams are listed.

\begin{figure}
\begin{center}
\includegraphics{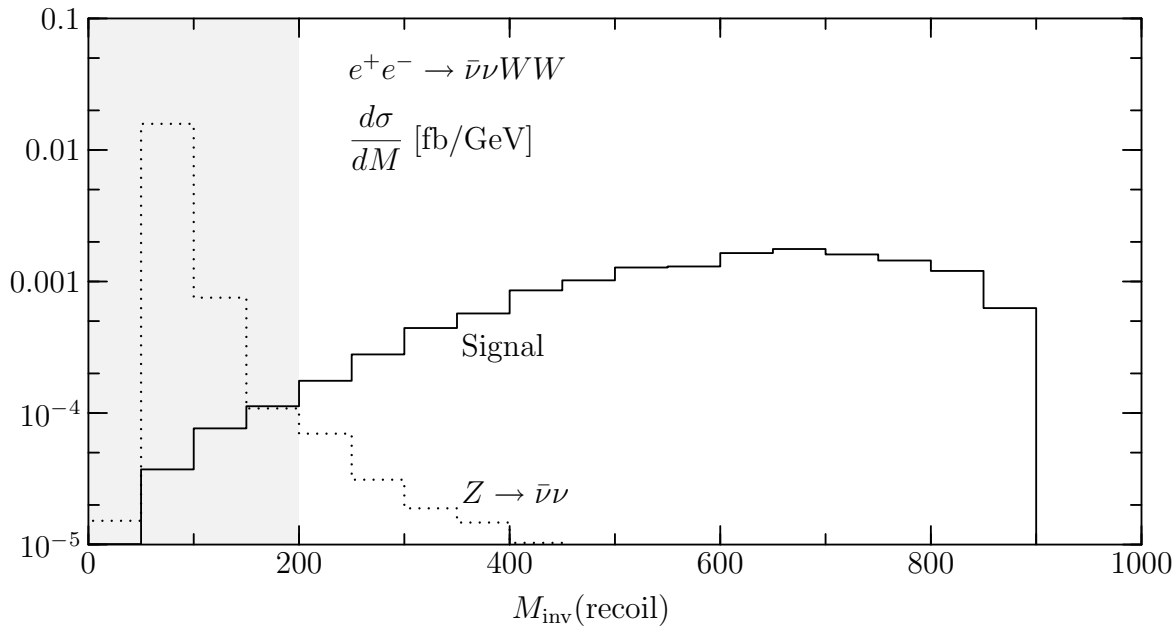}
%
%
\parbox{\captionwidth}{
\caption[]{\label{figstrongw}
  Distribution of the invariant mass of the produced vector bosons
  (here generically called $W$) 
  in the process $e^+e^-\to W^+W^-\bar\nu_e\nu_e$ (signal).
  The cut (shaded area) removes events in which the neutrinos are
  generated through $Z$ decays.}} 
\end{center}
\end{figure}

It is straightforward to go through the menus of {\tt CompHEP}, to enter
the different processes, and to produce {\tt FORTRAN} output for the
squared matrix elements.  This is then used in the numerical part of
{\tt CompHEP} to compute the cross sections.  The first run contains no
cuts, but computes all matrix elements corresponding to the generated
set of four-momenta.  In subsequent runs, cuts are applied that reduce
the background and isolate the signal. The data computed in the first
run are taken over so that the CPU time is significantly smaller in this
second step, as already mentioned in Section~\ref{subcomphep}. For
example, the cuts include a lower limit on the invariant mass of the
$\nu_e\bar{\nu_e}$ system, or the selection of central events in
combination with cuts on the transverse momentum of the $W^\pm$,
respectively, the $Z$ boson.  Fig.~\ref{figstrongw} shows how the number
of events containing neutrinos from $Z$ boson decay (c.f.
Fig.~\ref{figstrongwdia}($c$)) is reduced.  Figures of this kind can
easily be produced by using the data files generated by {\tt CompHEP}.

This is not the place to go into details concerning the physical
consequences for which we shall refer to~\cite{BHKPYZ98}.  The key
point, however, is that the couplings of the new interaction terms can
be related to the $2\to 2$ vector boson scattering amplitudes.  Thus,
the investigation of the corresponding total cross sections as well as
several other distributions like the one for invariant masses or
transverse momenta provide important tests on the mechanism responsible
for restoring unitarity at high energies.

%
%

%% file: delta_r.tex
%
%
%
\subsection[Higgs mass dependence of two-loop corrections to $\Delta
r$]{Higgs mass dependence of two-loop corrections to \bld{\Delta
    r}\label{sec:applic:deltar}}
As was already pointed out in the introduction, the dependence of
radiative corrections on yet undiscovered particles may be exploited in
order to gain information on the masses of these particles.  This has
been performed very successfully for the top quark and nowadays, as
$M_t$ is measured with reasonable accuracy, the same strategy is used to
pin down the mass of the Higgs boson.  An important quantity in this
respect is $\Delta r$, which comprises the radiative corrections to muon
decay. In this way it relates four fundamental parameters of the
Standard Model to each other --- the electromagnetic coupling $\alpha$,
the Fermi constant $G_F$ and the masses of the $W$ and $Z$ boson:
\begin{eqnarray}
M_W^2\left(1-{M_W^2\over M_Z^2}\right) &=& {\pi\alpha\over
    \sqrt{2}G_F}(1+\Delta r)\,.
\end{eqnarray}
$G_F$, $\alpha$ and $M_Z$ are very well measured quantities,
and the experimental accuracy of $M_W$ is expected to be
considerably improved within the near future by the current and upcoming
measurements at LEP2 and TEVATRON.

Therefore, precise knowledge of the $M_H$-dependence of $\Delta r$ is
highly desirable. The one-loop corrections to $\Delta r$ are known
analytically since long~\cite{MarSir80}, and two-loop corrections up to
next-to-leading order in $1/M_t$ were considered
in~\cite{DegGamVic96,DegGamSir97,DegGamPasSir98}.  It was the concern
of~\cite{BauWei97} to precisely account for the $M_H$ dependence of the
two-loop diagrams containing the top quark and the Higgs boson.  The
idea was to get a result valid for arbitrary values of $M_t$ and $M_H$
without performing an expansion in these masses.  This obviously leads
to two-loop multi-scale diagrams, their number being of the order of one
hundred.  Not only the large number of diagrams but also the complicated
tensor structure and the evaluation of the scalar two-loop integrals
makes the use of highly automated software indispensable.  A further
technical problem arises from the renormalization which has to be known
up to two loops in the electroweak sector of the Standard Model.  In
order to investigate the Higgs-mass dependece of $\Delta r$,
in~\cite{BauWei97} the subtracted quantity
\begin{eqnarray}
\Delta r(M_H) - \Delta r(M_H=65~{\rm GeV})
\end{eqnarray}
was considered which describes the change in the prediction for $\Delta
r$ when $M_H$ is varied.  The value for $\Delta r(M_H=65~{\rm GeV})$ was
taken from \cite{DegGamPasSir98}.

It turned out that the computation of the diagrams contributing to the
considered corrections to $\Delta r$ could be reduced to two-loop
tadpole integrals and two-point functions. This apparently is a task for
the {\tt Mathematica} programs {\tt FeynArts} and {\tt TwoCalc}.  The
various contributions to $\Delta r$ together with the corresponding
counter-terms were generated by means of {\tt FeynArts}.  Contraction of
Lorentz indices, evaluation of the Dirac algebra and the two-loop tensor
reduction was performed by {\tt TwoCalc}.  The resulting expressions
were split into finite and divergent pieces which explicitly
demonstrates the cancellation of poles.  The scalar one- and two-loop
integrals were reduced to one-dimensional integrals allowing a fast
numerical evaluation to high accuracy with the help of special {\tt C}
routines which themselves were fully integrated into the {\tt
  Mathematica} environment~\cite{Bauetal}.  CPU time for this
calculation added up to the order of a few days.

\begin{figure}
  \begin{center}
    \leavevmode
    \epsfxsize=10.cm
    \epsffile{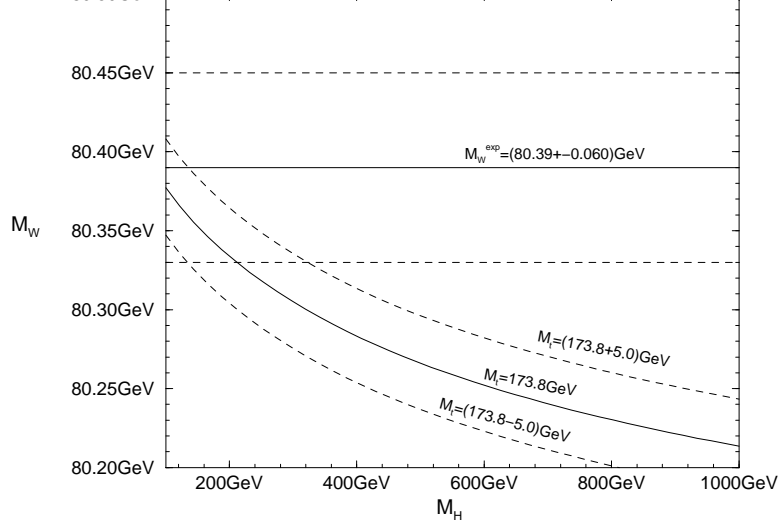}
    \hfill
    \parbox{\captionwidth}{
    \caption[]{\label{mwplotcmpl.eps}\sloppy
      Limits on the Higgs mass depending on the values of the top and
      $W$ mass~\cite{WeiBarc98}. The value of $\Delta r(M_H=65~{\rm
      GeV})$ was taken from~\cite{DegGamPasSir98}.
      \label{fig::MH-MW}
      }}
  \end{center}
\end{figure}

Instead of displaying the full result, we only note that the resulting
prediction for the $W$ mass agrees within a few MeV with the one of the
heavy-top expansion~\cite{DegGamPasSir98}, which is far
below the experimental precision at the moment. (For a more detailed
discussion see \cite{WeiBarc98} and \cite{GamBarc98}.) It is remarkable
that the top-induced corrections turn out to be maximal at the physical
value of the top quark mass within a range of $\pm 50$~GeV around this
physical value.  As shown in Fig.~\ref{fig::MH-MW}, the interrelation
between $M_W$ and $M_t$ favours relatively small values for $M_H$,
consistent with other indirect determinations using the radiative shift
in the weak mixing angle or the leptonic decay rate of the $Z$ boson.
Both of these quantities recently have been calculated with similar
techniques as described in this section~\cite{WeiRhein98}.

%% file: acknowledge.tex
\section*{Acknowledgements} 
First of all we would like to thank J.H.~K\"uhn for his kind advice and
numerous valuable suggestions and discussions.  We are grateful to the
following colleagues for carefully reading the manuscript (or parts of
it) and their expert advice on various subjects: P.~Baikov,
K.G.~Chetyrkin, P.~Gambino, T.~Hahn, W.~Kilian, B.A.~Kniehl,
K.~Melnikov, A.~Pukhov, A.~R\'etey, T.~van~Ritbergen, T.~Seidensticker,
V.A.~Smirnov, R.G.~Stuart, M.~Stuber, O.V.~Tarasov, and G.~Weiglein.
This work was supported in part by {\it Graduiertenkolleg
  ``Elementarteilchenphysik an Beschleunigern''} and {\it
  DFG-Forschergruppe ``Quantenfeldtheorie, Computeralgebra und
  Monte-Carlo-Simulationen''} at the University of Karlsruhe, and the
{\it Schweizer Nationalfond}.